\documentclass[12pt,preprint]{aastex}
\usepackage{enumerate}
\usepackage{float}
\usepackage{placeins}
\usepackage{amsmath}
\usepackage{natbib}
\usepackage{color}
\usepackage[utf8x]{inputenx}
\usepackage[english]{babel}
\usepackage{footnote}
\usepackage{pdflscape}

\newcommand{\HII}{H\,{\sc ii}~}

\newcommand\adndt{Atomic Data \& Nuclear Data Tables}
\newcommand\jphb{J. Phys. B}
\newcommand\cjp{Can. J. Phys.}

\def\4959_5007{[\ion{O}{3}]~$\lambda \lambda$4959,5007}
\def\OIII49595007{[\ion{O}{3}]~$\lambda \lambda 4959,5007$}
\def\ratioR23{([\ion{O}{2}]~$\lambda \lambda$3727,9 +
[\ion{O}{3}]~$\lambda\lambda$4959,5007)/H$\beta$}
\def\R23{${\rm R}_{23}$}
\def\dS23{${\rm S}_{23}$}
\def\OIIIl{[\ion{O}{3}]~$\lambda$5007}

\def\mum{$\mu$m}

\def\lOII{[\ion{O}{2}]~$\lambda \lambda$3726,9}

\def\NIIOII{[\ion{N}{2}]/[\ion{O}{2}]}
\def\ratioNIIOII{[\ion{N}{2}]~$\lambda$6584/[\ion{O}{2}]~$\lambda \lambda$3727,29}

\def\ratioS23{([\ion{S}{2}]~$\lambda \lambda$6717,31 +
[\ion{S}{3}]~$\lambda\lambda$9069,9532)/H$\beta$}

\def\Hb{{H$\beta$}}
\def\O4363{[{\ion{O}{3}}]~$\lambda$4363}
\def\OIII{[{\ion{O}{3}}]}

\defcitealias{Nicholls12}{NDS12}

\begin{document}

\title{New Strong Line Abundance Diagnostics for \HII Regions: Effects of $\kappa$-Distributed Electron Energies and New Atomic Data}
\shorttitle{put short title here}
\shortauthors{Dopita et al.}

\author{ Michael A. Dopita\altaffilmark{1}\altaffilmark{2}\altaffilmark{3}, Ralph S. Sutherland\altaffilmark{1}, David C. Nicholls\altaffilmark{1}, Lisa J. Kewley\altaffilmark{1}\altaffilmark{3} \& Fr\'ed\'eric P. A. Vogt\altaffilmark{1}}
\email{Michael.Dopita@anu.edu.au}
\altaffiltext{1}{Research School of Astronomy and Astrophysics, Australian National University, Cotter Rd., Weston ACT 2611, Australia }
\altaffiltext{2}{Astronomy Department, King Abdulaziz University, P.O. Box 80203, Jeddah, Saudi Arabia}
\altaffiltext{3}{Institute for Astronomy, University of Hawaii, 2680, Woodlawn Drive, Honolulu, HI 96822}

\begin{abstract}
Recently, \citet{Nicholls12}, inspired by \emph{in situ} observations of solar system astrophysical plasmas, suggested that the electrons in \HII regions are characterised by a $\kappa$-distribution of electron energies rather than by a simple Maxwell-Boltzmann distribution. Here we have collected together the new atomic data within a modified photoionisation code to explore the effects of both the new atomic data and the $\kappa$-distribution on the strong-line techniques used to determine chemical abundances in \HII regions. By comparing the recombination temperatures ($T_{\rm rec}$) with the forbidden line temperatures ($T_{\rm FL}$) we conclude that $ \kappa \sim 20$. While representing only a mild deviation from equilibrium, this is sufficient to strongly influence abundances determined using methods which depend on measurements of the electron temperature from forbidden lines. We present a number of new emission line ratio diagnostics which cleanly separate the two parameters determining the optical spectrum of \HII regions  - the ionisation parameter $q$ or $\cal{U}$ and the chemical abundance; 12+log(O/H). An automated code to extract these parameters is presented. Using the homogeneous dataset from \citet{vanZee98}, we find self-consistent results between all these different diagnostics. The systematic errors between different line ratio diagnostics are much smaller than was found in the earlier strong line work. Overall the effect of the $\kappa$-distribution on the strong line abundances derived solely on the basis of theoretical models is rather small. 
\end{abstract}


\section{Introduction}
Much of what we know - or believe we understand - about the chemical evolution history of the Universe depends upon the interpretation of the strong emission lines originating from \HII regions in distant galaxies. These emission lines enable us to investigate the metallicity evolution of the Universe as a whole \citep{Nagamine01,DeLucia04, Kobayashi07,Yuan13}. We may also measure the mass-metallicity relationship of disk galaxies (see \citet{Kewley08} and references therein), understand how chemical abundance gradients are formed and maintained \citep{Bothun84,Wyse85,Skillman89,Vila-Costas92,Zaritsky94,vanZee98} and discover how abundance gradients can be removed in galaxy interactions \citep{Kewley10,Rupke10,Torrey12, Rich12}. Chemical abundances also encode information about the history of star formation, mass infall, and radial mixing driven by viscous processes in galactic disks. 

The ``classical" technique to derive chemical abundances first uses the observed temperature, $Te$, derived from temperature sensitive line ratios such as \O4363/ \OIIIl in combination with the ratio of a strong forbidden line of the element of interest to a Balmer line of hydrogen to estimate the abundance of the
observed ion of a given atomic species. For those ions which do not produce an observable line in the optical, we estimate an ionisation correction factor (ICF) based upon theory to account for the abundance of the unobserved ions of the same atomic species. This $T_e + $ ICF technique has been very extensively used over the past 50 years since it was first devised by \citet{Aller59} and further developed by \citet{Peimbert67} and \citet{Peimbert69}. 

Later on, strong line techniques were developed which simply rely on the ratio of forbidden lines to the hydrogen recombination lines, a natural enough approach if we believe that such ratios encode information about the relative abundance of the ion considered compared to hydrogen. The most commonly used ratios are often referred to by their shorthand contractions; $R_2 =$\lOII/\Hb\ , $R_3 =$\OIIIl/\Hb\ , $R_{23}=R_2+R_3$ \citep{Pagel79}, and similar ratios involving S, for example $S_{23} =$ \ratioS23  \citep{Diaz00,Oey02}. The abundance calibration for these ratios may either be based purely on photoionisation models, or on an empirical calibration based upon alignment of the abundance scale to objects for which the $T_e + $ ICF technique has been used. Strong line techniques which depend on ratios involving hydrogen recombination lines suffer from a fundamental ambiguity, in that these line ratios decreases as both the low- and high- abundance ends, requiring some other technique to resolve which ``branch'' the observed \HII region lies.

Despite the best efforts of theoreticians, the degree of scatter among the different strong emission line diagnostics remains alarming (\emph{e.g.} \citet{Kewley02,Kewley08,LopezSanchez12}), and this scatter between the various strong line techniques is strongly dependent on the approach used to estimate abundances from such ratios. Some rely entirely upon the construction of theoretical \HII\ region models using stellar model atmospheres and photoionisation codes \citep{McGaugh91,Dopita00}. Others seek to calibrate $R_{23}$ or $S_{23}$ in terms of abundances derived from objects for which there are direct measurements of electron temperature $T_e$ from temperature sensitive line ratios such as \O4363/\OIIIl. This method, was pioneered by \citet{Pagel79} and \citet{Alloin79}. This approach was used by \citet{Diaz00}, and has been investigated in great detail in a series of papers by Pilyugin and his collaborators \citep{Pilyugin05,Pilyugin11} culminating  in the  ``counterpart" method of \citet{Pilyugin12}.  A slightly different approach was used by \citet{vanZee98}, who computed the oxygen abundances with the \citet{McGaugh91} method, using the \ion{N}{2} line to distinguish between the high- and low- abundance branches, and used this to propose a calibration of  \ion{N}{2}/H$\alpha$. This is  especially useful in the ``turnover region" between the high- and low- abundance branches. 

In addition to the uncertainties depending on whether a given \HII region lies on the high- or low- abundance branch, and those associated with the theoretical or observational calibrations, considerable uncertainty is introduced into the derived strong line abundances in those cases where the ionisation parameter is not explicitly solved for as a separate variable \citep{Pagel79,Alloin79,Zaritsky94,Pettini04}. The first to properly account for the effect of the ionisation parameter were \citet{Evans85,Evans86}, followed by \citet{McGaugh91}. Although many bright \HII regions are observed to have similar ionisation parameters, the observed scatter in $\cal{U}$ or $q$  introduces significant error into the abundance estimates.

Even when all of these effects are accounted for, there remains a significant offset between those strong line techniques based purely upon photoionisation models \citep{McGaugh91,Kewley02,Kobulnicky04}, and those based upon an empirical alignment of the strong line intensities to abundances derived in objects for which $T_e$ has been directly estimated \citep{Pilyugin01a,Pilyugin01b,Bresolin04,Pilyugin05,Pilyugin12}. The offset can be large, amounting to 0.3-0.5 dex, in the sense that the strong-line abundances estimated from a $T_e$ abundance calibration are systematically lower than those estimated from models. This is clearly highlighted in \citet{LopezSanchez12}. The cause of this offset is the same as the systematic difference seen between abundances derived from model-based strong line techniques, and those obtained by the traditional method of using the $T_e + $ ICF technique. 

Three possibilities have been advanced to explain the offset in abundances between the strong line and the $T_e + $ ICF techniques:
\begin{itemize}
\item The models predicting the strong lines not deliver the correct electron temperature either because they do not consider all the necessary physics, or because the physical data they use is incorrect or incomplete.
\item Fluctuations or gradients in the electron temperature systematically bias the estimate of the temperature derived from line ratios such as \O4363/ \OIIIl ~\citep{Peimbert69}.
\item The measured electron temperature suffers from systematic errors relating to the choice of the input atomic data used to derive it. The size of such errors was recently quantified by \citet{Nicholls13}.
\end{itemize}

In order to eliminate the possibility that the strong line techniques have some systematic error, \citet{LopezSanchez12} subjected a set of photoionisation models from the MAPPINGS code \citep{Sutherland93, Allen08} to a double-blind derivation of the abundances using the $T_e + $ICF technique. This demonstrated that, for those species for which the optical line emission arises principally in the low-ionisation zone of the \HII region, N, S and Cl, the abundances input into the models could be recovered through the  $T_e + $ICF analysis. However, for those species arising principally in the high-ionisation zone of the \HII region - O, Ne and Ar - a systematic shift of 0.2-0.3 dex is found. This is in the same sense as the offset observed for real \HII regions, the abundances estimated from a $T_e+$ICF calibration are systematically lower than those used in the photoionisation models. Thus, at least part of the disagreement between the two techniques is due to real temperature gradients which exist in the high-ionisation zones of the \HII regions.

These large-scale temperature gradients play a role analogous to the small-scale temperature fluctuations proposed by \citet{Peimbert67} and used by very many others since then \emph{e.g.} \citet{Esteban02,GarciaRojas07,PenaGuerrero12}. \citet{Kingdon95} attempted to reproduce temperature fluctuations in the context of detailed phototionisation models, and the whole thorny issue of their existence and theoretical justification was discussed by \citet{Stasinska04}. Both temperature gradients and temperature fluctuations tend to increase the $T_e$ estimated from the \O4363/ \OIIIl ~ratio, because the emissivity of the \O4363 line is biased towards the regions of higher $T_e$. This results in a systematic underestimate of the total chemical abundances. There are obvious physical causes for temperature gradients such as hardening of the radiation field through radiative transfer effects and through the appearance or disappearance of important coolant ions, or through suppression of cooling by collisional de-excitation. Likewise, one can imagine physical causes of small-scale temperature fluctuations (usually characterised by a parameter, $t^2$, the mean square fractional temperature fluctuation). These could result from local turbulent heating, or else local shocks induced by colliding flows from ionisation fronts. Such microphysics is not currently captured in photoionisation codes.

Recently, \citet{Nicholls12}, inspired by \emph{in situ} observations of astrophysical plasmas in and beyond the solar system, suggested that the electrons in \HII regions may be characterised by a $\kappa$-distribution of electron energies rather than by a simple Maxwell-Boltzmann distribution. Such distributions arise naturally in plasmas where there exist long-range energy transport processes \citep{Livadiotis11a,Livadiotis11b}. These ``hot tail" electron distributions can arise from plasma waves, magnetic re-connection, shocks, super-thermal atom or ion heating (as in a stellar wind \HII region interaction zone) or by fast primary electrons produced by photoionisation with X-ray or EUV photons. In many ways the physical processes which may drive a microscopic $\kappa$-distribution of electrons is similar to that which may generate macroscopic temperature fluctuations, and neither can be discounted \emph{a priori}.

 \citet{Nicholls12} demonstrated that a $\kappa$-distribution enhances the emissivity of the \O4363 line relative to \OIIIl, again resulting in a tendency for $T_e$ to be systematically overestimated compared to the Maxwell-Boltzmann distribution case. Similar considerations apply to other temperature-sensitive line ratios, as demonstrated by \citet{Nicholls13}.  A cause for concern is that the newer fully-relativistic close-coupling calculations of atomic term energies and collision strengths such as those by \citet{Palay12} provide a different absolute calibration from that used hitherto for these same temperature-sensitive ratios, as shown in \citet{Nicholls13}.
 
The three suggestions listed above suggest that there are grounds for supposing that abundances derived from either strong line techniques or from the $T_e + $ICF analysis may be in error. Likewise, the solution to the abundance discrepancy problem is likely to be found in a combination of one or more of these three effects, in addition to the temperature gradient issue already discussed above. We need to systematically take into account the newer atomic data and its effect on the photoionisation models before we can investigate the effect of the $\kappa$-distribution of electron energies in these photoionisation models, or investigate how temperatures derived from line sensitive line ratios are changed by use of either the new atomic data or by the  application of a $\kappa$-distribution. 

The purpose of the current paper is to provide the first systematic and quantitative study of the effect of the $\kappa$-distribution on the strong-line abundance diagnostics, not only at optical wavelengths, but also insofar as the strong UV and IR lines are concerned. The rest of the paper is organised as follows. In Section \ref{Code} we discuss what changes we have made to our photoionisation code to incorporate both the new atomic data and the  $\kappa$-distribution of electron energies. In Section \ref{Models} we explain the parameters of the photoionisation models used in the grid of theoretical \HII regions. In section \ref{Results} we present a reference catalog of \HII region models, varying the abundance set, the ionisation parameter, and the value of $\kappa$. For each of the 324 models, we give the computed line intensities and a complete set of ionic and recombination temperatures. In Section \ref{Kappa} we estimate the likely value of $\kappa$ using high-quality observations of galactic and extragalactic \HII regions. In Section \ref{UV_IR} we discuss the effect of the $\kappa$-distribution on the intensities of the strong emission lines in both the UV and the far-IR regions of the spectrum. In Section \ref{Diagnostics} we present the results of the line ratio diagnostics, and present a number of new line ratio diagrams which enable us to cleanly separate the effects of both the chemical abundance, 12+$\log$(O/H), and the ionisation parameter, $q$, from strong-line spectra of \HII\ regions. In Section \ref{Application} we compare the abundances derived for real data for \HII regions using these diagnostic line ratios,  provide a code to derive from observed strong line ratios the ionisation parameter and to provide the abundance for a plausible range of $\kappa$ values. Finally, we compare our derived abundances with earlier work on these same \HII regions, and provide a preliminary estimate of the effect that the $\kappa$-distribution has in producing a systematic offset between the strong-line and $T_e + $ICF techniques of deriving abundances.
 
\newpage
\section{The MAPPINGS IV Code} \label{Code}

We have modified the MAPPINGS code \citep{Sutherland93, Allen08} to incorporate new non-thermal ($\kappa$) electron energy excitation \citep{Nicholls12, Nicholls13} and to bring up to date the atomic data and the Maxwell averaged collision strengths.  The number of ionic species treated as full non-LTE multi-level ions has increased from 37 to 43.  The multi-level atoms are modelled using 3 to 9 levels depending on the ionic configuration. In particular - of particular relevance to \HII region modelling - the species \ion{C}{1}--\ion{C}{4}, \ion{N}{1}--\ion{N}{5}, \ion{O}{1}--\ion{O}{6}, \ion{Ne}{3}--\ion{Ne}{5}, \ion{S}{2}--\ion{S}{4} are now uniformly handled.  \ion{Ne}{2} is still treated as a two level atom for the purpose of computing its important 12.8$\mu$m transition.

\subsection{Energy levels and fundamental constants}

We have adopted the 2010 CODATA concordance on fundamental constants \citep{Mohr12}.  All multi-level atom energy level data are converted from derived energies in ergs to the more fundamental  wavenumbers, in cm$^{-1}$.  The cm$^{-1}$ values were taken uniformly from the 2012 values in the  NIST Atomic Spectroscopy Database v2 \citep{Kramida12}, and are now independent of constants such as $h$ or the  value of the electron volt.  While the effect of the change in the values of constants are small compared to the uncertainties in the level energies, when comparing values to boundary thresholds in the computation, more stable floating point representations lead to more stable execution of the code.  The real benefit of this change is to reduce (though not eliminate) systematic differences in the treatment of different ionic species, by adopting a more uniform set of atomic values, not readily possible in earlier decades.

\subsection{Transition probabilities, Aji}

In addition to a uniform source of energy level data, recent advances in atomic structure calculations \citep{Tachiev99, Tachiev00, Tachiev01, Tachiev02} now suggest that the theoretical transition probabilities have become very accurate, even for forbidden M2 quadrupole transitions, such as [\ion{O}{3}] $\lambda 5006.8$.  In a study of the \ion{O}{3} transitions, \cite{Froese09} found excellent agreement at the level of 10\% or better.  With the advent of the NIST MCHFD database based on these multi-configuration relativistic transition probability calculations, we are now able to use MCHFD transition probabilities uniformly for all the multi-level ions used in our models.

\subsection{Collision Strengths}

For many of the multi-level atoms, the data used in MAPPINGS III \citep{Sutherland93, Allen08} were merely translated to the new code format. However for the important species, new electron energy-averaged collision strengths ($\Upsilon$) were calculated from the original energy resolved collision strength data ($\Omega$) (either published or supplied as a private communication by the authors) The ions for which this treatment applies are listed in \citet{Nicholls13}. This approach enables two key features:
\begin{enumerate}
\item
We can convolve the original $\Omega$ data with a Maxwell-Boltzmann distribution to obtain energy-averaged collision strengths $\Upsilon$ values for any temperature and at any resolution desired, removing interpolation errors that would arise if we depended only on published tabular data, and 
\item
We are also able to convolve with $\kappa$ non-thermal distributions and obtain directly the $\Upsilon_\kappa$ values.
\end{enumerate}

In order to be able to rapidly evaluate both the Maxwell-Boltzmann averaged collision strength, $\Upsilon$, and the equivalent $\kappa$ -distribution averaged collision strength, $\Upsilon_{\kappa}$, we fit cubic spline functions in the following way:
\begin{enumerate}
\item
High resolution integrals as a function, $f(T)$,  of the $\Omega$ data convolved with the electron distributions were computed, along with the local second derivative $f''$ at each point.
\item
A fitting method akin to the one described in \citet[p.120 et seq.]{Press07} was employed except that we used the actual second derivative from the high-resolution integrals instead of a least squares fit to $f$ at the subset of nodes.
\item
By fixing the second derivative with the physical integral, the fitting procedure then became one of choosing the location of the spline nodes that minimised the difference between the spline fit and the high-resolution integral data, evaluating the spline at the data points, usually 1000-3000 points.  With optimal manual adjustment, the global error between the spline with 17 points and the high-resolution data was less than 1e-3 and generally $\ll$ 5e-4 RMS, achieving approximately third order accuracy, and better than that in some regions.
\item
The temperature coordinates were normalised in a fashion similar to that used in CHIANTI 7.1 \citep{Dere97, Landi12} and originally proposed by \cite{Burgess92}, but here we use a more direct scaled temperature coordinate $x = T/(T + T_C)$, where $T$ is the temperature and $T_C$ is a characteristic temperature, chosen for each transition so that the main features of the Upsilon curve are well modelled.  $T_C$ is not directly related to the threshold energy, as used by \cite{Burgess92} and CHIANTI, but to structure in the $\Upsilon$ curve with energies characteristic of the major resonances in the underlying $\Omega$ data.
\end{enumerate}

This transform has the property of scaling the temperature to $0 \le x \le 1$, and by including spline nodes at or very near $x$ = 0 and 1, ensures that the cubic spline interpolation is very stable at extreme temperature values when transformed back to a physical temperature scale, eliminating the well known instability of cubic spline extrapolation.  In the atomic data fitting procedure, every spline fit to every transition was plotted and evaluated. 

\subsection{Multi-Level Atoms}

The high resolution Maxwell-averaged collision strength data used in \citep{Nicholls13} were adopted for \ion{O}{2} (5 levels), \ion{O}{3} (6 levels)), \ion{N}{2} (6 levels), \ion{S}{2} (5 levels) and \ion{S}{3} (6 levels).  For \ion{Ne}{3}, \ion{Ne}{4} and \ion{Ne}{5}, the spline fits given in CHIANTI 7.1 were transformed into the slightly different coordinate system used here. Lithium-like species, \ion{C}{4} (3 levels), \ion{N}{5} (3 levels) and \ion{O}{6} (3 levels) were fit from low resolution tabular data given in the literature. Other important multi-level species include \ion{C}{1} (5 levels), \ion{C}{2} (5 levels), \ion{C}{3} (5 levels), \ion{N}{1} (5 levels),  \ion{N}{3} (5 levels), \ion{N}{4} (4 levels), \ion{O}{1} (5 levels), \ion{O}{4} (5 levels), \ion{O}{5}, \ion{S}{1}, and \ion{S}{4} where less detailed data were used, but include temperature dependent data. Table \ref{t1} lists all the sources of data we have used in this work.

\begin{table}[htbp]
\caption{Literature sources used for collision strength data.}
\begin{center}
\begin{tabular}{ll}
\\ \hline \hline
Ion & Reference \\
\hline 
C I & Pequignot, D., \& Aldrovandi, S.~.M.~.V., 1976, \aap, 50, 141 \\ 
C II & Tayal, S.~S., 2008, \aap, 486, 629 \\
C III & Berrington, K.~A. et al., 1985, \adndt, 33, 195 \\
C III & Berrington, K.~A. et al., 1989, \jphb, 22, 665 \\ 
C IV & Liang, G.~Y. \& Badnell, N.~R., 2011, \aap, 528, A69 \\
N I & Tayal, S.~S., 2000, \adndt, 76, 191 \\
N I & Tayal, S.~S., 2006, \apjs, 163, 207 \\
N II & Tayal, S.~S., 2011, \apjs, 195, 12 \\
N III & Stafford, R.P., Bell, K.L. \& Hibbert, A., 1994, \mnras, 266, 715 \\
N IV & Ramsbottom, C.A., Berrington, K.A., Hibbert, A. \& Bell, K.L., \physscr, 1994, 50, 246 \\
N V & Liang, G.~Y. \& Badnell, N.~R., 2011, \aap, 528, A69 \\
O I & Bell, K.L., Berrington, K.A. \& Thomas, M.R.J., 1998, \mnras, 293, L83 \\
O I & Zatsarinny, O., \& Tayal, S.~S. 2003, \apjs, 148, 575 \\
O II & Tayal, S.~S., 2007, \apjs, 171, 331 \\
O III & Palay, E. et al., 2012, \mnras, 423, 35 \\
O IV & Blum, R.~D. \& Pradhan, A.~K., 1992, \apjs, 80, 425 \\
O V & Berrington, K.A., 2003, private communication, 13-Mar-03 \\
O V & Bhatia, A.~K. \& Landi, E., 2012, \adndt, in press \\
O VI & Liang, G.~Y. \& Badnell, N.~R., 2011, \aap, 528, A69 \\
Ne III & Landi, E. \& Bhatia, A.~K., 2005, \adndt, 89, 139 \\
Ne IV & Ramsbottom, C.~A., Bell, K.~L. \& Keenan, F.~P., 1998, \mnras, 293, 233 \\
Ne V & Badnell, N.~R. \& Griffin, D.~C., 2000, J.Phys.B, 33, 4389 \\
Si III & Galavis, M.~E., Mendoza, C.\& Zeippen, C.~J., 1995, \aaps, 111, 347\\
Si III & Galavis, M.~E., Mendoza, C.\& Zeippen, C.~J., 1998, \aaps ,133, 245 \\
Si III & Mendoza C., \& Zeippen C.~J., 1982, \mnras, 199, 1025 \\
S II & Tayal S.~S. \& Zatsarinny O., 2010, \apjs, 188, 32 \\
S III & Hudson, C.~E., Ramsbottom, C.~A. \& Scott, M.~P., 2012 \apj, 750, 65 \\
Ca V & Galavis, M.~E., Mendoza, C. \& Zeippen, C.~J., 1995, \aaps, 111, 347\\
Fe II & Nussbaumer, H, \& Storey, P.~J., 1980, \aap, 89, 308 \\
Fe II & Nussbaumer, H. \& Storey, P.~J., 1988, \aap, 193, 327 \\
\hline \hline
\end{tabular}
\end{center}
\label{t1}
\end{table}
\FloatBarrier
\subsection{Collisional excitation rates}
The collisional excitation rate from energy level 1 to 2, $R_{12}$, depends on the collision strength $\Omega_{12}(E)$ and the energy $E$ \emph{c.f.} \citet{Nicholls12};
\begin{equation}\label{e1}
R_{12}=n_e N_1  \frac{h^2}{8 \pi m_e g_1}\int\limits_{E_{12}}^\infty\frac{\Omega_{12}(E)}{\sqrt E}\ f(E) \mathrm{d}E\ 
\end{equation}
where $h$ is the Planck constant, $m_e$ the mass of the electron $g_1$ the statistical weight of the lower level, $f(E)$ the energy distribution function, $N_1$ the number density of atoms in the ground state and $n_e$ is the electron density.

The collisional excitation rate from level 1 to level 2 for a Maxwell-Boltzmann (M-B) distribution is given by
\begin{equation}\label{e2}
R_{12}(\mathrm{M-B})=n_e N_1 \frac{h^2 }{4 \pi^{3/2} m_e g_1} \left(k_B T_U \right)^{-3/2} \int\limits_{E_{12}}^\infty \Omega_{12}(E)\ \exp \left[-\frac{E}{k_BT_U}\right] \mathrm{d}E\ ,
\end{equation}
and for a $\kappa$-distribution, the corresponding rate is:
\begin{equation}\label{e3}
R_{12}(\kappa)=n_e N_1 \frac{h^2}{4 \pi^{3/2} m_e g_1} \frac{\Gamma(\kappa+1)}{(\kappa-\frac{3}{2})^{3/2}\Gamma(\kappa-\frac{1}{2})} \left({k_BT_U}\right)^{-3/2}  \int\limits_{E_{12}}^\infty \frac{\Omega_{12}(E)}  {(1 + E/[(\kappa-\frac{3}{2}) k_BT_U)]^{\kappa + 1}} \mathrm{d}E\ .
\end{equation}
where $E_{12}$ is the energy gap between levels 1 and 2, $g_1$ is the statistical weight of the lower state, and $\Gamma$ is the gamma function.

If the detailed collision strengths, $\Omega (E)$, are known, equations \ref{e2} and \ref{e3} can be integrated numerically, and the $\kappa$ collisional excitation rate can be expressed in terms of the M-B collisional excitation rate.

As a first order approximation, we can assume that the collision strength for excitations from level 1 to 2, $\Omega_{12}$, is independent of energy. For this case the ratio of the rates of collisional excitation from level 1 to level 2 for a $\kappa$ distribution can be expressed analytically \citepalias{Nicholls12} as:
\begin{equation}\label{e4}
\dfrac{R_{12}(\kappa)}{R_{12}(\mathrm{M-B})}=  \frac{\Gamma(\kappa+1)}{(\kappa-\frac{3}{2})^{3/2}\Gamma(\kappa-\frac{1}{2})}\left(1-\frac{3}{2\kappa} \right)\exp \left[\frac{E_{12}}{k_BT_U}\right] \left( 1+\dfrac{E_{12}}{(\kappa-\frac{3}{2})k_BT_U)} \right) ^{- \kappa}\ .
\end{equation}

This equation can be evaluated analytically as a series of concave ``banana curves'' (see \citep{Nicholls12}, Figure 5).

When collision strengths are computed, in some cases \emph{only} the ``effective collision strengths'', $\Upsilon_{M-B} (T)$, computed assuming M-B equilibrium electron energies are published:
\begin{equation}\label{e5}
\Upsilon_{M-B}(T) = \dfrac{\int\limits_{E=E_{12}}^\infty \Omega_{12}(E)  \exp\left(\frac{-E}{kT}\right){\rm d}\left(\frac{E}{kT}\right)}{\int\limits_{E=E_{12}}^\infty \exp\left(\frac{-E}{kT}\right){\rm d}\left(\frac{E}{kT}\right)}\ .
\end{equation}

Where data on the detailed energy dependence of $\Omega$ are not available, a reasonable  approximation for the effective collision strengths $\Upsilon_\kappa$ for a $\kappa$ distribution can be calculated in terms of the  equilibrium effective collision strengths $\Upsilon_{M-B}$ using equation \ref{e4}:
\begin{equation}\label{e6}
\Upsilon_\kappa = \dfrac{R_{12}(\kappa)}{R_{12}(\mathrm{M-B})} \Upsilon_{M-B} .
\end{equation}

Equation \ref{e6} allows us to compute the $\kappa$ dependence of collisional excitation in terms of that for an equilibrium energy distribution, even where complete data on $\Omega$ is not available.

In the revised MAPPINGS code, as described above, where detailed data for $\Omega$ are available (see \citet{Nicholls13}), we compute the effective collision strengths $\Upsilon$ for temperatures between $10^3$ and $10^7$K, and express the effective $\kappa$ collision strengths $\Upsilon_\kappa$ in terms of the equilibrium $\Upsilon_{M-B}$s.  Where only M-B-averaged effective collision strengths are available, we compute $\kappa$ values using equation \ref{e6}.

\begin{figure}[htpb]
\includegraphics[scale=0.5]{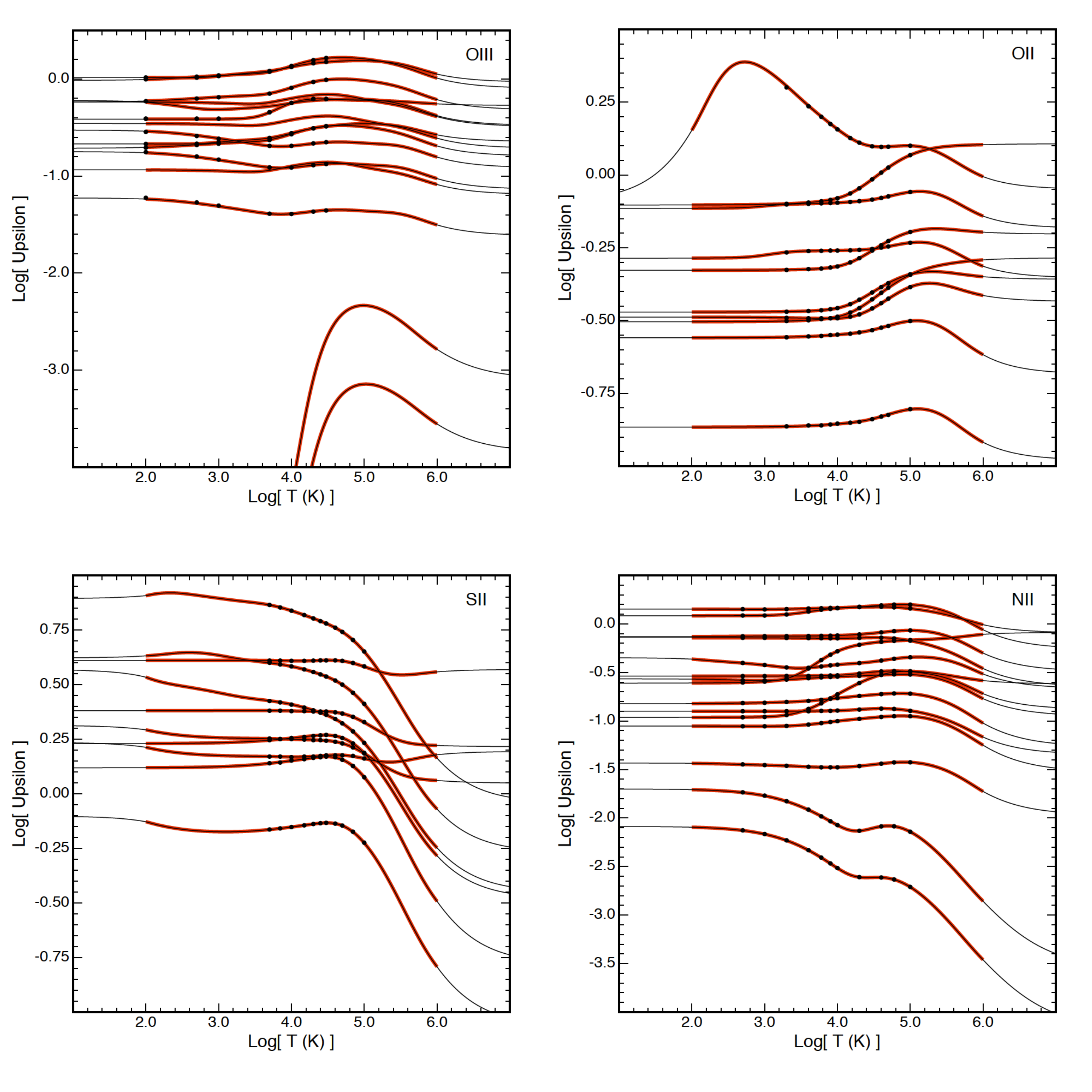}
\caption{The effective energy-averaged collision strengths, $\Upsilon$, for a Maxwell-Boltzmann electron energy distribution computed for the lowest transitions for \ion{O}{3}, \ion{O}{2}, \ion{S}{2} and \ion{N}{2}.  The dots are the values published in the literature; the thick red lines are our computations made at high energy resolution using equation \ref{e6}; and the thin black lines are our spline fits to this high resolution data. Note that our interpolation scheme provides well-behaved asymptotes as both the high- and low-temperature limits.}\label{fig1}
\end{figure}

To demonstrate the accuracy of the procedures used, in Figure \ref{fig1} we show the equilibrium effective collision strengths for the lowest 10 or 15 transitions for the ions \ion{O}{3}, \ion{O}{2}, \ion{S}{2} and \ion{N}{2}.  The dots are the values as published in the literature (see Table \ref{t1}); the thick red lines represent the high temperature resolution computations using equation \ref{e6}; and the thin black lines are the spline fits to the high resolution data, calculated as described earlier.

\begin{figure}[htpb]
\includegraphics[scale=0.8]{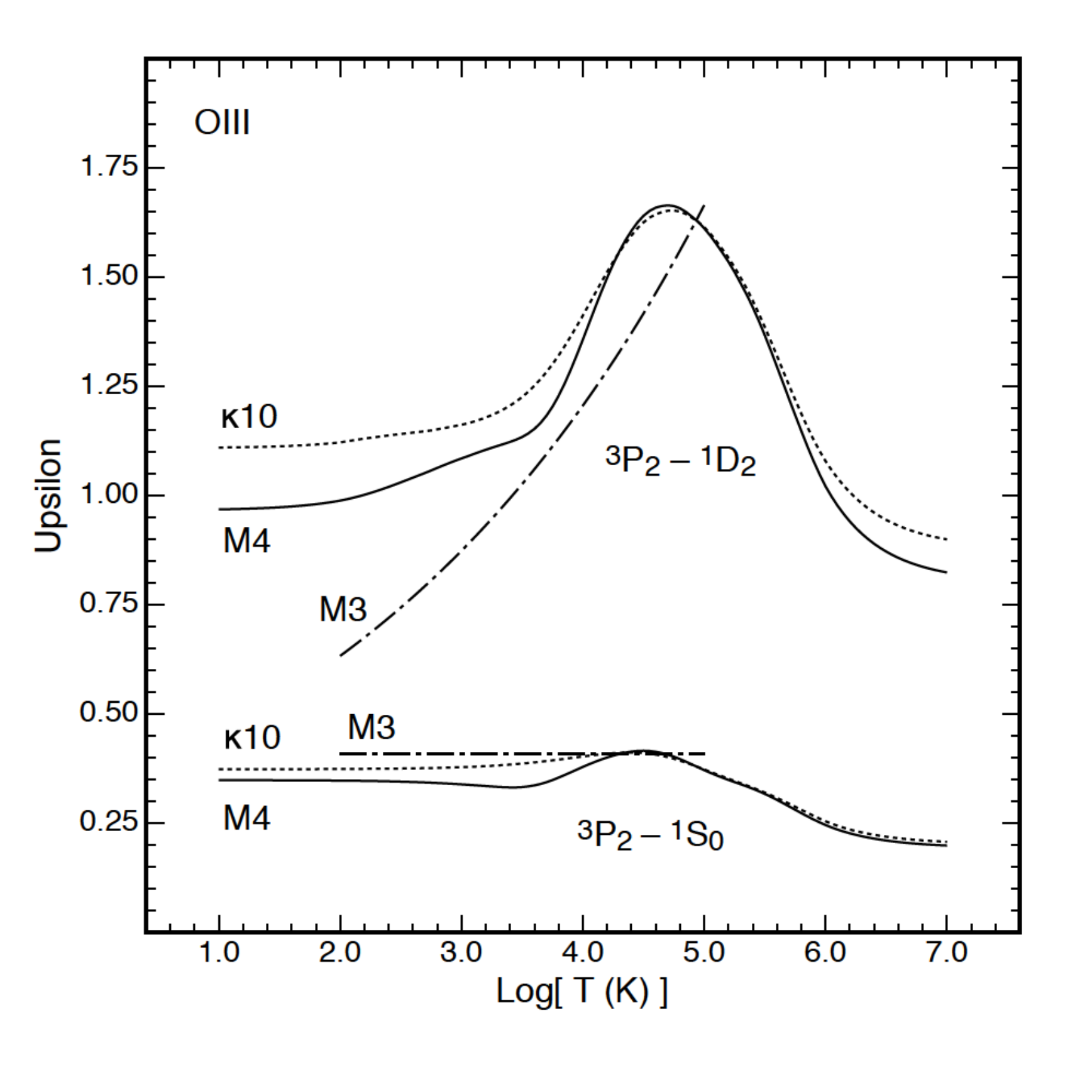}
\caption{Effective collision strengths for equilibrium (M3 and M4) and $\kappa$=10 electron energy distributions ($\kappa10$), for the $^3$P$_2$-$^1$D$_2$ and $^3$P$_2$-$^1$S$_0$ transitions in [\ion{O}{3}]. M3 indicates the effective collision strengths used in the previous version of MAPPINGS, M4 shows the new values. The differences are greatest at low temperature in the  $^3$P$_2$-$^1$D$_2$ transition, producing the greatest effect in high-abundance \HII regions.}\label{fig2}
\end{figure}

In Figure \ref{fig2} we show the effective collision strengths for the $^3$P$_2$-$^1$D$_2$ and $^3$P$_2$-$^1$S$_0$ transitions in [\ion{O}{3}].  M3 indicates the collision strengths used in the previous version of MAPPINGS.  M4 indicates the latest versions.  $\kappa$10 shows the effective collision strengths for a non-equilibrium electron energy distribution with $\kappa$=10.  While the  $^3$P$_2$-$^1$D$_2$ values are reasonably similar between 10$^{3.5}$ and 10$^4$ K, at lower temperatures the divergence is considerable between the older version and the new MAPPINGS data.  Extrapolating the older MAPPINGS data above 10$^5$K is likely to give severe errors.  These differences are likely to produce significant effects in models of X-ray ionized nebulae, or for models of the emission spectrum of material entering shock fronts.

\FloatBarrier

\section{The Model Grid}\label{Models}
\subsection{Abundance Set and Dust Physics}
The solar abundance set is taken from \citet{Grevesse10}, and the depletion factors for each element are updated from \citet{Dopita05} using the data from \citet{Kimura03}. These are listed in Table \ref{Z_table}. The elemental depletion results from condensation of the heavy elements onto dust grains. The treatment of dust grain composition, size distribution and absorption properties adopted here is essentially identical to that used in MAPPINGS 3, and is described in detail by \citet{Dopita05}. Suffice it to say here that within the ionised region, our dust model has silicate grains following a \citet{MRN} size distribution and a population of small carbonaceous grains. The polycyclic aromatic hydrocarbon molecules are assumed to be destroyed within the ionised \HII region (although they are present in the photodissociation regions around the periphery of the \HII region). The effects of radiation pressure acting on dust  \citep{Dopita02}, and photoelectric heating due to grain photoionisation \citep{Dopita00b} are fully taken into account in the models.

\begin{figure}[htpb]
\includegraphics[scale=1.0]{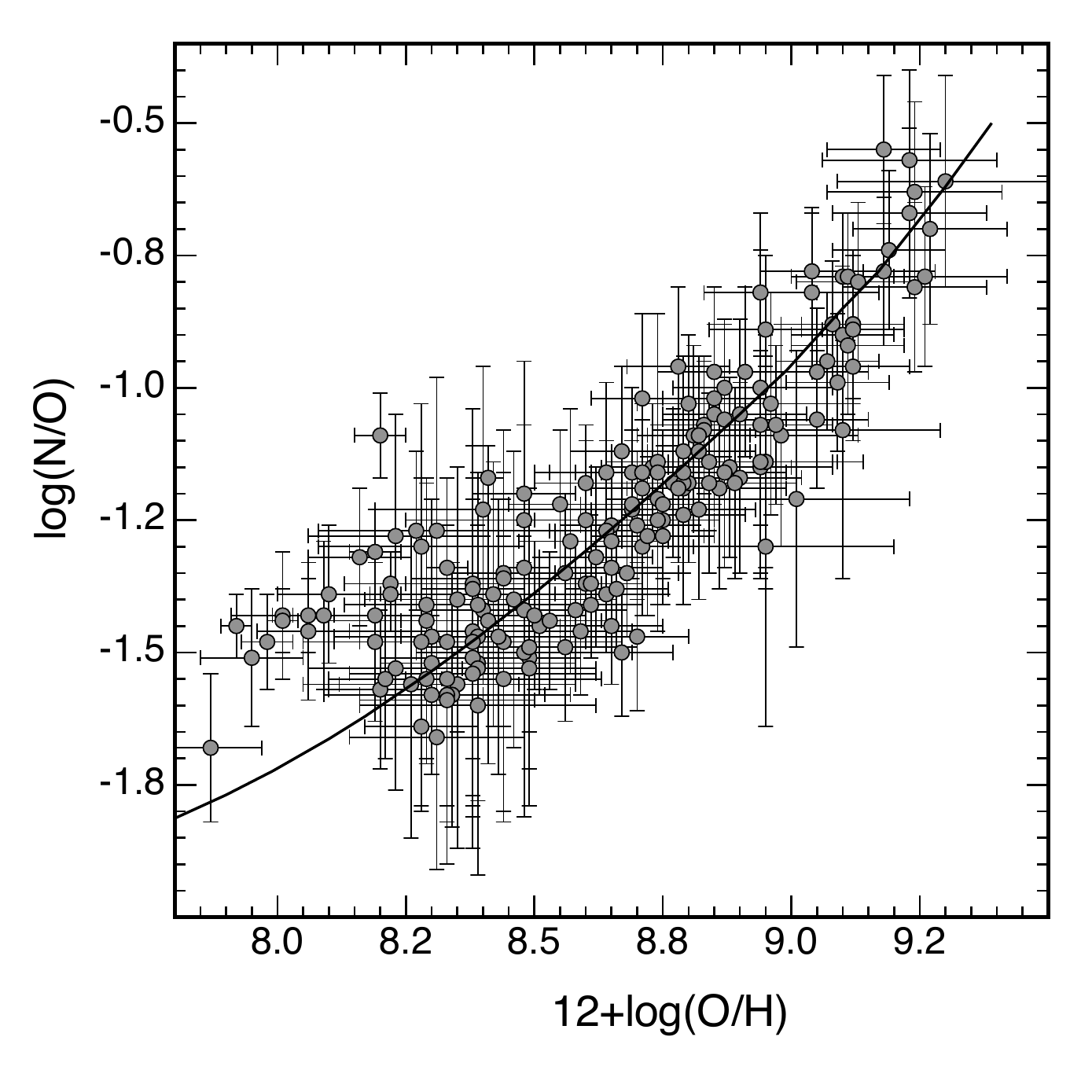}
\caption{The relationship between the oxygen abundance; $12+\log(\rm O/\rm H)$, and the N/O ratio for the \HII regions observed by \citet{vanZee98}. The functional relationship adopted in the models is indicated as a solid line, and is tabulated in Table \ref{Z_CN}. The accuracy of this calibration is central to the accuracy of the new strong-line diagnostics developed in this paper. Note that the increased scatter at the low abundnce end may pose an issue in the determination of the chemical abundances of low-abundance \HII regions.}\label{fig3}
\end{figure}

A perennial problem with fitting the spectrum of \HII regions over a wide range of abundances is the question of how to deal with the N and C abundances. Both of these elements contain a primary nucleosynthetic contribution as well as a secondary nucleosynthetic source which becomes important at higher abundance. In \citet{Dopita00} the transition from primary to secondary element was treated as more or less abrupt, but this does conform to the extensive observationally derived data of \citet{vanZee98} (for the N/O ratio) or to the data of \citet{Garnett04}, and references therein, for both N/O and C/O. In this work, we have adopted an empirical smooth function variation in both N/H and C/H as a function of O/H. This is listed in Table \ref{Z_CN}, and the fit for the adopted N/O ratio as a function of O/H is shown in figure \ref{fig3}, by comparison with the \citet{vanZee98} dataset. Note the increased scatter at the low abundance end, which may be of some importance to the accuracy of the strong line abundance diagnostics developed in this paper for $12+ \log({\rm O/H}) < 8.4$.

For helium, we adopt a similar prescription as used by \citet{Dopita02}, which provides a good fit to the He abundances observed in \HII regions. This has a primary production of He added to the primordial He abundance. By number of atoms,
\begin{equation*}
\frac{\rm He}{\rm H} = 0.0737+0.024\frac{Z}{Z_{\odot}}.
\end{equation*}

\FloatBarrier
\subsection{Stellar model atmospheres}
The stellar model atmospheres are based upon the STARBURST 99 code of \citet{SB99}. Here we have assumed a Salpeter initial mass function; $dN/dm \propto m^{-2.35}$, with a lower mass cutoff of $0.1M_{\odot}$ and an upper mass cutoff of $120M_{\odot}$,  as described in \citet{Dopita00}. We have used the \citet{Lejeune97} model atmospheres. For stars with strong winds we switch to the  \citet{Schmutz92} extended model atmospheres using the prescription of \citet{Leitherer95}. We assume that typical \HII regions are excited by a cluster with continuous star formation extending over 4 Myr. This approximation agrees with observed \HII regions since there is a strong bias towards observing the younger \HII regions, which have in general higher densities, much higher emissivities and larger absolute H$\alpha$ fluxes \citep{Dopita06a}. The models also provide a good approximation to the typical age spread of the stars observed in luminous clusters exciting bright \HII regions \citep{Beccari10,DeMarchi11}. 

We have elected to use the earlier STARBURST 99 models used by \citet{Dopita00}, as they provide a harder radiation field than the models generated by more recent versions of the code. These newer models incorporate  fully self-consistent radiatively-driven atmospheres, but they generate an EUV radiation field which is rather too soft to reproduce the \HII region sequence \citep{Dopita06b}. The most likely reason for this is that the stellar winds of OB stars are clumpy rather than smooth, which assists the escape of EUV photons.

A small difficulty with the STARBURST 99 models is that the ``solar" metallicity models do not correspond to the \citet{Grevesse10} abundance set.  For oxygen in particular, $12+\log(\rm O/\rm H)$ default solar abundance has changed from 8.86 to 8.69 -  nearly a factor of two lower. Thus, for most of the abundance sets that we would like to compute, the corresponding STARBURST 99 models are missing. To account for this, we generated stellar model atmospheres by linear interpolation of the logarithmic fluxes at any given frequency between the nearest adjacent STARBURST 99 models in metallicity, by assuming that the logarithm of the flux varies in proportion to  $12+\log(\rm O/\rm H)$. This allows us to construct a more finely sampled grid of models in which the stellar and the nebular chemical abundances are effectively identical, with the exception of the case  $0.05 Z_{\odot}$, where the atmosphere used is simply the lowest that can be obtained from the STARBURST 99 code, corresponding to $12+\log(\rm O/\rm H) \sim 7.56$ rather than 7.39.

\section{A reference catalog of \HII region models}\label{Results}
We have run a grid of spherical isobaric \HII region models at 5.0, 3.0, 2.0, 1.0, 0.5, 0.3, 0.2, 0.1 and $0.05 Z_{\odot}$, where the solar abundance corresponds to  $12+\log(\rm O/\rm H) = 8.69$. The full set of elemental abundances for solar abundance are listed in Table \ref{Z_table}. The models were terminated when more than 95\% of the hydrogen has recombined, and the temperature has fallen to less than 2000K. The pressure in the models was set at $P/k = 10^5$, typical of bright \HII regions in external galaxies. The density of these models, $n \sim 10$cm$^{-3}$, is fixed by the pressure, and is typical of giant extragalactic \HII regions. We do not need to consider models of different densities since, at the densities typically encountered in these \HII regions, collisional de-excitation is unimportant.

The ionisation parameter, $\log(q)$ \footnote{The dimensionless ionisation parameter $\cal{U}$ measures the ratio of the density per unit volume of ionising photons to the particle (atom plus ion) number density. In this paper, we use the alternative definition,  $q$, which is defined as the ratio of the number of ionising photons impinging per unit area per second divided by the gas particle number density. The transformation between the two definitions is simply ${\cal{U}} = q/c$, $c$ being the speed of light.} was fixed by its value at the inner boundary of the \HII region. For each set of abundances  $\log(q)$ ran from 8.5 down to 6.5 in steps of 0.25. Because these are spherical models, the radial divergence of the radiation field and attenuation of the radiation field by absorption in the ionised plasma within the models is important, especially at high ionisation parameter. By contrast, the low $\log(q)$ models approximate to a thin shell of ionised gas. As an example, for the $\log(q)=8.5$ model at solar abundance, the mean ionisation parameter in the ionised hydrogen is only $\log\left<q\right>=7.95$, while for the $\log(q)=6.5$ model at solar abundance, the mean ionisation parameter in the ionised hydrogen is $\log\left<q\right>=6.42$.

The full  $\log(q); Z$ grid was run for several different values of $\kappa=$ 10, 20, 50 and $\infty$ (which corresponds to the standard Maxwell-Boltzmann case), giving a complete family of 324 models covering the three independent variables which control the strong-line emission spectrum. The lines relevant to the abundance diagnostics, as well as those relevant for measuring electron temperatures in the nebulae are tabulated in Tables \ref{table_4} (the `blue' lines) and \ref{table_5} (which lists the lines in the red and near-IR portions of the spectrum). In these Tables, all line intensities are expressed as a fraction of the H$\beta$ intensity, to four significant figures. In the original models, the line fluxes are computed down to any intensity, but the spectral line list for each model gives lines with $F > 10^{-6}$ that of  H$\beta$. This allows an accurate computation of the effective forbidden line temperatures down to $T_e \sim 3000$K.

In order to compute the effective forbidden emission line temperatures, $T_{FL}$,  for the ions \ion{O}{3}, \ion{Ar}{3}, \ion{S}{3}, \ion{O}{2}, \ion{N}{2}, and \ion{S}{2}we have used the integrated line fluxes given by the models along with the fitting formulae given by \citet{Nicholls13}, which were obtained using the same atomic data. The temperature sensitive ratios used are as follows:  [\ion{O}{3}] $\lambda4363/\lambda5007$,  [\ion{Ar}{3}] $\lambda5192/\lambda7751$,  [\ion{S}{3}] $\lambda6312/\lambda9069$,  [\ion{O}{2}] $\lambda7318,30/\lambda3727,9$,  [\ion{N}{2}] $\lambda5755/\lambda6584$,  and [\ion{S}{2}] $\lambda4068,76/\lambda6731$.

In addition to the forbidden line temperatures, we have computed the effective recombination temperatures from the average ionic kinetic (or internal energy) temperatures, $T_{\rm U}$ in the H$^+$ and He$^+$ zones. In a $\kappa$-distribution, the Maxwell-Boltzmann ``core'', which determines the energy distribution of the recombining electrons, and hence the recombinations occur at a lower effective temperature, $T_{\rm  rec}$, than the internal energy temperature; $T_{\rm  rec}=T_{\rm U}(1-3/2\kappa)$ \citep{Nicholls12,Nicholls13}. We must therefore correct the average ionic kinetic temperatures given by the code by this factor. 

\begin{figure}[htpb]
\includegraphics[scale=0.85]{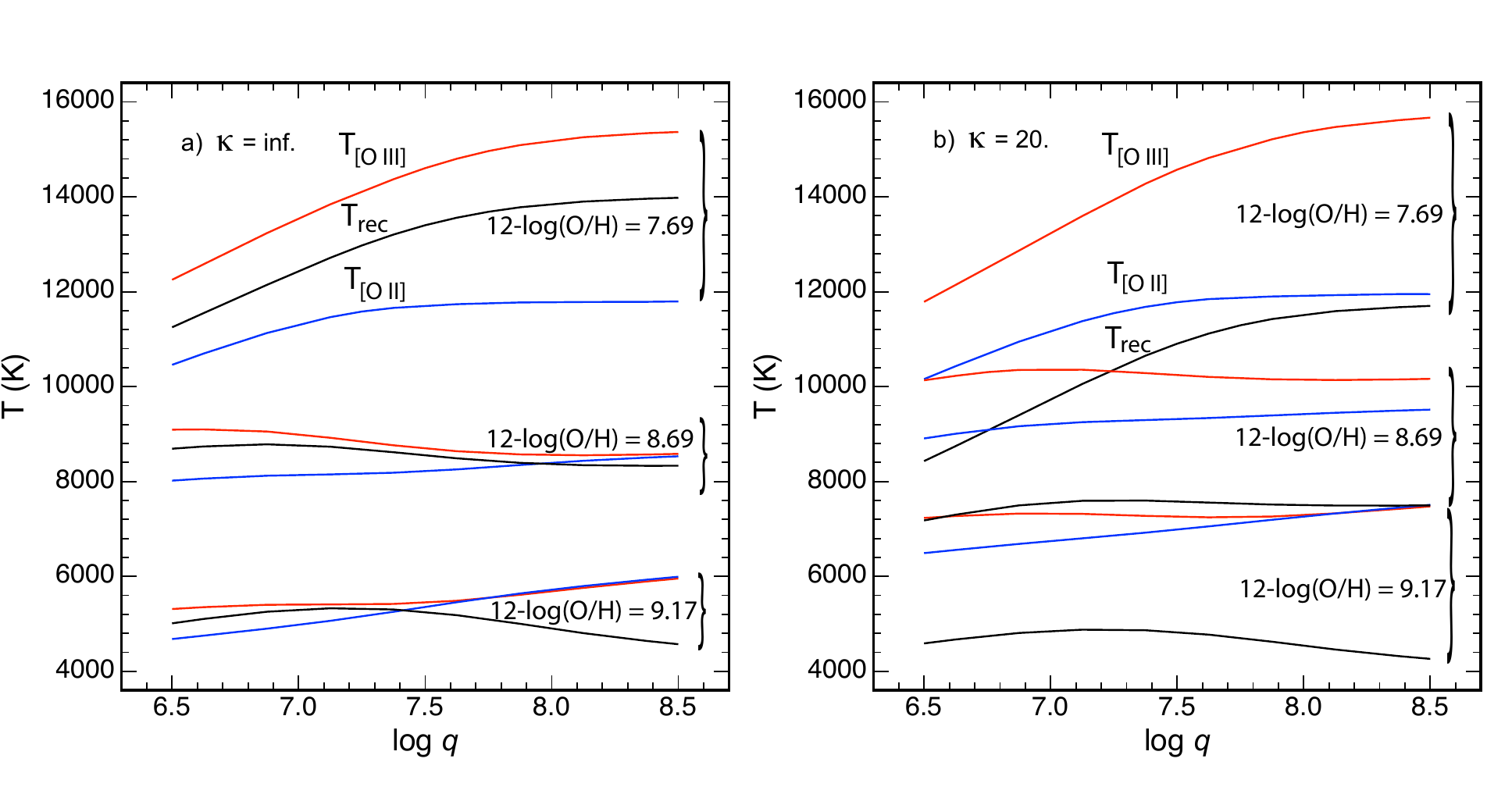}
\caption{ The relationship between the forbidden line and recombination line temperatures as a function of $\log{q}$ and chemical abundance. The left-hand panel is for the models with a Maxwell-Boltzmann electron distribution, and the right hand panel is for the case $\kappa = 20$. The red lines show  the temperature in the high-ionisation zone of the \HII region, $T_{\rm [O III]}$, the blue lines  the temperature in the low-ionisation zone of the \HII region, $T_{\rm [O II]}$, and the black lines show the hydrogen recombination temperature $T_{\rm  rec}$. For all models, the effect of $\kappa$ is to raise  $T_{\rm [O III]}$, and lower $T_{\rm  rec}$. However,  $T_{\rm [O II]}$ may either be higher (at high $Z$), or slightly lowered (at low $Z)$.}\label{fig4}
\end{figure}

The complete list of nebular temperature estimates is given in Table \ref{table_6}. Figure \ref{fig4} shows the way in which both metallicity, ionisation parameter and $\kappa$ affect the measured line temperatures ($T_{\rm [O III]}$ and $T_{\rm [O II]}$), as well as the hydrogen recombination temperature ($T_{\rm  rec}$). Some general observations can be made about this. First, as is well-known, low-abundance \HII regions have higher temperatures than high-abundance \HII regions. Second, the electron temperature is usually positively correlated with $\log{q}$. Third, $T_{\rm [O III]}$ is generally higher than $T_{\rm [O II]}$.  

Figure \ref{fig4} also brings out the way in which $\kappa$ influences the effective (as measured) forbidden emission line temperatures, $T_{FL}$,  and the hydrogen recombination temperature $T_{\rm  rec}$. In the presence of a $\kappa$ distribution, $T_{\rm  rec}$ is always lowered, while the $T_{FL}$ is usually (but not always) raised. The lower  $T_{\rm  rec}$ will result in generally stronger recombination lines. This may in turn lead to an overestimate of the chemical abundances derived from recombination lines when temperatures derived from the forbidden lines are used to interpret these. 

\FloatBarrier

\section{What is the likely value of $\kappa$?}\label{Kappa}
As Figure \ref{fig4} shows, the most sensitive dependence with $\kappa$ is found in the difference between the forbidden line temperatures, $T_{FL}$,  and the hydrogen recombination temperature $T_{\rm  rec}$. The helium recombination temperature could equally well be used in the place of the hydrogen recombination temperature, since the difference between the two is small for all models; see Table \ref{table_6}.

The high-quality \'echelle spectroscopy of galactic and extragalactic \HII regions shows that there is indeed a systematic a difference between $T_{FL}$ and $T_{\rm  rec}$.  Here, we have collected the data the galactic \HII regions from  M 42 \citep{E04}, NGC 3576 \citep{G-R04}, S311 \citep{G-R05}, M20 \& NGC 3602 \citep{G-R06}, M 8 \& M 17 \citep{GarciaRojas07}. For the extragalactic  \ion{H}{2} regions we have data for 30 Dor  \citep{Peimbert03}, NGC595 \citep{L-S07}, NGC 595, NGC 604, VS 24, VS 44, NGC 2365 and K 932 \citep{E09}. However, instead of using the forbidden line temperatures given by these authors, we have used the measured fluxes, and made an independent estimate of the temperatures implied by these using the \citet{Nicholls13} analytic equations. This ensures that the derived temperatures are consistent with the new atomic data used here. Noting that the helium recombination temperatures and the hydrogen recombination line temperatures are very similar, we have used the average of these (when available) to estimate $T_{\rm  rec}$. In the absence of either we have used the other to give our estimate of $T_{\rm  rec}$.  The error bars on the temperatures are assumed to be the same as given by the authors cited above.

It should be noted that, in all cases, the temperatures are derived for only a small patch of the total \HII region. These `pencil-beam' observations do not therefore provide a temperature which is representative of the nebula as a whole. This may be a problem when comparing the data with the models.

\begin{figure}[htpb]
\includegraphics[scale=0.8]{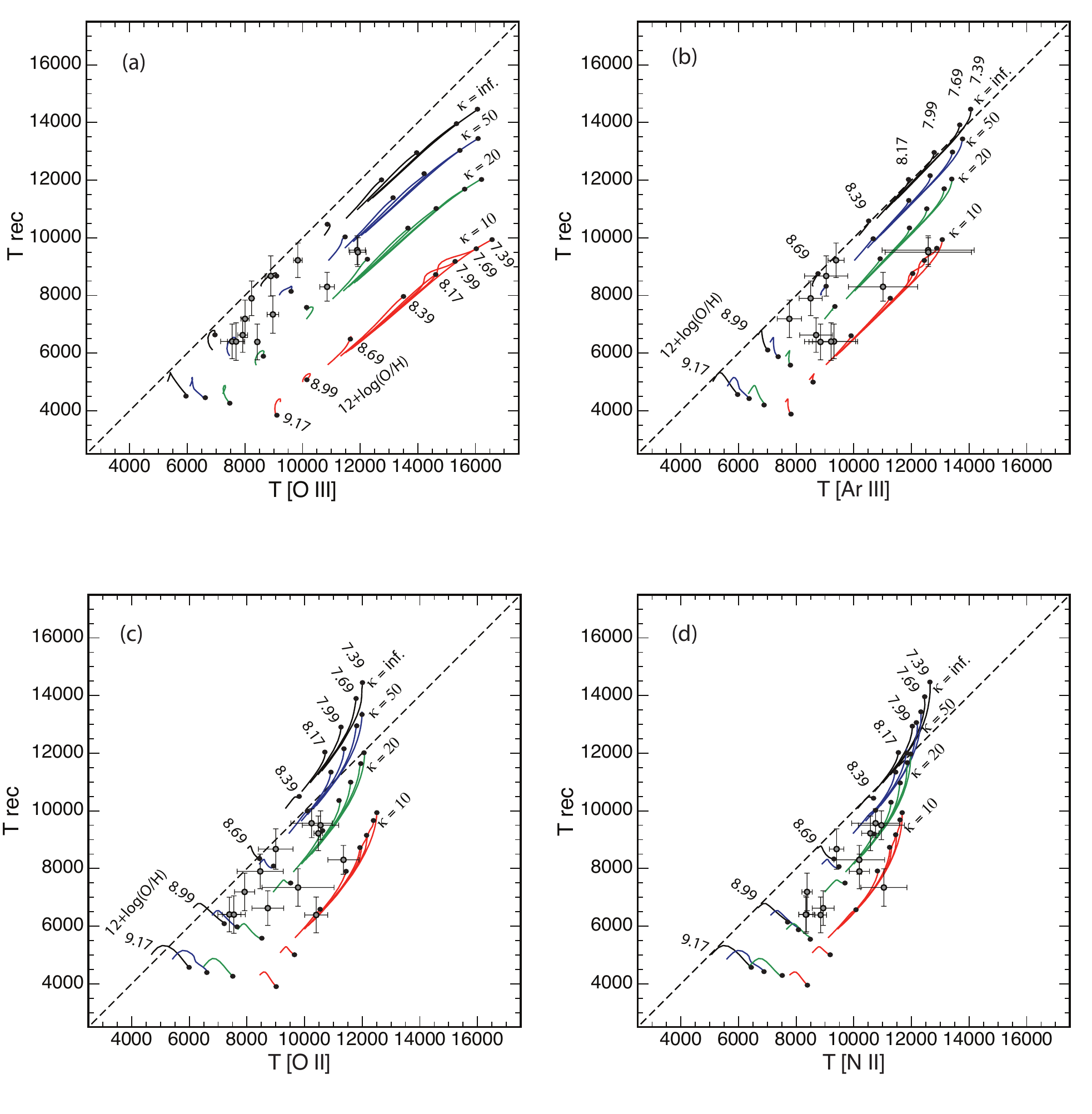}
\caption{ The relationship between the forbidden line and recombination line temperatures. The observations for the galactic and extragalactic \HII regions derived from the high-quality \'echelle data cited in the text are shown with error bars. The models at each abundance, and for each set of $\log{q}$ are shown as `worms' in which the head represents $\log{q} = 8.5$ and the tail $\log{q} = 6.5$. Each value of $\kappa$ is coded by color, as indicated. Note the difference in behavior between the ions arising in the high-ionisation zones of the \HII region; \ion{O}{3} and \ion{Ar}{3} (panels (a) and (b)), and those arising from the low-ionisation regions ; \ion{O}{2} and \ion{N}{2} (panels (c) and (d)). From this figure, we conclude that the best-fit value of $\kappa \sim 20 $, or possibly somewhat higher.}\label{fig5}
\end{figure}

With this data we have constructed Figure \ref{fig5}, which shows forbidden line temperatures plotted for two ions arising in the high-ionization zones of \HII regions, \ion{O}{3} and \ion{Ar}{3}, and for two ions arising in the low-ionization zones,  \ion{O}{2} and \ion{N}{2}, compared with the recombination temperatures as derived above. In general the $\kappa$- distribution models provide a better fit than the Maxwell-Boltzmann case.  There may well be an intrinsic scatter in the $\kappa$ appropriate to individual \HII regions. However, the data is mostly consistent with a fairly moderate $\kappa \sim 20$, or somewhat higher. This represents only a very mild deviation from the Maxwell-Boltzmann case, but is sufficient to significantly affect our estimates of forbidden line and recombination temperatures. In the analysis which follows we will adopt $\kappa \sim 20$.

\FloatBarrier

\section{Effect of $\kappa$ on UV and IR Lines}\label{UV_IR}
\subsection{The UV lines of carbon}
The UV spectrum of \HII regions is rather sparse \citep{Garnett95a}, and as a consequence only a few observations have been obtained in this wavelength region, mostly by Garnett and his collaborators \citep{Garnett95b,Garnett99}. The strongest lines are generally the \ion{C}{3}] $\lambda \lambda 1906,9$ doublet and the \ion{C}{2}] $\lambda2326$ multiplet.  \ion{Si}{3}] $\lambda \lambda 1883,92$  are sometimes seen, as well as the [\ion{O}{2}] $\lambda \lambda 2470$ line.

\begin{figure}[htpb]
\includegraphics[scale=0.75]{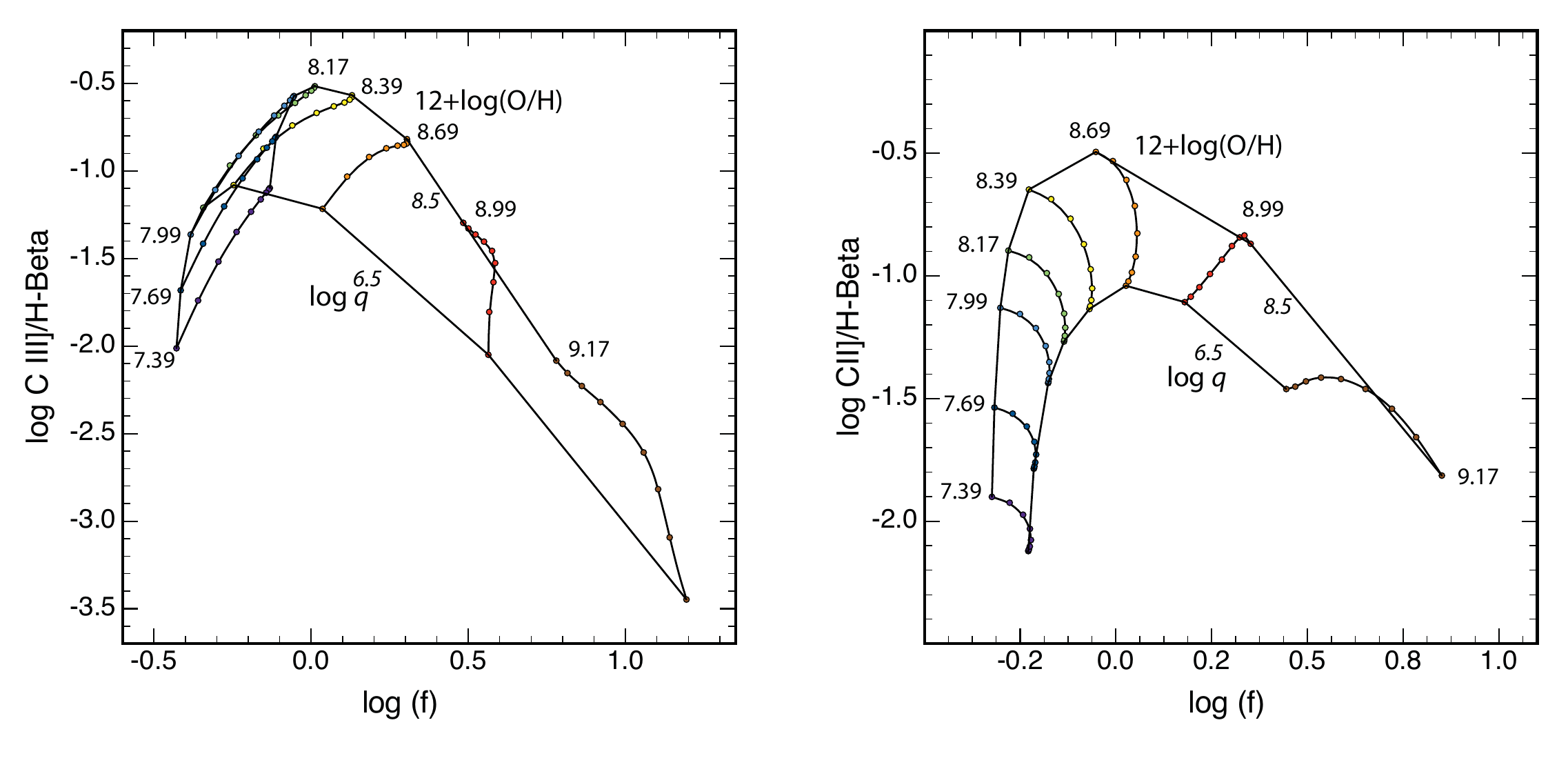}
\caption{The effect of $\kappa = 20$ on the  \ion{C}{3}] $\lambda \lambda 1906,9$ doublet (left) and the \ion{C}{2}] $\lambda2326$ multiplet (right). The vertical axis is the predicted line flux relative to H$\beta$ for $\kappa = \infty$, and the horizontal axis is the factor, $f$, by which the line intensity has to be multiplied to give the predicted intensity for  $\kappa = 20$. The effect of  $\kappa$ is most evident at high abundance, where line intensities may be enhanced by as much as a factor of ten. }\label{fig6}
\end{figure}

Here we will consider only the effect of the $\kappa-$distribution on the important carbon lines. The effect on the \ion{Si}{3}] $\lambda \lambda 1883,92$ lines is essentially the same as for the \ion{C}{3}] $\lambda \lambda 1906,9$ doublet, and the effect on the  [\ion{O}{2}] $\lambda \lambda 2470$ line is similar to that of the \ion{C}{2}] $\lambda2326$ multiplet. In Figure \ref{fig6} we plot the enhancement in the line intensity for $\kappa = 20$ compared with $\kappa = \infty$, against the predicted line flux relative to H$\beta$ computed for models with $\kappa = \infty$. The enhancement of line intensity is very significant for metallicities greater than two times solar. The reason for these large correction factor is that the temperature of the \HII region is low, and the hot tail in the $\kappa$ electron energy distribution becomes very important. However, there are no UV measurements for \HII regions in this metallicity range, so the computations cannot be checked. For lower metallicities, the corrections are much smaller, and the carbon lines are generally affected by less than a factor of two. Nonetheless, these corrections may well be important where UV lines are being used to estimate the C/O ratio.

\subsection{The IR lines}
By contrast with the UV, the effect of of  $\kappa$ on the IR lines is very weak indeed. At the high abundance end, the intensities of the strong lines are weakened by only $2-4$\%, even at  $\kappa = 10$. The effect becomes more noticeable for low abundance \HII regions, with line intensities being weakened by as much as 25\%. This is easily understood by reference to the ``banana curves'' of \citet{Nicholls12,Nicholls13}, which show that only weak $\kappa-$related changes are expected in the excitation rates of the IR lines for typical nebular temperatures.

\begin{figure}[htpb]
\includegraphics[scale=0.7]{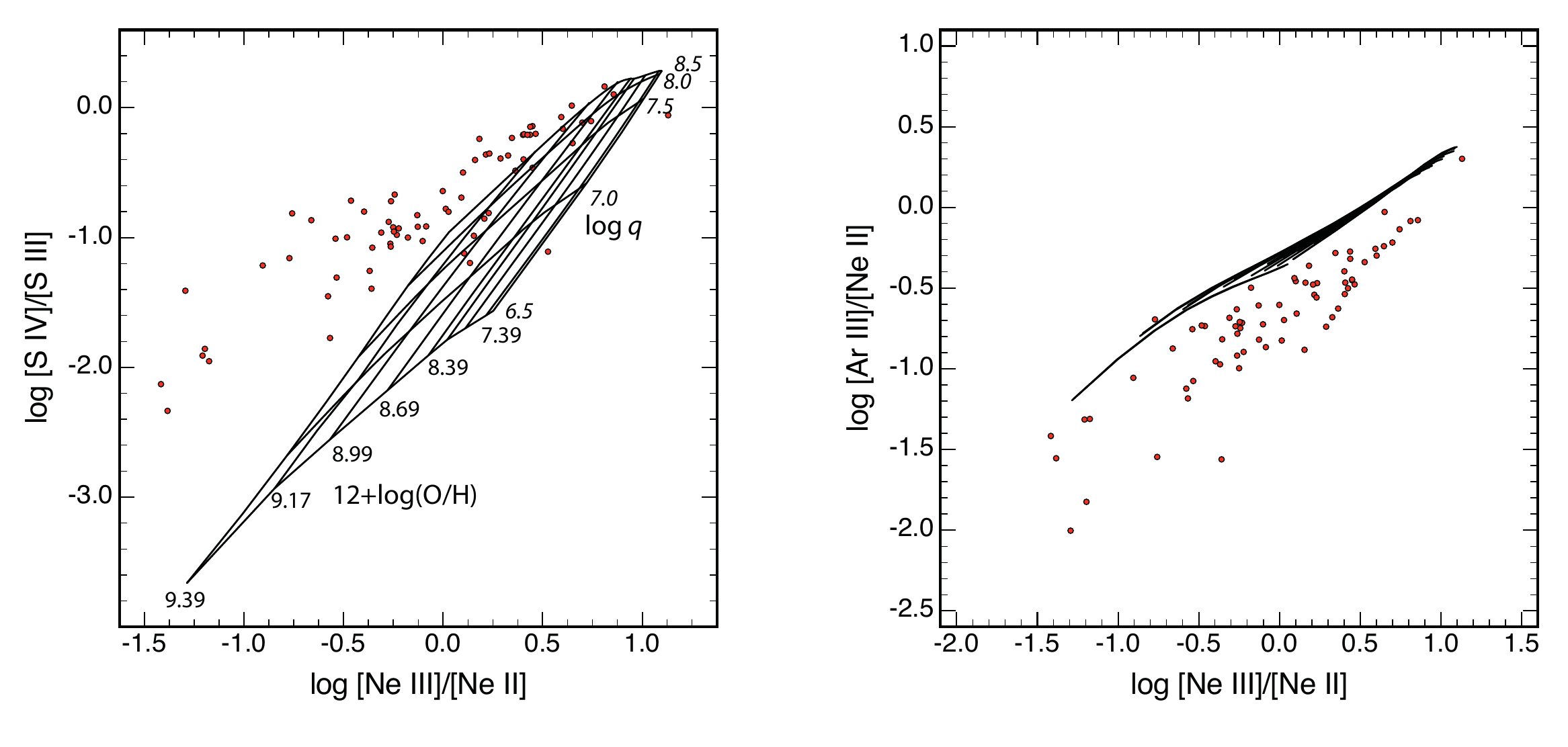}
\caption{The mid-IR diagnostics,  [\ion{Ne}{3}] 15.56 \mum/[\ion{Ne}{2}] 12.81 \mum\ \emph{vs.} [\ion{S}{4}] 10.51 \mum/[\ion{S}{3}] 18.71 \mum\ (left),  and [\ion{Ne}{3}] 15.56 \mum/[\ion{Ne}{2}] 12.81 \mum\ \emph{vs.} /[\ion{Ar}{3}] 8.99 \mum\ / [\ion{Ne}{2}] 12.81\mum\ (right) plotted for $\kappa=\infty$. The points represent the \citet{Giveon02} data for Galactic and Magellanic Cloud \HII regions. The probable cause of the offset between theory and observation is discussed in the text.}\label{fig7}
\end{figure}

More problematic is the fact that the models do no give a very good description of the observed spectrum of \HII regions. \citet{Snijders07}  compared the Mappings models with the observations of \citet{Giveon02}, and devised two new mid-IR diagnostics based upon the excitation-dependent ratios  [\ion{Ne}{3}] 15.56 \mum/[\ion{Ne}{2}] 12.81 \mum, [\ion{S}{4}] 10.51 \mum/[\ion{S}{3}] 18.71 \mum, [\ion{S}{3}] 18.71 \mum/[\ion{Ne}{2}] 12.81 \mum\ and [\ion{S}{4}] 10.51 \mum/[\ion{Ar}{3}] 8.99 \mum. In Figure \ref{fig7} we plot one of these; the  [\ion{Ne}{3}] 15.56 \mum/[\ion{Ne}{2}] 12.81 \mum\ \emph{vs.} [\ion{S}{4}] 10.51 \mum/[\ion{S}{3}] 18.71 \mum\ and a new one, [\ion{Ne}{3}] 15.56 \mum/[\ion{Ne}{2}] 12.81 \mum\ \emph{vs.} /[\ion{Ar}{3}] 8.99 \mum\ / [\ion{Ne}{2}] 12.81 \mum, along with the \citet{Giveon02} data. Here the grids are shown only for $\kappa=\infty$ - the grids for $\kappa=20$ fully overlay these.

The offsets between the theory and the observation require explanation. First, let us consider the possibility that the intensity of the [\ion{Ne}{2}] 12.81 \mum\ line is in error. If this were the case, the grid on the LHS of Figure \ref{fig7} would move closer to the observations. However, the effect on the RHS diagram would simply be to translate the curve along the 45 degree line, leading to no improvement here. Likewise, the [\ion{Ne}{3}] 15.56 \mum\ line intensity cannot be the source of the problems, since changing this lines translates both grids sideways in the same direction, so the fit of one is improved only at the expense of the other. 

A more likely explanation is that the  [\ion{S}{4}] 10.51 \mum/[\ion{S}{3}] 18.71 \mum\  ratio is in error, particularly at the high abundance end. \citet{Snijders07} showed that a much better fit between observations and theory can be obtained if the mean effective age of the exciting clusters is somewhat older than we have used here. This allows an appreciable population of Wolf-Rayet stars to develop. These stars have higher effective temperatures, can excite the species with higher ionisation potentials and so produce more \ion{S}{4} ions in the nebula, resulting in a stronger [\ion{S}{4}] 10.51 \mum\ line. To explain the offset in the [\ion{Ar}{3}] 8.99 \mum\ / [\ion{Ne}{2}] 12.81\mum\ ratio, we would have to appeal to errors in either the atomic data for the [\ion{Ar}{3}]  line or else errors in the charge-exchange rates which strongly affect the ionisation balance of Ar. All of these issues point to the need for more work in refining the predictions of the models in the IR.

\begin{figure}[htpb]
\includegraphics[scale=0.7]{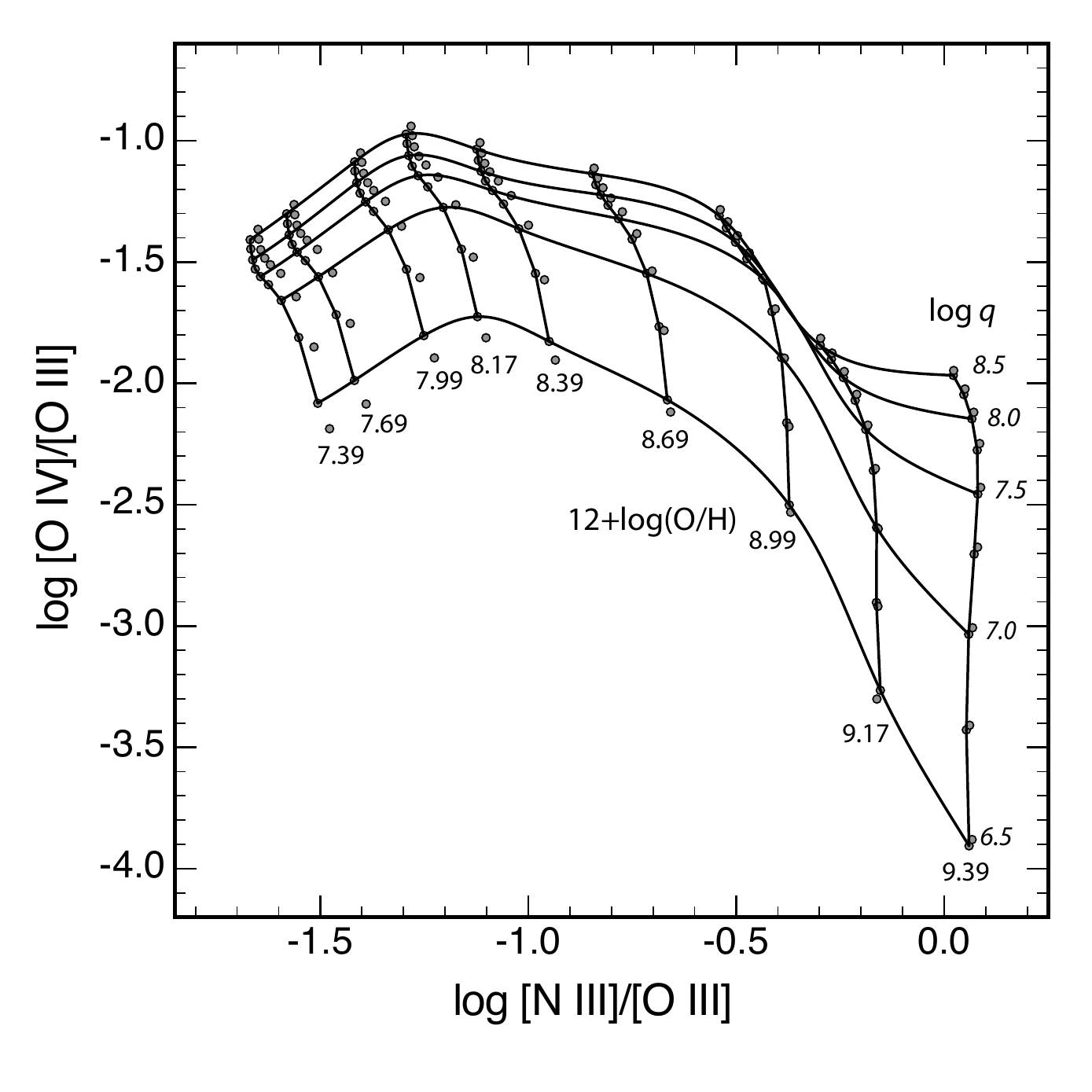}
\caption{A far-IR diagnostic for abundance and ionisation parameter, [\ion{N}{3}] 57.34 \mum/[\ion{O}{3}] 51.81 \mum\ \emph{vs.} [\ion{O}{4}] 25.89 \mum/[\ion{O}{3}] 51.81 \mum\ . The grid is shown for $\kappa=\infty$, and gray circles represent the $\kappa=20$ case. The effect of $\kappa$ on this diagnostic is very small.}\label{fig8}
\end{figure}

In the far-IR, we have identified a very promising abundance - ionisation parameter diagnostic. This is shown in Figure \ref{fig8} where we plot  [\ion{N}{3}] 57.34 \mum/[\ion{O}{3}] 51.81 \mum\ \emph{vs.} [\ion{O}{4}] 25.89 \mum/[\ion{O}{3}] 51.81 \mum\ as delivered by the models. This grid is little affected by $\kappa$, and provides a very clean separation between the abundance and the ionisation parameter.


\FloatBarrier

\section{Strong Line Ratio Diagnostics}\label{Diagnostics}
\subsection{Veilleux \& Osterbrock Diagnostics}
Following the pioneering work of \citet{BPT}, \citet{VO87} (VO87) exploited the utility of diagnostics based upon the ratio of the strong red lines; [\ion{N}{2}]/H$\alpha$, [\ion{S}{2}]/H$\alpha$ and [\ion{O}{1}]/H$\alpha$ plotted against the ratio \OIIIl/H$\beta$. These ratios have the great advantage of using lines close together in wavelength, so that the reddening correction for dust is negligible. They noted that the \HII regions are confined to a rather narrow strip on these diagrams, while Seyferts and LINERS lay systematically respectively above and to the right of the \HII region sequence. These diagrams are therefore of great utility in determining the mode of excitation for photoionised objects. The permitted \HII region and starburst region was established in \citet{Kauffmann06} and the formal division between the starburst, Seyfert and LINER zones on these diagrams was established empirically by \citet{Kewley06} using the very extensive SDSS spectrophotometry.

A number of theoretical attempts to reproduce the narrow \HII region and starburst sequence have been made. \citet{Dopita00} demonstrated that the narrowness of the the sequence could be understood (in part) as a result of the folding of the $\log(Z):\log(q)$ surface which ensures that \HII regions having a wide range in these parameters occupy the same region of the diagnostic diagram. \citet{Dopita06b} demonstrated that the ionisation parameter, the hardness of the ionising spectrum and the metallicity are tightly correlated, and provided a theoretical explanation of why this should be the case. Both of these effects clearly play a role in making for a tight \HII region and starburst sequence. The fundamental difficulty with the theoretical models is that the `fold' in the surface was not the correct shape, presumably due to a combination of errors or incompatibilities in the stellar atmospheres used, issues with the atomic data used, or errors in the modelling procedure, especially in the treatment of the geometry of the \HII region and in the treatment of dust absorption in the \HII region. \citet{Dopita06b} made an attempt to take into account the time evolution of the spectrum of the \HII regions as the stellar cluster ages and as the nebular shell expands. This analysis showed that the more recent STARBURST 99 atmospheres were definitely too soft in their EUV spectra, as a consequence of the use of non-clumpy stellar winds. This analysis also showed that most extragalactic \HII regions are observed when they are very young ($\lesssim 2$ Myr); shortly after the absorbing placental dust cloud is dispersed, when the pressure -- and hence the emissivity -- in the \HII region is high, and before the central stars fade in their EUV photon production rate. All of these factors militate to making them easily observable while very young. 

\begin{figure}[htpb]
\includegraphics[scale=1.0]{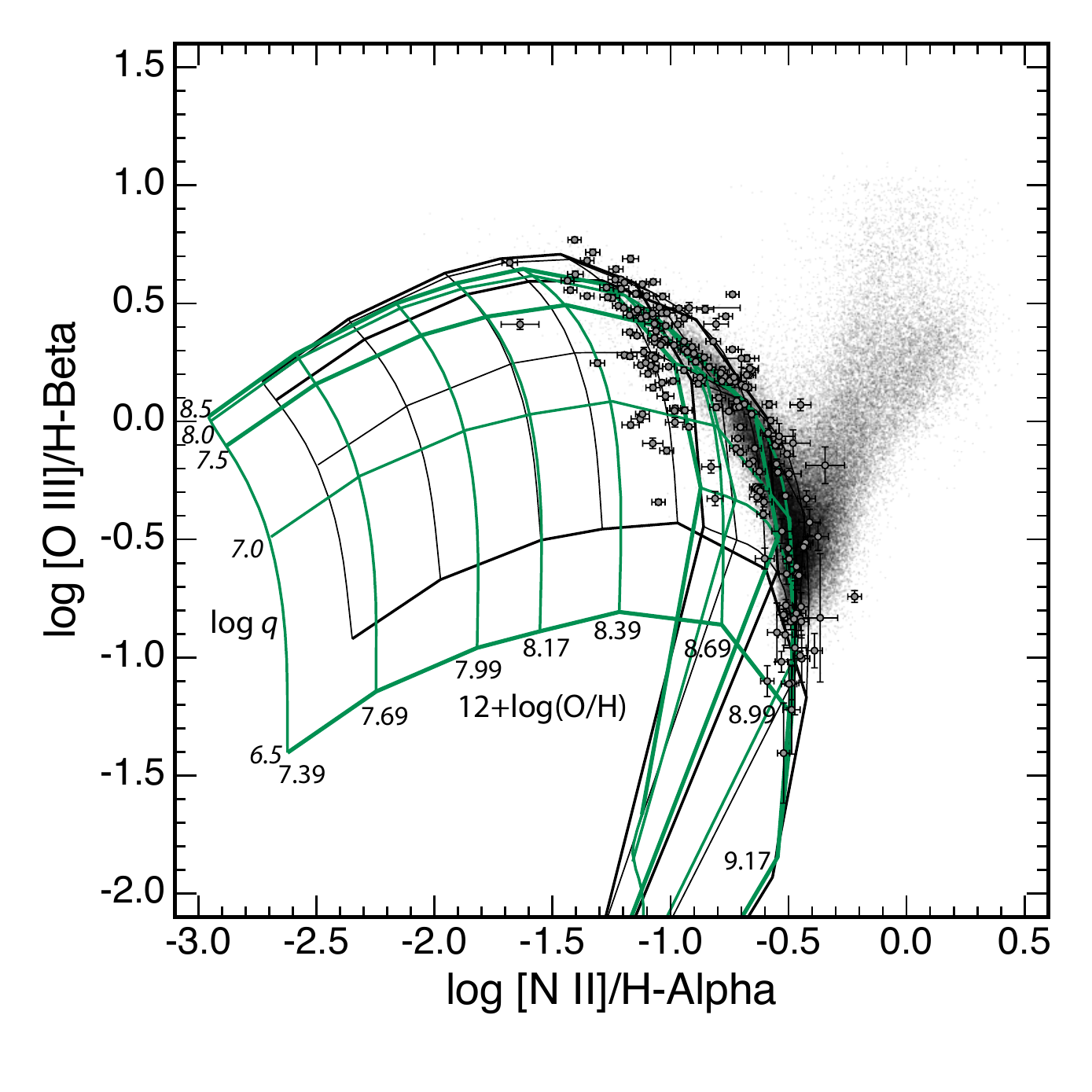}
\caption{ The VO87 plot of  [\ion{N}{2}]$\lambda 6584$/H$\alpha$ vs. \OIIIl/H$\beta$. The grey dots represent the SDSS dataset as used by \citet{Kewley06}, while the points with error bars are from the \citet{vanZee98} dataset. The delineation of the two AGN sequences in the upper right-hand side of the diagram, the Seyferts (upper) and LINERS (lower) is very clear on this plot. The models grids on this and all subsequent diagnostic diagrams are shown for two values of kappa; $\kappa = \infty$ (black lines) and $\kappa = 20$ (green lines). Note that the effect of $\kappa$ is relatively small in this diagnostic.}\label{fig9}
\end{figure}

\begin{figure}[htpb]
\includegraphics[scale=1.0]{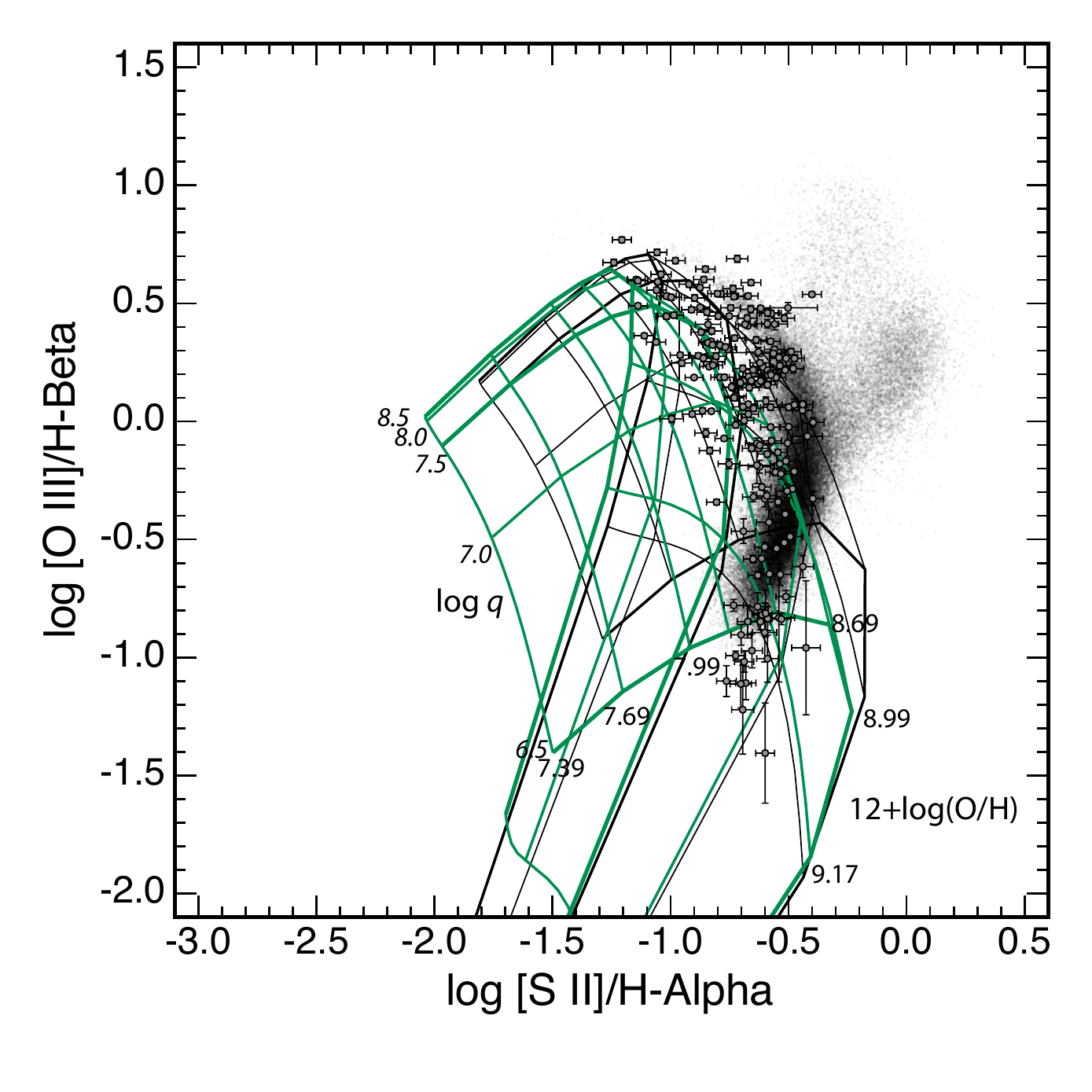}
\caption{ As figure \ref{fig9} but for the  VO87 plot of  [\ion{S}{2}]$ \lambda \lambda 6717,31$/H$\alpha$ vs. \OIIIl/H$\beta$. The models seem to predict slightly too weak  [\ion{S}{2}] line intensities.}\label{fig10}
\end{figure}
\FloatBarrier

The models presented here represent a great improvement upon the earlier work of our group, although we should note in passing that the models of \citet{Stasinska06b} provide a rather good description to the upper envelope of the \HII region-like points on these diagrams. In Figure \ref{fig9}, we present the VO87 plot of  [\ion{N}{2}]$\lambda 6584$/H$\alpha$ vs. \OIIIl/H$\beta$. This figure should be compared to Figure 2 of \citet{Dopita00}, showing the improvement in the fit of the theory compared with the observations. Important contributors to this improvement are the proper treatment of the EUV absorption dust, improved atomic data, the use of spherical models, and the fact that the nebular abundance set is now fully compatible with the stellar model atmospheres.

 In Figure \ref{fig10}, we present the VO87 plot for  [\ion{S}{2}]$\lambda \lambda 6717,31$/H$\alpha$ vs. \OIIIl/H$\beta$. Once again there is a great improvement in the fit between theory and observation \emph{c.f.}  Figure 3 of \citet{Dopita00}. However, the  [\ion{S}{2}] lines are perhaps about 0.1 dex too weak. At the high abundance end, the observed sequence is best explained by the \HII regions having a roughly constant $\log(q)$, roughly between 7.0--7.5. This is confirmed in the many diagnostics presented below.

We have not attempted to provide the third diagnostic,  [\ion{O}{1}]$\lambda6300$/H$\alpha$ vs. \OIIIl/H$\beta$. This is because the  [\ion{O}{1}] line arises in a very narrow zone close to the ionisation front, where shocks and non-equilibrium heating may well be important. \citet{Dopita97} noted that shocks have a significant effect on this line ratio even when the ratio of mechanical energy to photon energy flux is as small as $10^{-3}$. We therefore regard our computations of this line as much more unreliable than the other two. However, the intensity of the line is given in Table \ref{table_5}, if the theoretical values are required by the reader for any reason.

\begin{figure}[htpb]
\includegraphics[scale=1.0]{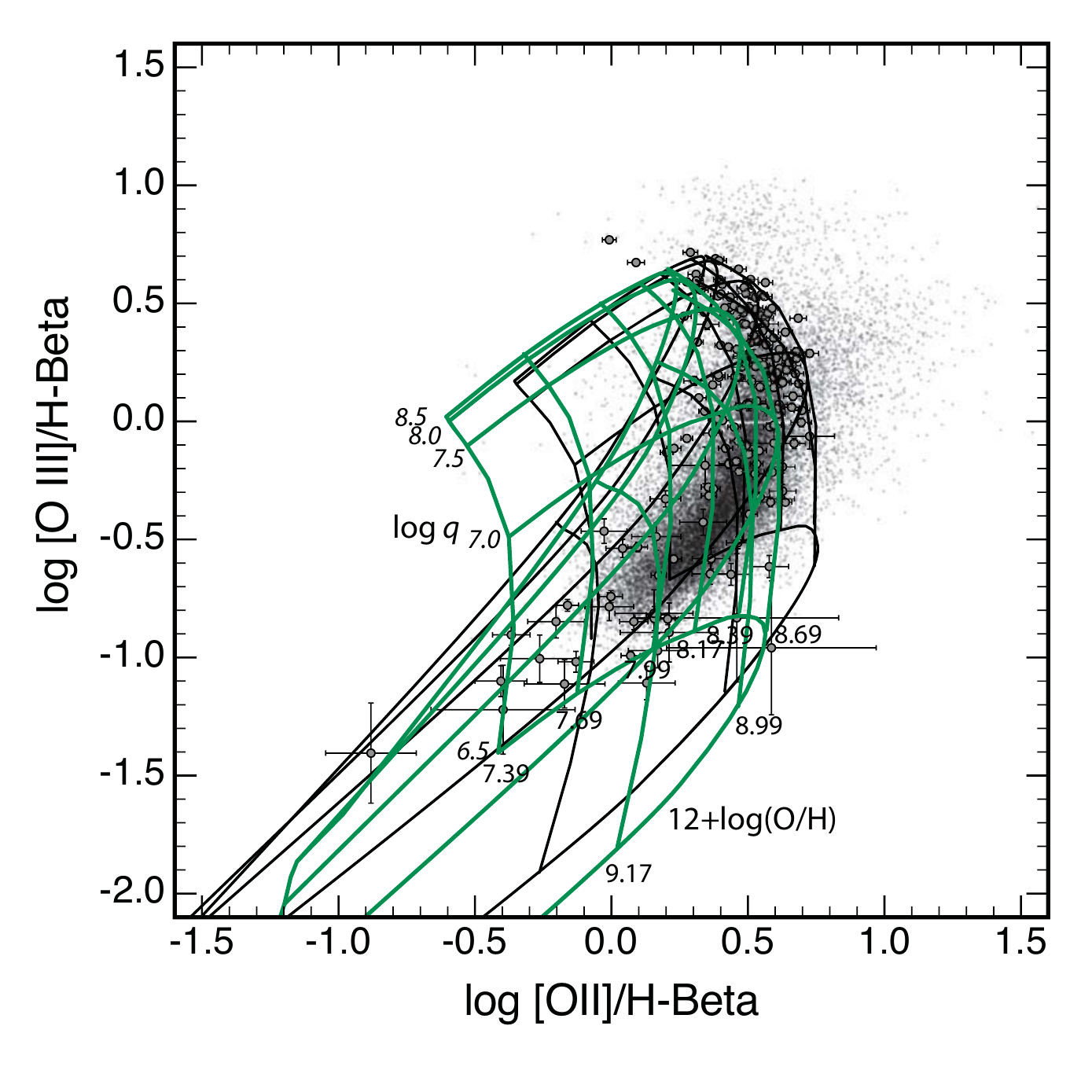}
\caption{ As figure \ref{fig9} but for  [\ion{O}{2}]$\lambda \lambda3727,9$/H$\beta$ vs. \OIIIl/H$\beta$. }\label{fig11}
\end{figure}

Closely related to the above diagnostics is that of  [\ion{O}{2}]$\lambda \lambda3727,9$/H$\beta$ vs. \OIIIl/H$\beta$. or completeness, this ratio is shown in Figure \ref{fig11}. The observational errors on the $x-$ axis are somewhat greater because of uncertain reddening corrections.
\FloatBarrier

\subsection{Excitation-Dependent Diagnostics}
\begin{figure}[htpb]
\includegraphics[scale=1.0]{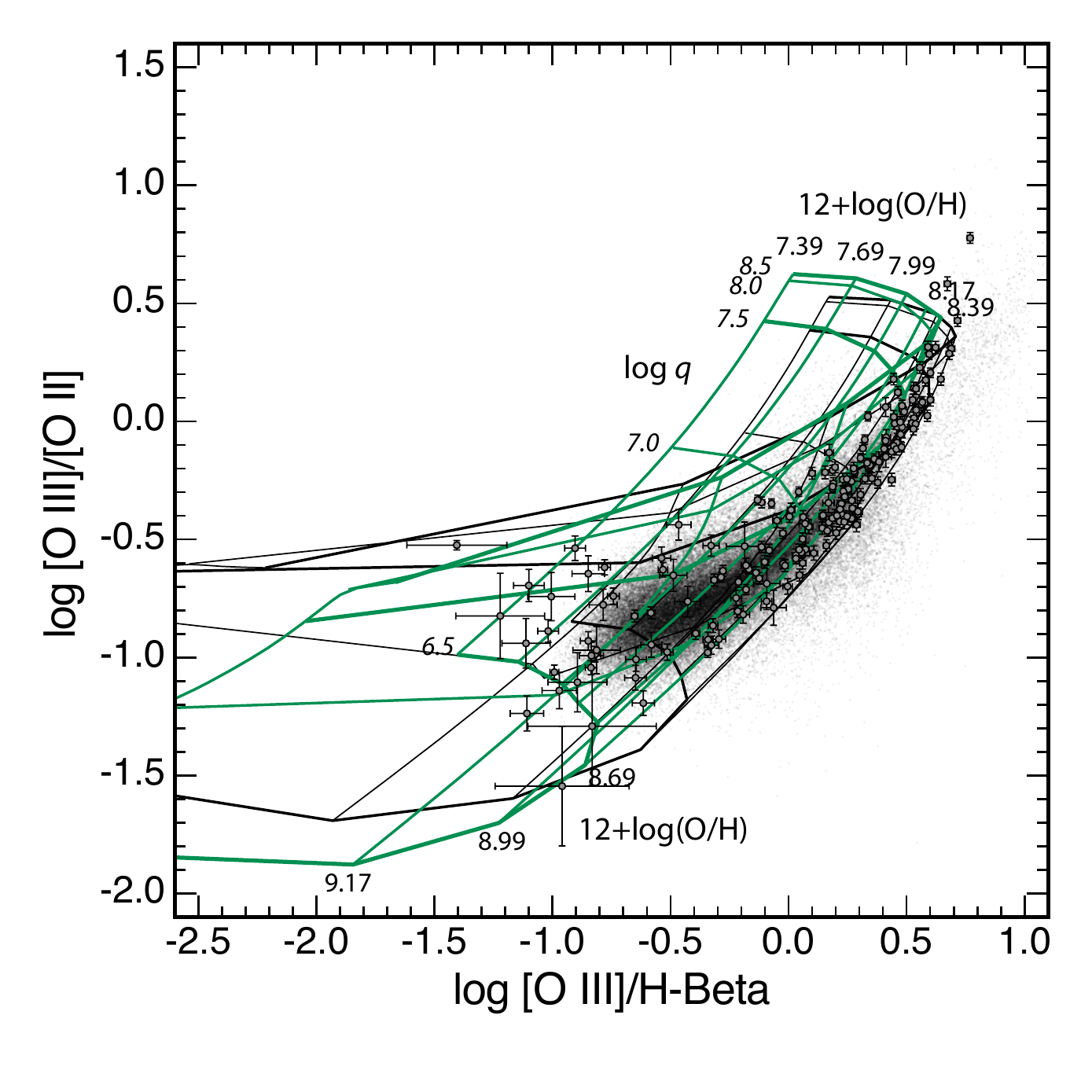}
\caption{ The excitation-sensitive ratios   [\ion{O}{3}]$\lambda5007$/[\ion{O}{2}]$\lambda \lambda 3727,9$ and  \OIIIl/H$\beta$ plotted against each other. This is a transposed version of one of the BPT diagnostics. Clearly, \OIIIl/H$\beta$ shows a rather greater sensitivity to the chemical abundance, but both depend upon a mixture of both 12+$\log$(O/H) and $\log q$. Neither can be used alone to estimate the excitation.}\label{fig12}
\end{figure}

 \citet{BPT} were the first to emphasise the importance of nebular excitation, measured by ratios such as [\ion{O}{3}]$\lambda5007$/[\ion{O}{2}]$\lambda3727,9$ in distinguishing and separating the various modes of ionisation commonly observed in nature (\HII regions, planetary nebulae (PNe), power-law ionised or shock excited). Within a given class of object, such ratios are also sensitive to the ionisation parameter. For \HII regions,  \citet{BPT} showed that \OIIIl/H$\beta$ is also correlated with  [\ion{O}{3}]$\lambda5007$/[\ion{O}{2}]$\lambda \lambda 3727,9$, as it is separately in the case of PNe. In Figure \ref{fig12}, we show how well these two line ratios track each other. As it stands, this plot is not very useful. It neither effectively separates 12+$\log$(O/H) from $\log q$, nor does it reveal the AGN as a separate branch.

\begin{figure}[htpb]
\includegraphics[scale=1.0]{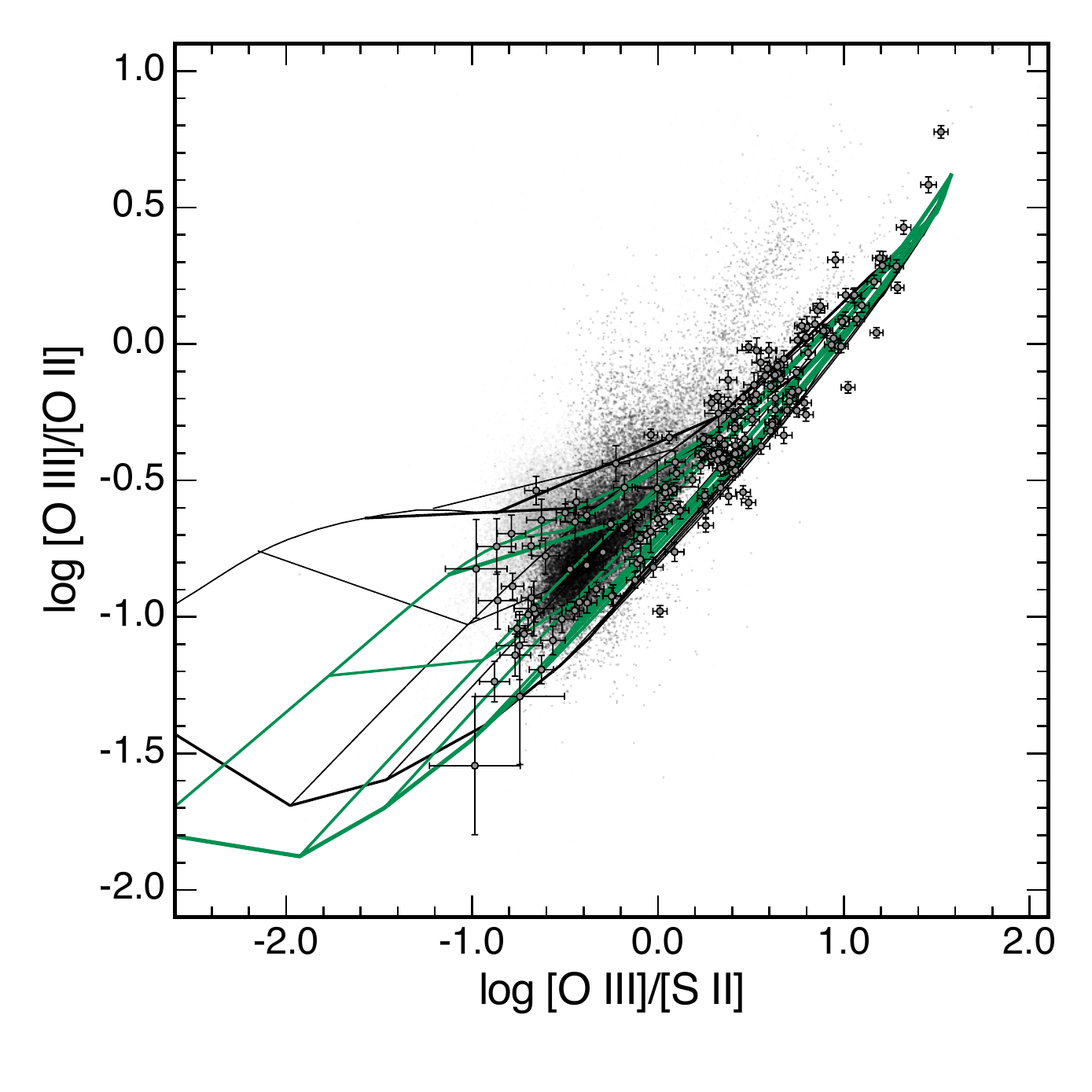}
\caption{ The excitation-sensitive ratio [\ion{O}{3}]$\lambda5007$/[\ion{S}{2}]$\lambda6717,31$ plotted against a second  excitation-sensitive ratio [\ion{O}{3}]$\lambda5007$/[\ion{O}{2}]$\lambda \lambda 3727,9$. The two ratios show good sensitivity to ionisation parameter at abundances less than solar, and the two AGN sequences are now clearly distinguished.  Due to the great degeneracy of the theoretical curves on this plot, we have not attempted to label the individual curves. Suffice it to note that the high abundance objects are located in the lower left-hand corner of the plot.}\label{fig13}
\end{figure}

A better excitation-dependent diagnostic is obtained if we substitute the excitation-dependent ratio [\ion{O}{3}]$\lambda5007$/[\ion{S}{2}]$\lambda \lambda 6717,31$ for  \OIIIl/H$\beta$, as shown in Figure \ref{fig13}. Both ratios provide a similar sensitivity to ionisation parameter at abundances less than solar, and in this abundance range the abundance sensitivity is also weak. Clearly we can substitute [\ion{S}{2}]$\lambda \lambda6717,31$ in place of [\ion{O}{2}]$\lambda \lambda 3727,9$, if necessary. This is useful if reddening corrections are uncertain, or if the nebular spectra obtained do not extend much below H$\beta$.

\begin{figure}[htpb]
\includegraphics[scale=1.0]{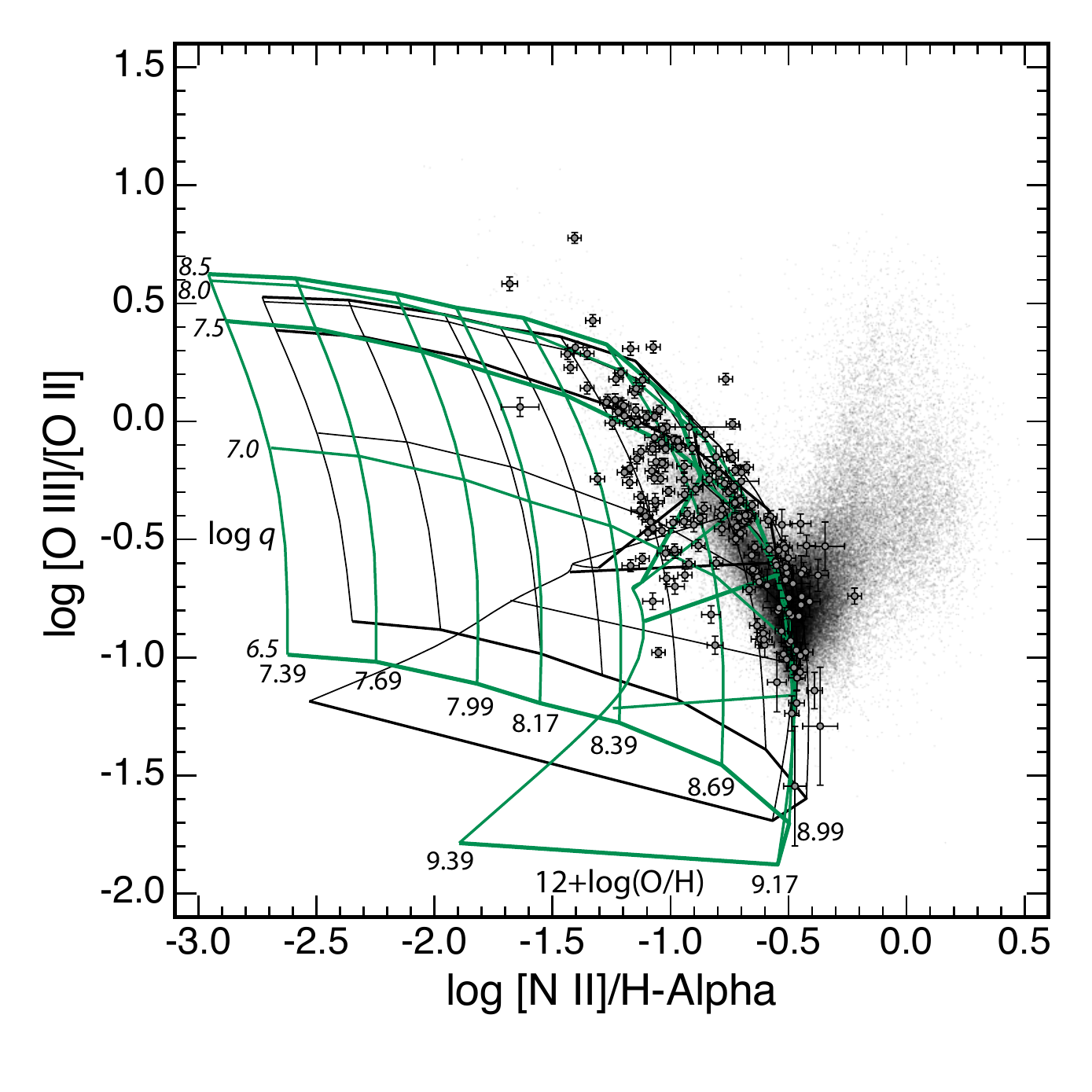}
\caption{  [\ion{N}{2}]$\lambda 6584$/H$\alpha$ vs.[\ion{O}{3}]$\lambda5007$/[\ion{O}{2}]$\lambda \lambda 3727,9$. This is one of the BPT diagnostics, with axes transposed. Both this diagram and Figure \ref{fig9} provide a clean separation of the \HII region sequence from the Seyfert and LINER branches. However, this figure might prove more useful in separating the Seyfert and LINER branches than Figure \ref{fig9}.}\label{fig14}
\end{figure}

Given the similar sensitivity of both \OIIIl/H$\beta$ and [\ion{O}{3}]$\lambda5007$/[\ion{O}{2}]$\lambda \lambda 3727,9$ ratios to both excitation and chemical abundance, we examine the substitution of the second for the first in the \citet{BPT} and \citet{VO87} diagnostic diagram in Figure \ref{fig14}. This diagram is, in fact, another  \citet{BPT} diagram (with transposition of the axes). Again, as in Figure \ref{fig9} , there is a clean separation of the AGN excited objects from the narrow \HII region sequence.

\FloatBarrier

\subsection{Abundance-Sensitive Diagnostics}

\subsubsection{The R$_{23}$ Diagnostic}
It has been traditional to use the ratio of a forbidden line to a hydrogen recombination line in the quest to determine chemical abundance. However, all such ratios are two-valued in terms of abundance. As a consequence, a great deal of effort has been expended in identifying the appropriate `branch' a particular \HII region lies upon in diagnostics such as the $R_{23}$ ratio, \ratioR23 \citep{Pagel79}, or in similar ratios such as $S_{23} =$ \ratioS23  \citep{Diaz00,Oey02}.  

For the $R_{23}$ ratio, in particular, the use of both the visible forbidden line of [\ion{O}{3}] and the UV  [\ion{O}{2}] together mixes two regions of different \HII region temperatures, and makes the maximum of this line ratio very broad in terms of abundance. This makes the actual abundance very difficult to estimate in the region of the maximum, and it becomes critical to have a good estimate of the ionisation parameter to remove the residual sensitivity of the $R_{23}$ ratio to this parameter. This point was emphasized by \citet{McGaugh91}

\begin{figure}[htpb]
\includegraphics[scale=1.0]{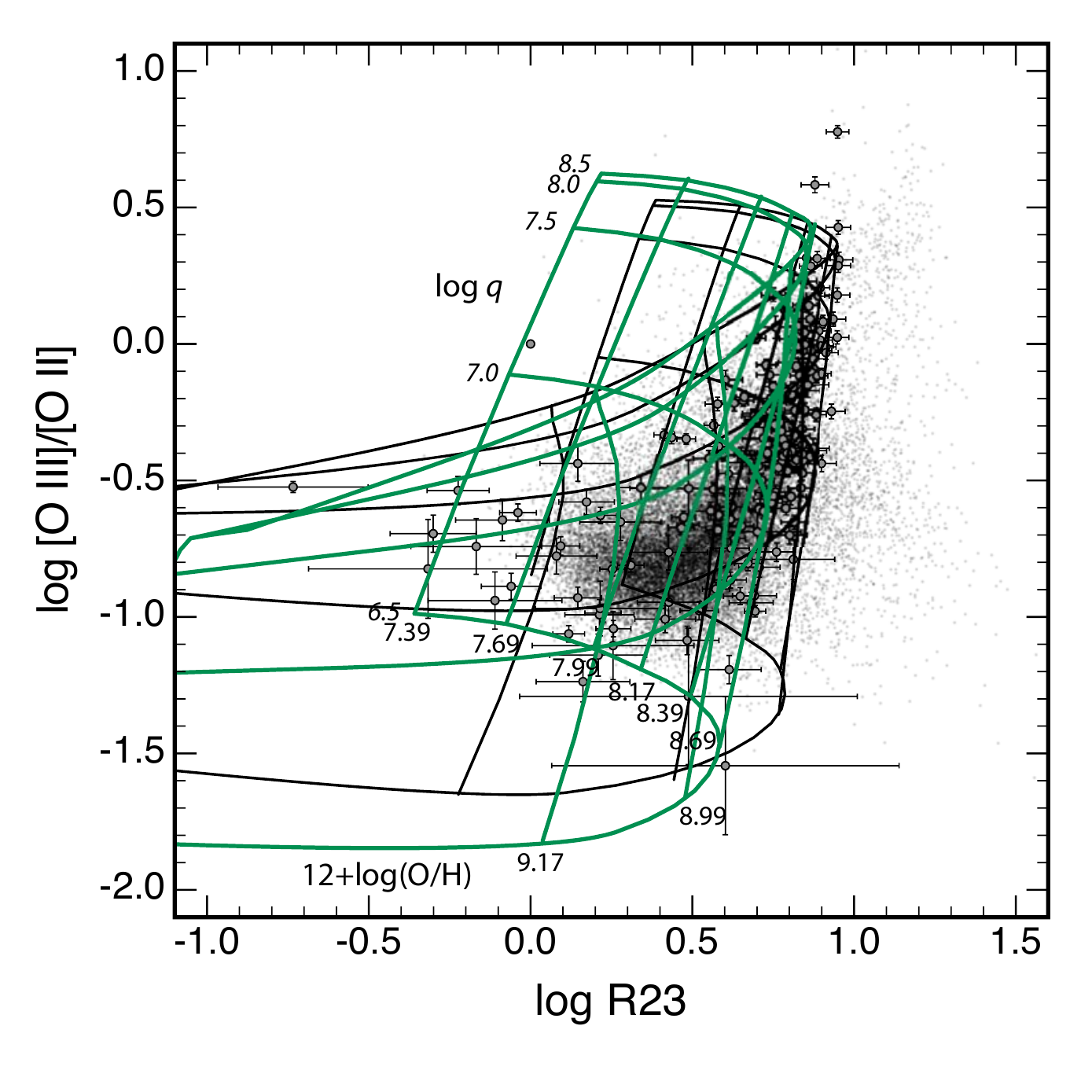}
\caption{ The $R_{23}$ ratio, \ratioR23 \citep{Pagel79} plotted against the excitation-sensitive  [\ion{O}{3}]$\lambda5007$/[\ion{O}{2}]$\lambda \lambda 3727,9$ ratio.  This diagram graphically illustrates the serious problems associated with any attempt to derive the chemical abundance from the  $R_{23}$ ratio alone. We strongly recommend against the use of this ratio as an abundance diagnostic.}\label{fig15}
\end{figure}

These problems become  very evident in Figure \ref{fig15}, in which we plot the  $R_{23}$ ratio, \newline  \ratioR23 \citep{Pagel79}, against the excitation-sensitive  [\ion{O}{3}]$\lambda5007$/[\ion{O}{2}]$\lambda \l3727,9$ ratio, as was done by \citet{McGaugh91}. First, the ratio is only weakly dependent on abundance for a broad range of abundance; $8.3 > 12 + \log({\rm O/H}) > 9.0$, approximately. Second, in this range the sensitivity to the ionisation parameter is as great as to the abundance. Elsewhere, the ratio is two-valued leading to the associated problems of determining whether the abundance solution lies on the low- or high- abundance branches. All these issues apply whether or not the electrons have a $\kappa-$distribution. In conclusion therefore, we strongly recommend against use of the  $R_{23}$ ratio, \ratioR23 in attempts to determine chemical abundances, and advise observers to treat any such attempts with a great deal of caution.

\FloatBarrier

\subsubsection{Ratios of Strong Forbidden Lines} 
Unlike the  $R_{23}$  diagostic, and others which use the ratio of a forbidden line to a hydrogen recombination line, a number of purely forbidden line ratios are known to vary monotonically with abundance, such as the \ratioNIIOII\  ratio \citep{Dopita00}, or the [\ion{Ar}{3}]$\lambda7135$/[\ion{O}{3}]$\lambda5007$ and the [\ion{S}{3}]$\lambda9069$/[\ion{O}{3}]$\lambda5007$ ratios \citep{Stasinska06a}.  Th use of such pairs of forbidden line ratios to unambiguously separate the abundance and the ionisation parameter was pioneered by \citet{Evans85,Evans86}. It seems  it seems somewhat surprising that ratios such as these have since not been much more widely employed for strong line abundance diagnostics. Perhaps this is the result of a natural psychological pressure to include hydrogen if an attempt is being made to determine the abundance of a heavy element with respect to hydrogen.

\begin{figure}[htpb]
\includegraphics[scale=0.7]{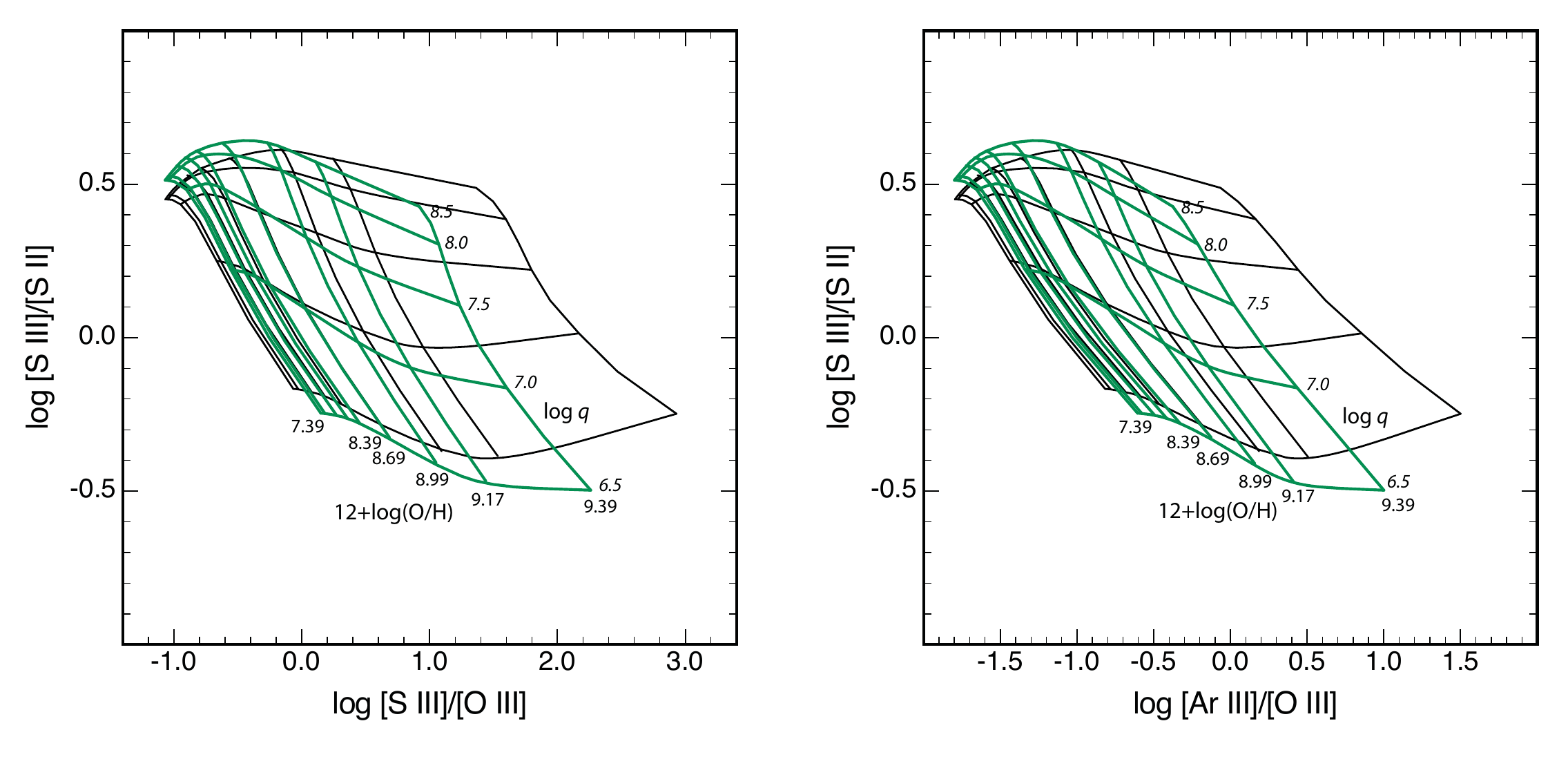}
\caption{ The \citet{Stasinska06a} abundance-sensitive ratios plotted against the excitation-sensitive [\ion{S}{3}]$\lambda9069$/[\ion{S}{2}]$\lambda6717,31$ ratio. As before, the black grids correspond to $\kappa = \infty$ and the green labelled grids are for $\kappa = 20$. The data sets we are using do not give the  [\ion{S}{3}]$\lambda9069$ line fluxes, and so cannot be plotted on these diagrams. Note that the behaviour of both abundance indicators is similar, being relatively insensitive to abundance below $12 + \log({\rm O/H}) < 8.0$, and showing considerable sensitivity to ionisation parameter. }\label{fig16}
\end{figure}

In Figure \ref{fig16} we show the two ratios used by \citep{Stasinska06a}. Here we have used the excitation-sensitive [\ion{S}{3}]$\lambda9069$/[\ion{S}{2}]$\lambda6717,31$ ratio as the prime ionisation parameter diagnostic. Both abundance indicators are similar, being relatively insensitive to abundance below $12 + \log({\rm O/H}) < 8.0$, and showing considerable sensitivity to ionisation parameter. The data sets we are using do not have the  [\ion{S}{3}]$\lambda9069$ line fluxes, and so cannot be plotted on these diagrams. In addition, the excitation-sensitive [\ion{O}{3}]$\lambda5007$/[\ion{S}{2}]$\lambda6717,31$ ratio cannot be substituted on the $y-$axis, as the ratios then become degenerate. 

As pointed out by  \citep{Stasinska06a}, these ratios work because they (indirectly) measure the electron temperature in the high-ionisation zone of the \HII region. At high abundances the temperature of this zone changes rapidly with abundance, giving the observed sensitivity at the high abundance end, but at the low-abundance end of the scale, the temperature sensitivity to abundance is much weaker.

\FloatBarrier

\subsubsection{\NIIOII\ as an abundance diagnostic}
\begin{figure}[htpb]
\includegraphics[scale=1.0]{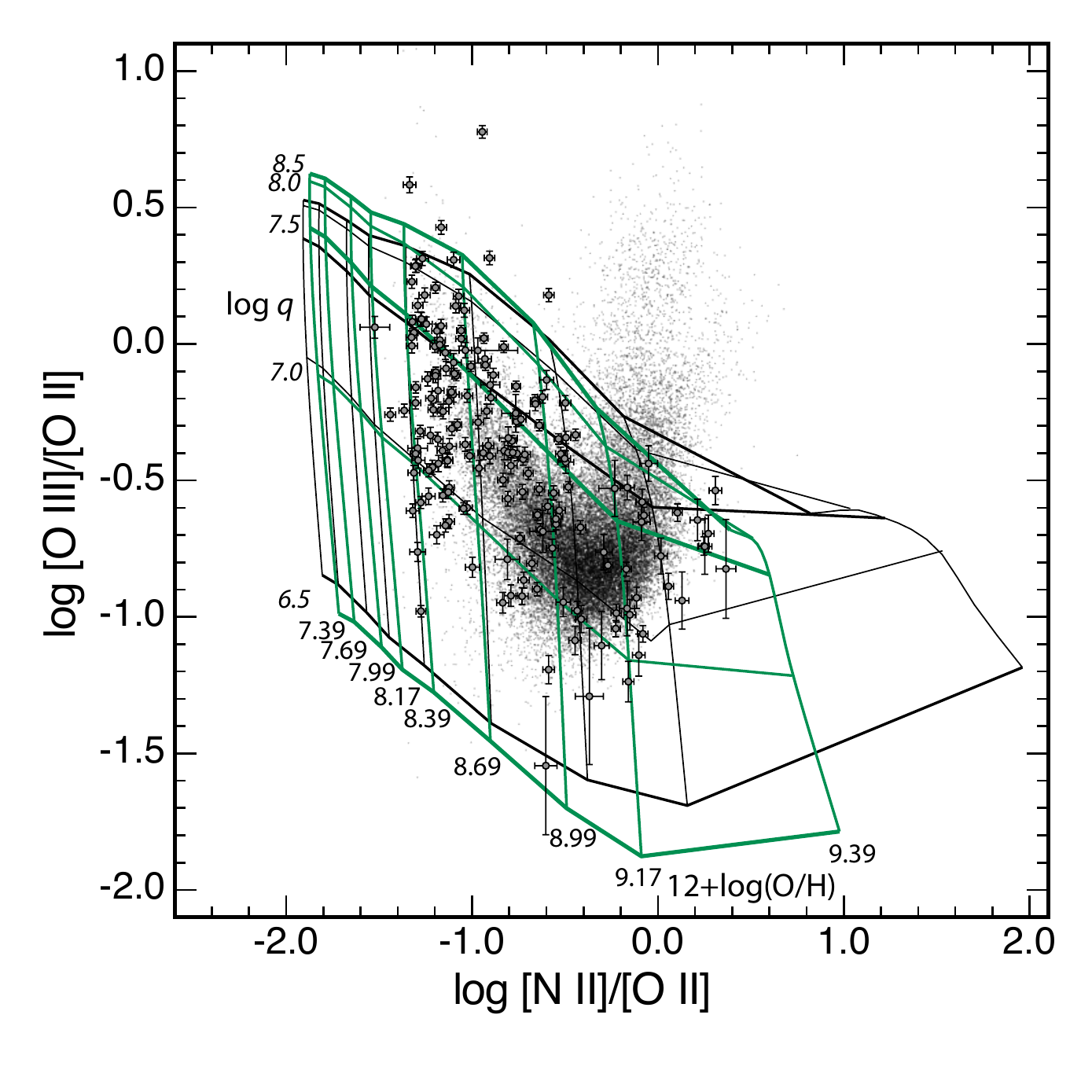}
\caption{ The \citet{Dopita00} diagnostic diagram. This clearly separates the abundance from the ionisation parameter. Note that the SDSS galaxies and the \citet{vanZee98} sample of \HII regions have a relatively restricted range of abundance parameters. Also, the AGN sequence is quite distinct in this diagnostic.}\label{fig17}
\end{figure}

\citet{Dopita00} separated the effects of abundance and ionisation parameter by using  \ratioNIIOII\  as the prime abundance diagnostic, and using the excitation-sensitive [\ion{O}{3}]$\lambda5007$/[\ion{O}{2}]$\lambda \lambda 3727,9$ ratio as the prime ionisation parameter diagnostic. Our re-computation of this diagnostic is shown in Figure \ref{fig17}. The  \ratioNIIOII\ ratio is particularly sensitive to abundance for two reasons. First, nitrogen has a large secondary component of nucleosynthesis at high abundance, see Figure \ref{fig3}, which ensures an increase of  \NIIOII , and second, the nebular electron temperature falls systematically as the abundance increases. This ensures that collisional excitations of the [\ion{O}{2}]$\lambda \lambda 3727,9$ lines are quenched at the high abundance end of the scale.  

 Several points are to be noted. First, the effect of the $\kappa$-distribution on the implied chemical composition is small, but nonetheless significant. In general, the $\kappa$-distribution leads to systematically higher derived chemical abundances. Likewise, the $\kappa$-distribution tends to systematically decrease the derived ionisation parameters. The most significant difference occurs in the high-$q$ low-$Z$ regime. Second, the vertical scatter of the observational points  on this figure can be ascribed to intrinsic variability in $\log q$ between different \HII regions. Most of the \citet{vanZee98} \HII regions have $\log q$ in the range 7.0--7.5, with the high-$q$ outliers tending to be associated with low-abundance \HII regions. Third, the SDSS galaxies display a systematically smaller abundance spread than the \citet{vanZee98} sample, consistent with the fact that the SDSS spectra are heavily weighted towards the central regions of galaxies, while most of the \citet{vanZee98} \HII regions are located in the spiral arms, which have lower oxygen abundance due to the presence of galactic abundance gradients. A few of the \citet{vanZee98} \HII regions have extremely high abundances. These \HII regions are located in very luminous disk galaxies such as NGC 1068, NGC 1637 and NGC 3184. Lastly, the SDSS galaxies clearly show the AGN branches emerging in the vertical direction from the main cloud of galaxies. This implies that most of the AGN in the Local Universe are associated with super-solar chemical abundances; 12+$\log$(O/H)$ \gtrsim 9.0$.
 
\begin{figure}[htpb]
\includegraphics[scale=1.0]{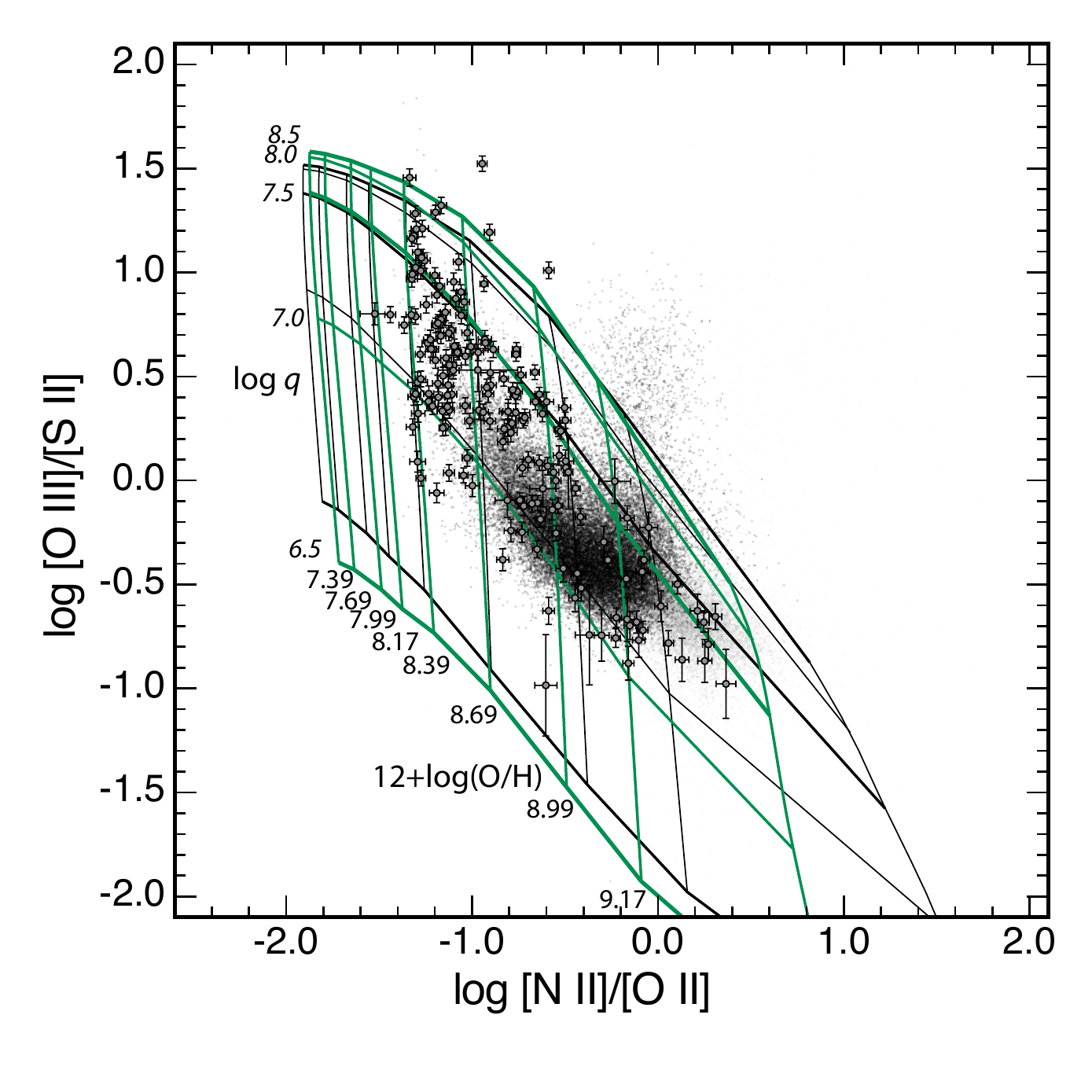}
\caption{ As Figure \ref{fig17}, above, but substituting  [\ion{O}{3}]$\lambda5007$/[\ion{S}{2}]$\lambda6717,31$ in the place of  [\ion{O}{3}]$\lambda5007$/[\ion{O}{2}]$\lambda \lambda 3727,9$. The [\ion{O}{3}]/[\ion{S}{2}] ratio is more sensitive to abundance, but some of the scatter is reduced because the [\ion{O}{3}]/[\ion{S}{2}] ratio is much less sensitive to reddening corrections than the [\ion{O}{3}]/[\ion{O}{2}] ratio. The ionisation parameter is more closely confined to $7.6\gtrsim \log(q)  \gtrsim 6.9$. The two grids agree closely as to the abundance of the \HII regions.}\label{fig18}. 
\end{figure}

 We had already demonstrated in Figure \ref{fig13}, above, that either  [\ion{O}{3}]$\lambda5007$/[\ion{O}{2}]$\lambda \lambda 3727,9$ or [\ion{O}{3}]$\lambda5007$/[\ion{S}{2}]$\lambda6717,31$ can provide good excitation diagnostics. Figure \ref{fig18} shows the effect of making this substitution in the \citet{Dopita00} diagnostic which we have just discussed. The abundances implied by both diagnostics agree closely, as they should, since only the  \ratioNIIOII\  is sensitive to abundance. However, the scatter in the inferred $\log(q)$ is reduced in  Figure \ref{fig18}. This is almost certainly because the reddening corrections and their associated errors are much smaller for the  [\ion{O}{3}]/[\ion{S}{2}] ratio than for the  [\ion{O}{3}]/[\ion{O}{2}] ratio. It is now evident that the observed range of ionisation parameter for either the spiral arm \HII regions or the SDSS nuclear spectra is rather restricted; most objects are located in the narrow band $6.9 \lesssim \log(q)  \lesssim 7.6$.
 
\begin{figure}[htpb]
\includegraphics[scale=1.0]{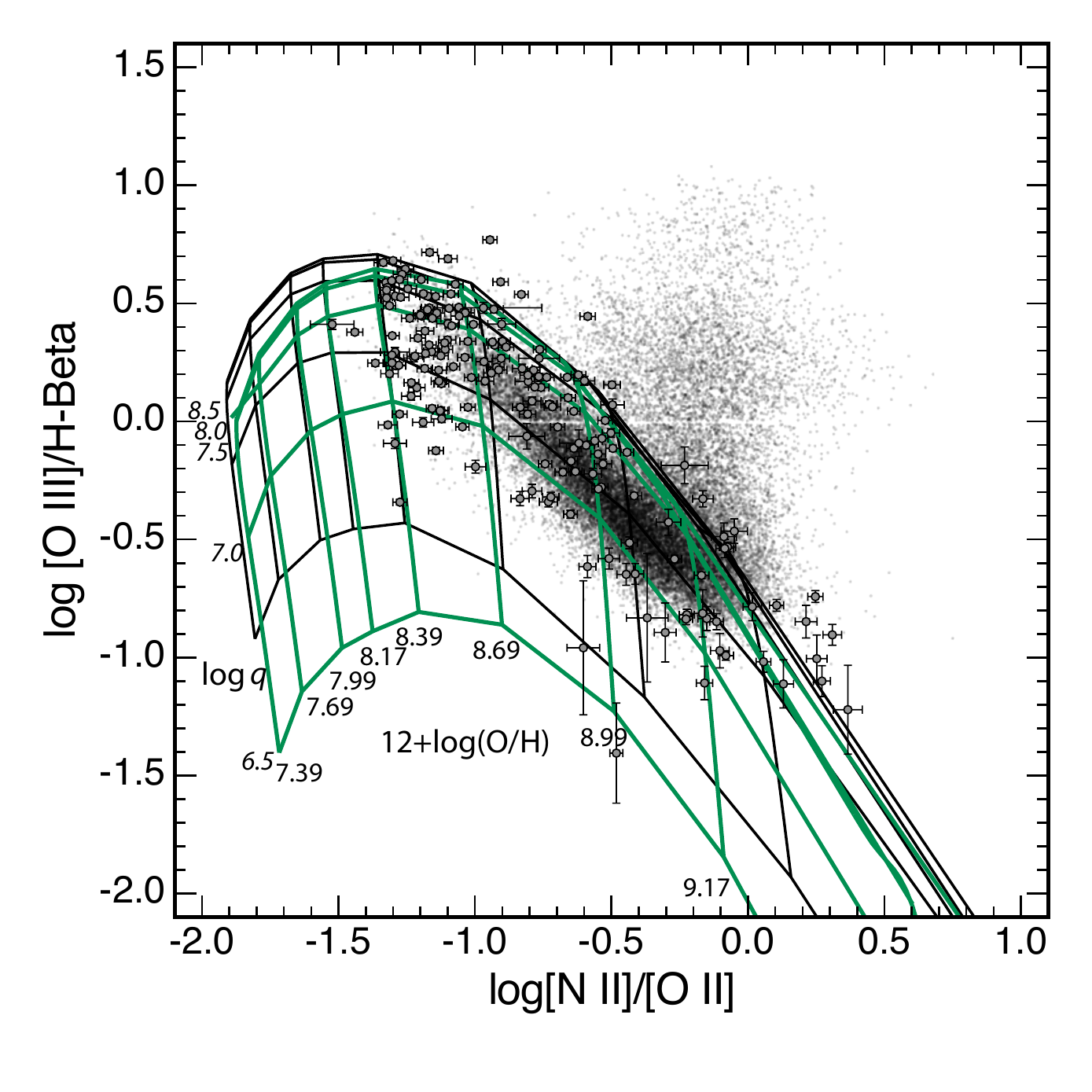}
\caption{ As Figure \ref{fig17}, above, but substituting  [\ion{O}{3}]$\lambda5007$/H$\beta$ in the place of  [\ion{O}{3}]$\lambda5007$/[\ion{O}{2}]$\lambda \lambda 3727,9$. This is not so useful to determine $\log(q)$, but the sharp upper boundary of the theoretical models suggests that this diagram is very useful for distinguishing AGN and transitional types from the normal star-forming galaxies.}\label{fig19}. 
\end{figure}

 If we try to use the excitation-sensitive [\ion{O}{3}]$\lambda5007$/H$\beta$ on the $y$-axis, we obtain the diagnostic diagram in Figure \ref{fig19}. This is degenerate in terms of the ionisation parameter over a wide range for higher values of $\log(q)$. This diagram is probably better suited to separate the AGN galaxies from the normal star-forming galaxies, despite the fact that it distinguishes between the LINER and Seyfert sequences rather poorly.  
\FloatBarrier

\FloatBarrier

 \subsubsection{ [\ion{N}{2}]/[\ion{S}{2}] as an abundance diagnostic}
\begin{figure}[htpb]
\includegraphics[scale=1.0]{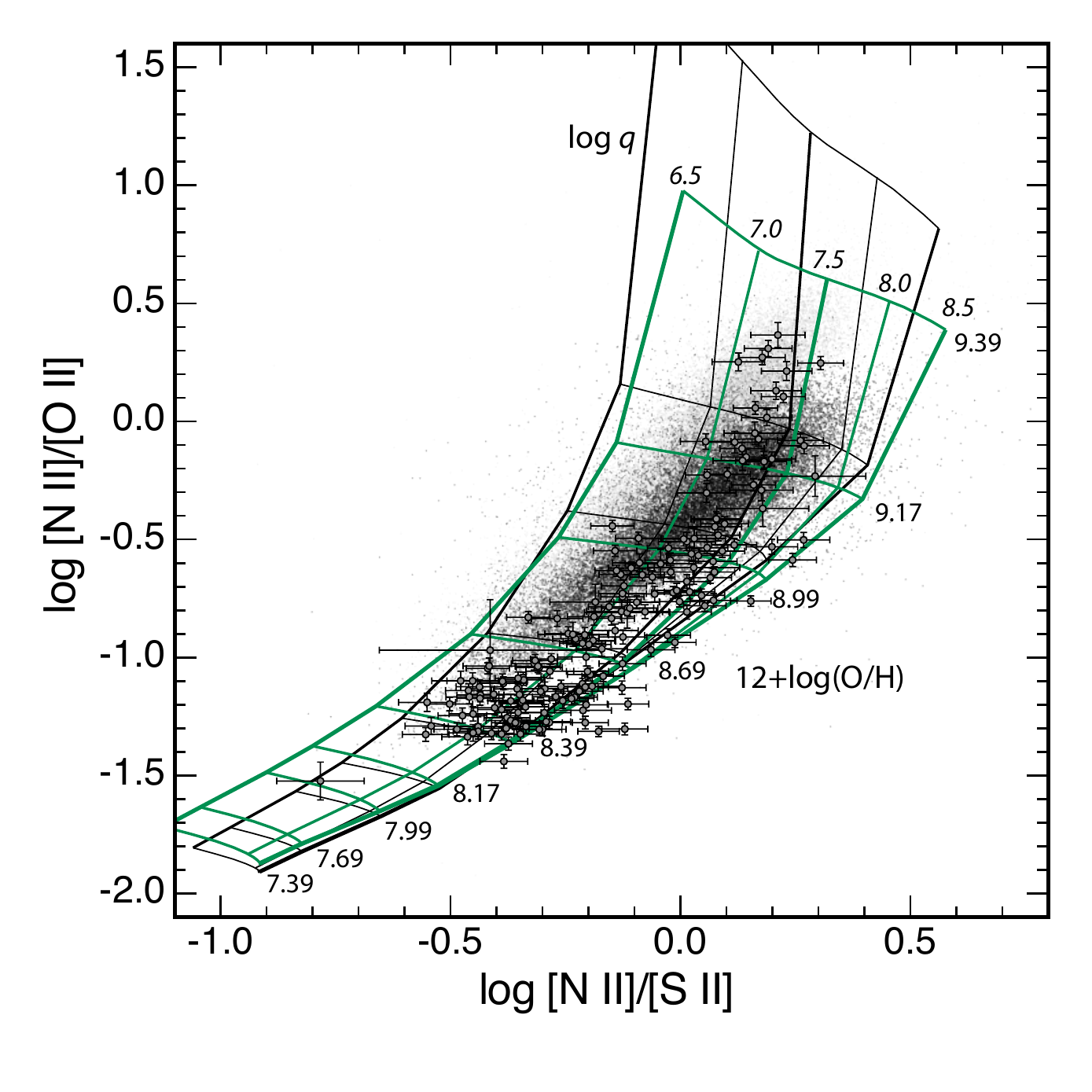}
\caption{ [\ion{N}{2}]/[\ion{S}{2}] and [\ion{N}{2}]/[\ion{O}{2}] compared as abundance diagnostics. Clearly both are sensitive to abundance, but for [\ion{N}{2}]/[\ion{S}{2}]  the sensitivity is weaker, and there is a greater sensitivity to $\log(q)$. The AGN sequence is not distinguished in this diagnostic diagram.}\label{fig20}. \end{figure}

Given that  \ratioNIIOII\ is the ratio of an element formed in intermediate mass stars to a standard $\alpha$-process element, it is reasonable to ask whether another $\alpha$-process element could be substituted for oxygen. An obvious candidate to use is [\ion{S}{2}]$\lambda6717,31$. In Figure \ref{fig20} we compare [\ion{N}{2}]$\lambda6584$/[\ion{S}{2}]$\lambda6717,31$ to the \ratioNIIOII\  ratio. Both ratios are sensitive to abundance, except that for [\ion{N}{2}]$\lambda6584$/[\ion{S}{2}]$\lambda6717,31$ there is a greater sensitivity to the ionisation parameter, which is a consequence of the mis-match of the ionisation potential of  \ion{S}{2} as compared to either  \ion{N}{2} or  \ion{O}{2}. In addition, the sensitivity of the ratio to abundance is less, because the ratio of collisional excitation rates of  [\ion{N}{2}] and  [\ion{S}{2}] is a very weak function of nebular temperature.

The great advantage in the use of [\ion{N}{2}]/[\ion{S}{2}] as an abundance diagnostic is that reddening corrections are negligible, facilitating an accurate determination of the line ratio. To minimise the reddening corrections in the determination of the excitation, an obvious choice is to use the [\ion{O}{3}]/[\ion{S}{2}] ratio, since the \ion{S}{2} line is close to H$\alpha$, the \ion{O}{3} line is close to H$\beta$, and the intrinsic Balmer Decrement is well defined. In this context, we should note that our models provide a systematically higher H$\alpha$/H$\beta$ ratio than the standard Case B recombination value, as can be seen in Table \ref{table_5}. This is a result of an important contribution of collisional excitation from the metastable $2^1{\rm S}_{1/2}$ level to the H$\alpha$ line flux, since there is a large resonance just above threshold in the collisional cross section of this line.

\begin{figure}[htpb]
\includegraphics[scale=1.0]{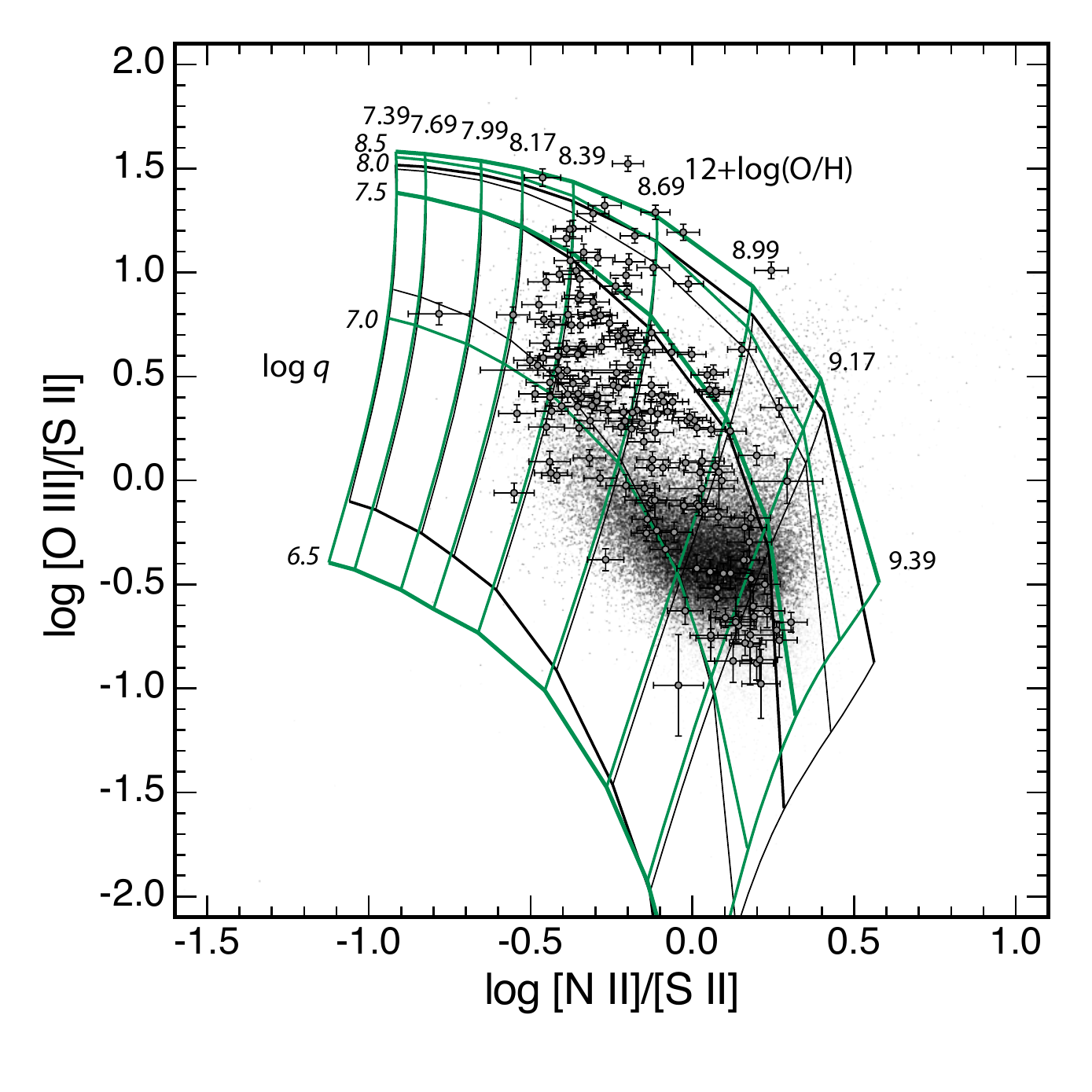}
\caption{ [\ion{N}{2}]/[\ion{S}{2}] vs. [\ion{O}{3}]/[\ion{S}{2}]. This new diagnostic diagram is valuable for several reasons. First, it provides an excellent separation of $\log(q)$ and 12+$\log$(O/H). Second, the reddening corrections are simple to make. Third, only a limited spectral coverage is required.}\label{fig21}
\end{figure}

Figure \ref{fig21} shows [\ion{N}{2}]/[\ion{S}{2}] vs. [\ion{O}{3}]/[\ion{S}{2}]. This new diagnostic is very useful, since it provides an excellent discrimination between 12+$\log$(O/H) and $\log(q)$ over the full range of both these parameters. Comparing with Figure \ref{fig18}, it is clear that observational data provide very similar solutions for both of these parameters. In addition, spectra of limited wavelength coverage and poor spectrophotometric calibration can be used to provide robust solutions for both 12+$\log$(O.H) and $\log(q)$. Finally, for this diagnostic it hardly matters whether the $\kappa$-distribution applies or not, since both theoretical grids overlap almost perfectly.

\begin{figure}[htpb]
\includegraphics[scale=1.0]{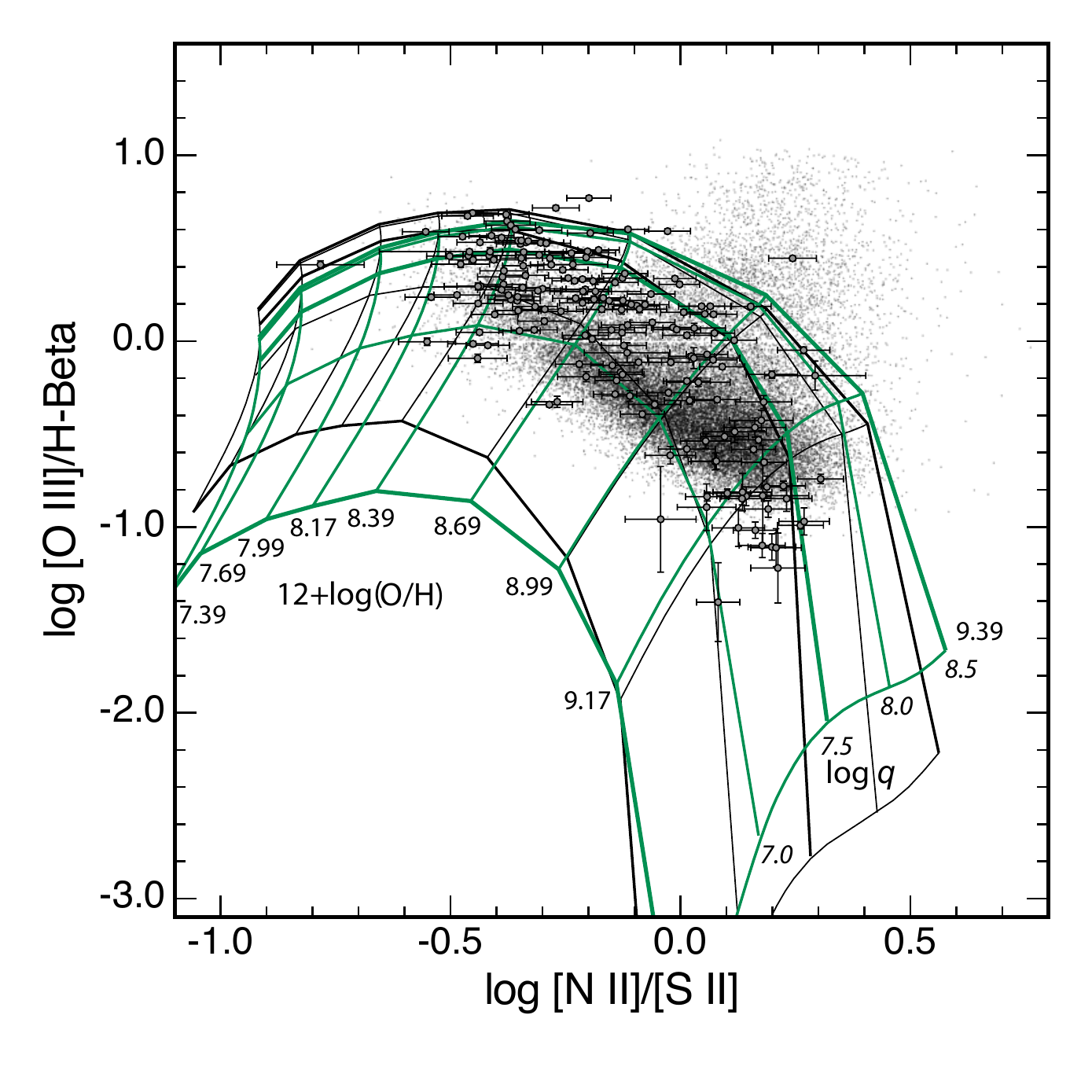}
\caption{ [\ion{N}{2}]/[\ion{S}{2}] vs. [\ion{O}{3}]/H$\beta$. This new diagnostic diagram is useful for the same reasons as Figure \ref{fig21}, except perhaps at the high ionisation parameter, low abundance limit.}\label{fig22}
\end{figure}

A useful diagnostic is also obtained if we substitute  [\ion{O}{3}]/H$\beta$ for  [\ion{O}{3}]/[\ion{S}{2}] as our excitation-dependent diagnostic ratio. This is shown in Figure \ref{fig22}. This suffers a little from the issues of Figure \ref{fig19} in that it is not very capable of distinguishing $q$ at the high ionisation parameter limit. This tendency is more marked at the low-abundance end. However, the AGN sequence is well distinguished, and like Figure \ref{fig21}, it has the advantage that spectra of limited wavelength coverage and poor spectrophotometric calibration can be used to provide a good solution.

\FloatBarrier

 \subsubsection{ The \citet{Alloin79} abundance diagnostic}
\begin{figure}[htpb]
\includegraphics[scale=0.85]{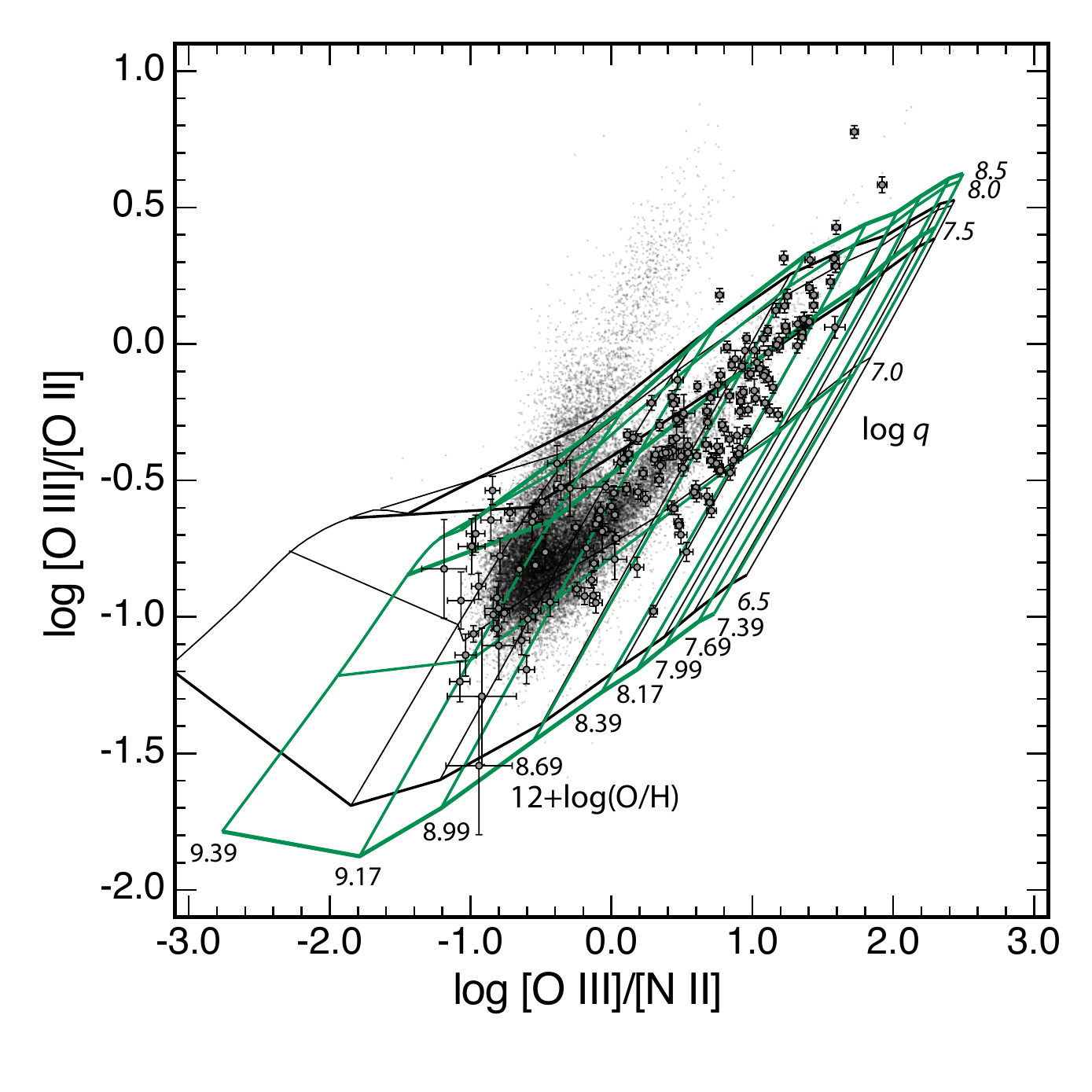}
\caption{ The \citet{Alloin79} abundance-sensitive diagnostic  [\ion{O}{3}]/[\ion{N}{2}] plotted against the excitation-dependent [\ion{O}{3}]/[\ion{O}{2}] ratio. Both ratios are also sensitive to $\log q$, but provide sufficient sensitivity to both abundance and ionisation parameter to make this a useful diagnostic diagram. In addition, the AGN branch is quite distinct. }\label{fig23}
\end{figure}

 \citet{Alloin79} suggested that the ratio [\ion{O}{3}]/[\ion{N}{2}] could provide a good abundance diagnostic since they demonstrated a good correlation between this ratio and the measured electron temperature, and in turn the electron temperature is inversely correlated with chemical composition (\emph{c.f.} Figure \ref{fig4}, above). However, as we have amply demonstrated above, no one ratio provides either a clean abundance diagnostic or a clean ionisation parameter diagnostic. All are sensitive to both parameters. Figure \ref{fig23} brings out this point. Here, we have plotted [\ion{O}{3}]/[\ion{N}{2}] against the the excitation-dependent [\ion{O}{3}]/[\ion{O}{2}] ratio. This figure also provides a clean separation of the abundance from the ionisation parameter over the full range of these parameters, and shows very little sensitivity to the value of $\kappa$.
 
\FloatBarrier

\section{Applying the Abundance Diagnostics}\label{Application}
\subsection{Self-Consistency}
Before applying the new diagnostics, it is mandatory to check them for self-consistency. For this purpose, we have selected the four best diagnostics on the grounds that they should adequately separate the two parameters, $\log(q)$ and 12+$\log$(O/H), and be sensitive to these over the full range of both parameters. Bearing in mind that many of the diagnostics are not truly independent, since they employ the same line ratios in at least one axis, we have selected a subset of four, two based upon [\ion{N}{2}]/[\ion{O}{2}], and two based upon [\ion{N}{2}]/[\ion{S}{2}]:
\begin{enumerate}
\item{ [\ion{N}{2}]/[\ion{O}{2}] vs.  [\ion{O}{3}]/[\ion{O}{2}]}
\item{ [\ion{N}{2}]/[\ion{O}{2}] vs.  [\ion{O}{3}]/[\ion{S}{2}]}
\item{ [\ion{N}{2}]/[\ion{S}{2}] vs.  [\ion{O}{3}]/H$\beta$, and} 
\item{ [\ion{N}{2}]/[\ion{S}{2}] vs.  [\ion{O}{3}]/[\ion{S}{2}].}
\end{enumerate}

\begin{figure}[htpb]
\includegraphics[scale=0.8]{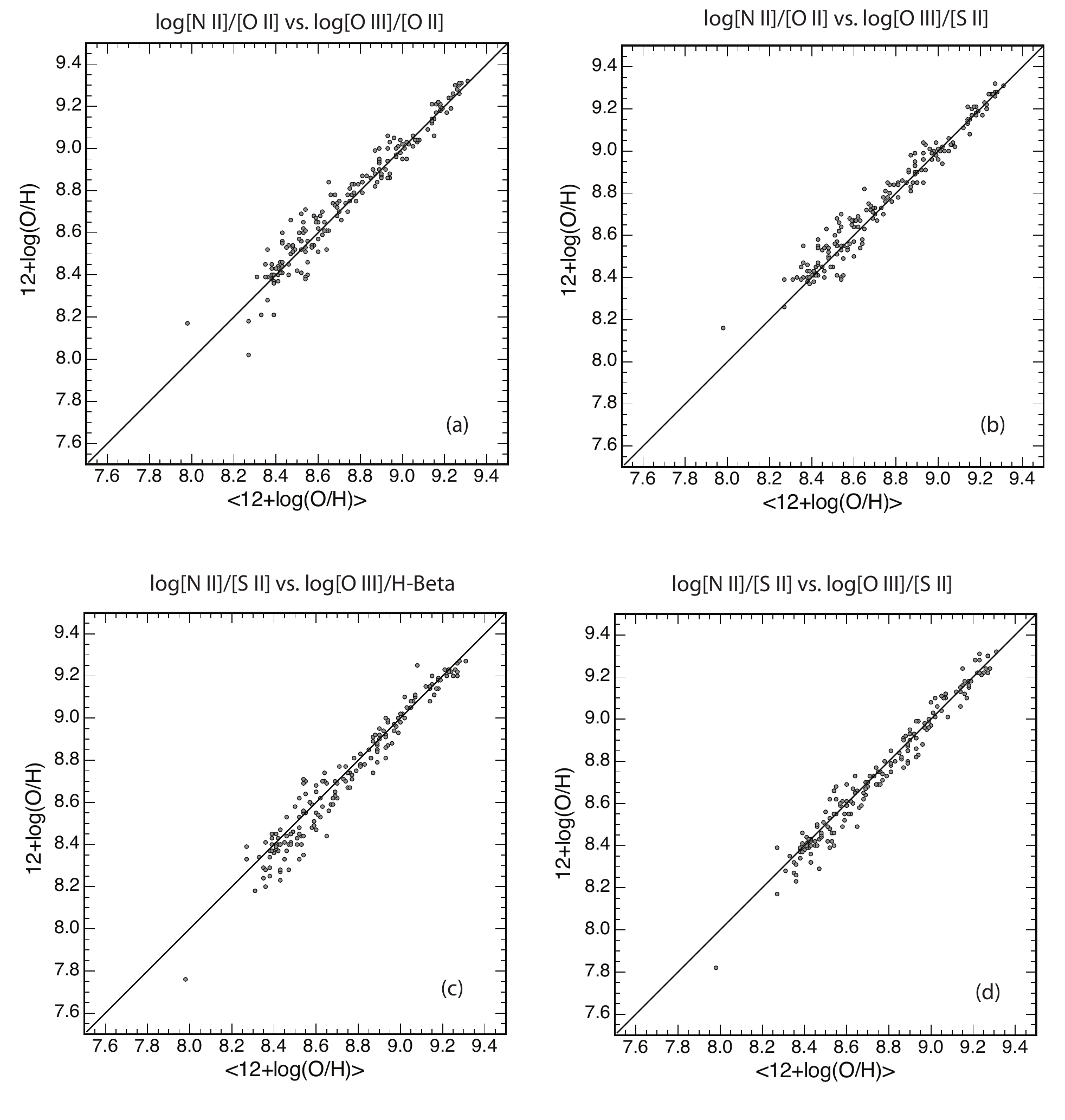}
\caption{ Chemical abundances derived for the \citet{vanZee98} \HII regions using each of the four diagnostics given on the label, plotted against the average of all four.}\label{fig24}
\end{figure}

\begin{figure}[htpb]
\includegraphics[scale=0.8]{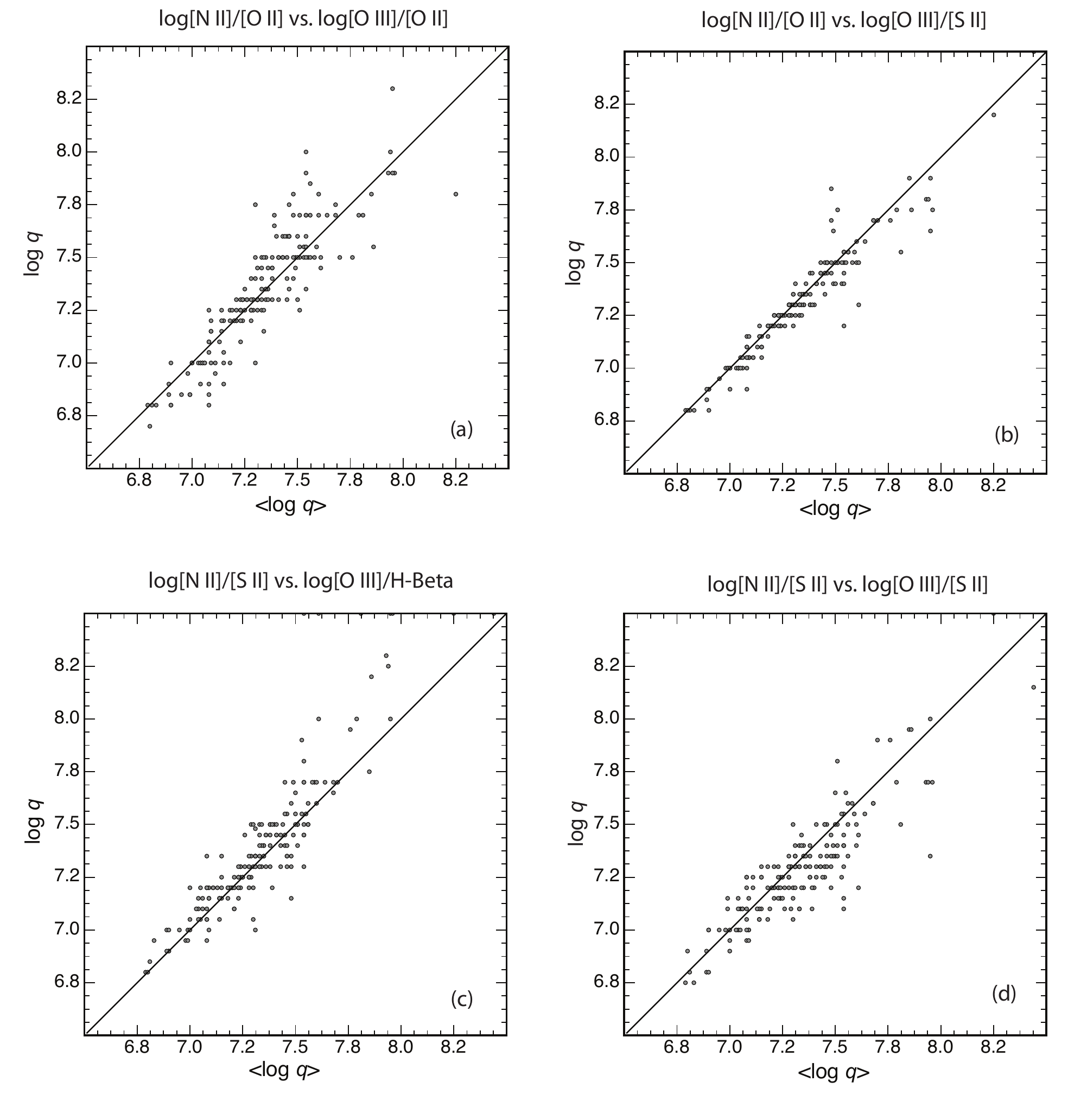}
\caption{ As figure \ref{fig24}, but for the derived ionisation parameter.}\label{fig25}
\end{figure}

For the observational test set we have used the homogeneous \citet{vanZee98} observations, which cover a wide abundance range of \HII regions in several galaxies. For each of these, we have graphically solved for the implied oxygen abundance (to the nearest 0.01dex) and for the ionisation parameter (to the nearest 0.1dex) using our four diagnostics. A value of $\kappa = 20$ was assumed, on the basis of the discussion in Section \ref{Kappa}. We then formed a global average for each of the parameters using all for diagnostics. The result is shown in Figure \ref{fig24} for the chemical abundances, and in Figure \ref{fig25} for the ionisation parameter.

It is clear that all four methods are in remarkably close agreement with each other. For those abundance diagnostics involving [\ion{O}{2}], the scatter is somewhat larger, presumably reflecting increased photometric and reddening correction errors. As expected, the [\ion{N}{2}]/[\ion{S}{2}] vs.  [\ion{O}{3}]/[\ion{S}{2}] diagnostic gives the smallest scatter. There appears to be little or no systematic difference in derived abundance as a function of abundance for any of the diagnostics.

For the ionisation parameter, [\ion{N}{2}]/[\ion{O}{2}] vs.  [\ion{O}{3}]/[\ion{S}{2}] gives the smallest scatter. For  [\ion{N}{2}]/[\ion{O}{2}] vs.  [\ion{O}{3}]/[\ion{O}{2}] a small systematic trend towards higher derived $\log(q)$ at higher $q$ is apparent. [\ion{N}{2}]/[\ion{S}{2}] vs.  [\ion{O}{3}]/[\ion{S}{2}] gives the greatest scatter in derived $\log(q)$. However these effects are small, and the global solution using all four diagnostics appears to be robust.
\FloatBarrier

\subsection{Comparison with \citet{Kewley02}}
Given that the \citet{Kewley02} work was based upon an earlier version of the MAPPINGS code, it is interesting to see how our new abundance diagnostics compare with that earlier work. The main changes in the code that have occurred in the eleven years since are:
\begin{itemize}
\item{A proper match of the stellar and nebular abundances.}
\item{Use of \citet{Grevesse10} revised abundance set and new CN abundance variations.}
\item{Inclusion of the effect of radiation pressure.}
\item{Use of spherical geometry rather than plane parallel geometry.}
\item{Improved atomic data (as described above)}
\item{Inclusion of the possibility of $\kappa$-distributed electrons.}
\end{itemize}

\begin{figure}[htpb]
\includegraphics[scale=0.8]{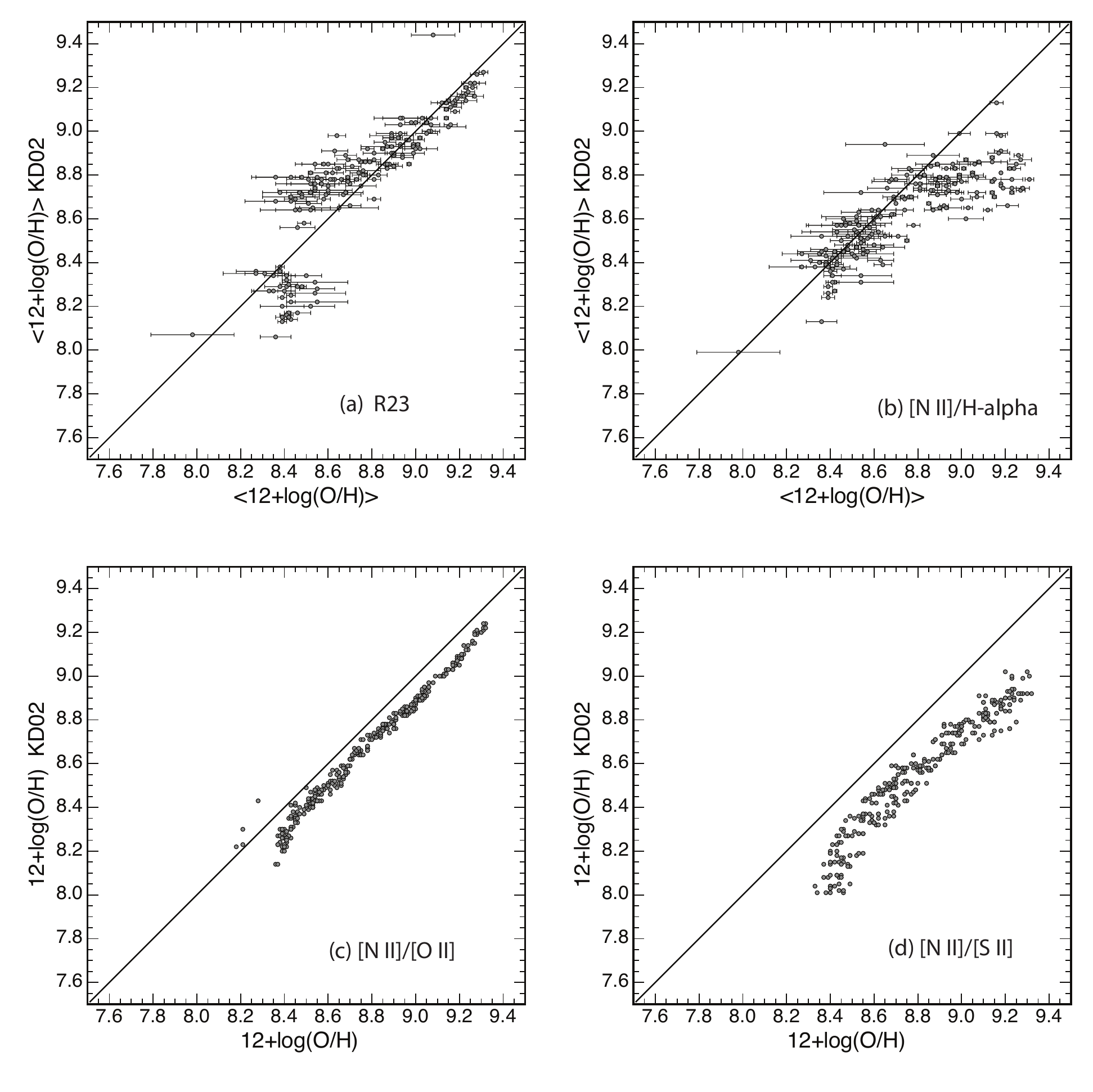}
\caption{ The abundances derived from the \citet{vanZee98} \HII regions using the \citet{Kewley02} abundance diagnostics, compared with those of this paper. In panels (a) and (b) we compare the $R_{23}$ method and the  [\ion{N}{2}]/H$\alpha$ method with the mean of our abundance diagnostics. The error bars show the internal RMS dispersion for our abundance estimates. In panels (c) and (d) we use the [\ion{N}{2}]/[\ion{O}{2}] vs.  [\ion{O}{3}]/[\ion{O}{2}]  and the  [\ion{N}{2}]/[\ion{S}{2}] vs.  [\ion{O}{3}]/[\ion{O}{2}]  diagnostics from the  \citet{Kewley02} paper, compared with these same diagnostics from this paper.
}\label{fig26}
\end{figure}

Again, we have used the homogeneous \citet{vanZee98} observations to facilitate this comparison, reducing the data with the procedure described in \citet{Kewley02}. This provides four abundance diagnostics, based on, respectively,  the $R_{23}$ calibration, the [\ion{N}{2}]/[\ion{O}{2}] vs.  [\ion{O}{3}]/[\ion{O}{2}] diagnostic, the [\ion{N}{2}]/[\ion{S}{2}] vs.  [\ion{O}{3}]/[\ion{O}{2}] and the [\ion{N}{2}]/H$\alpha$ ratio. The results are shown in Figure \ref{fig26}. 

The correlation between our abundances and the \citet{Kewley02} $R_{23}$ abundances is good at the high abundance end. However, the scatter is large in the region $8.2 \lesssim 12+\log({\rm O/H}) \lesssim 8.7$, where the  $R_{23}$ indicator is almost insensitive to O/H. The  [\ion{N}{2}]/H$\alpha$ method produces large scatter, with a systematic offset at the high abundance limit. As expected, the [\ion{N}{2}]/[\ion{O}{2}] vs.  [\ion{O}{3}]/[\ion{O}{2}] diagnostics agree very well with one another. The systematic offset can be largely ascribed to the re-calibration of the N/O abundance with respect to O/H (which provides both an offset and the curvature seen at low abundance) and the change in the stellar EUV spectra. However, the [\ion{N}{2}]/[\ion{S}{2}] vs.  [\ion{O}{3}]/[\ion{O}{2}] diagnostic shows both a large ($\sim 0.3$dex) offset and marked curvature.

\begin{figure}[htpb]
\includegraphics[scale=1.0]{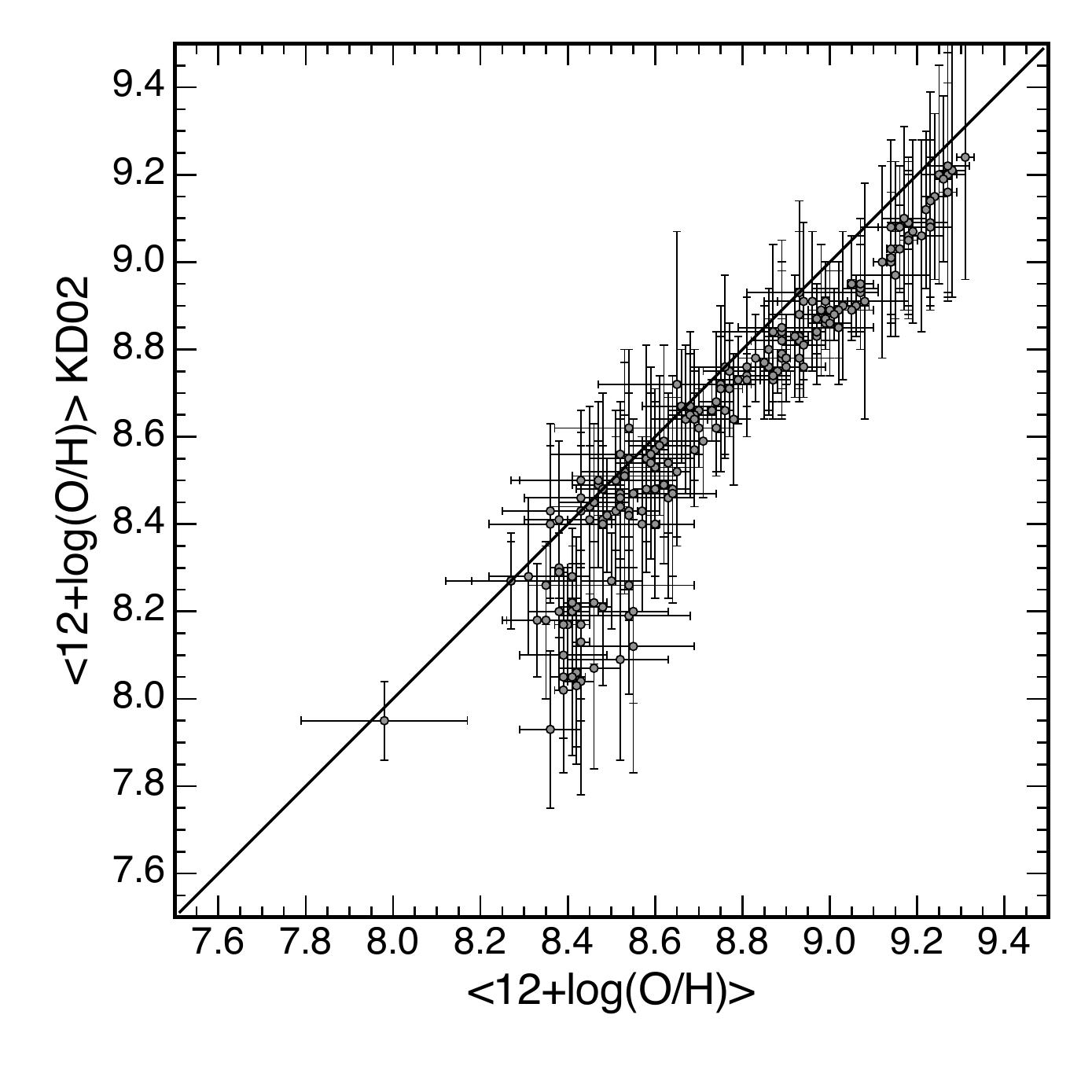}
\caption{ The mean of the  \citet{Kewley02} diagnostics plotted against the mean of the diagnostics used in this paper. The error bars are the RMS scatter of the four diagnostics used to create the mean in each case. The systematic offset can be largely ascribed to the offset of the [\ion{N}{2}]/[\ion{O}{2}] and [\ion{N}{2}]/[\ion{S}{2}] diagnostics.}\label{fig27}
\end{figure}

The average of all four \citet{Kewley02} diagnostics is given in Figure \ref{fig27}. The overall correlation is good, and this implies that previous results on the chemical composition of galaxies based on the \citet{Kewley02} diagnostics do not need revision except perhaps at the low- and high-abundance extremes.
\FloatBarrier

\subsection{Comparison with \citet{vanZee98}}
\begin{figure}[htpb]
\includegraphics[scale=1.0]{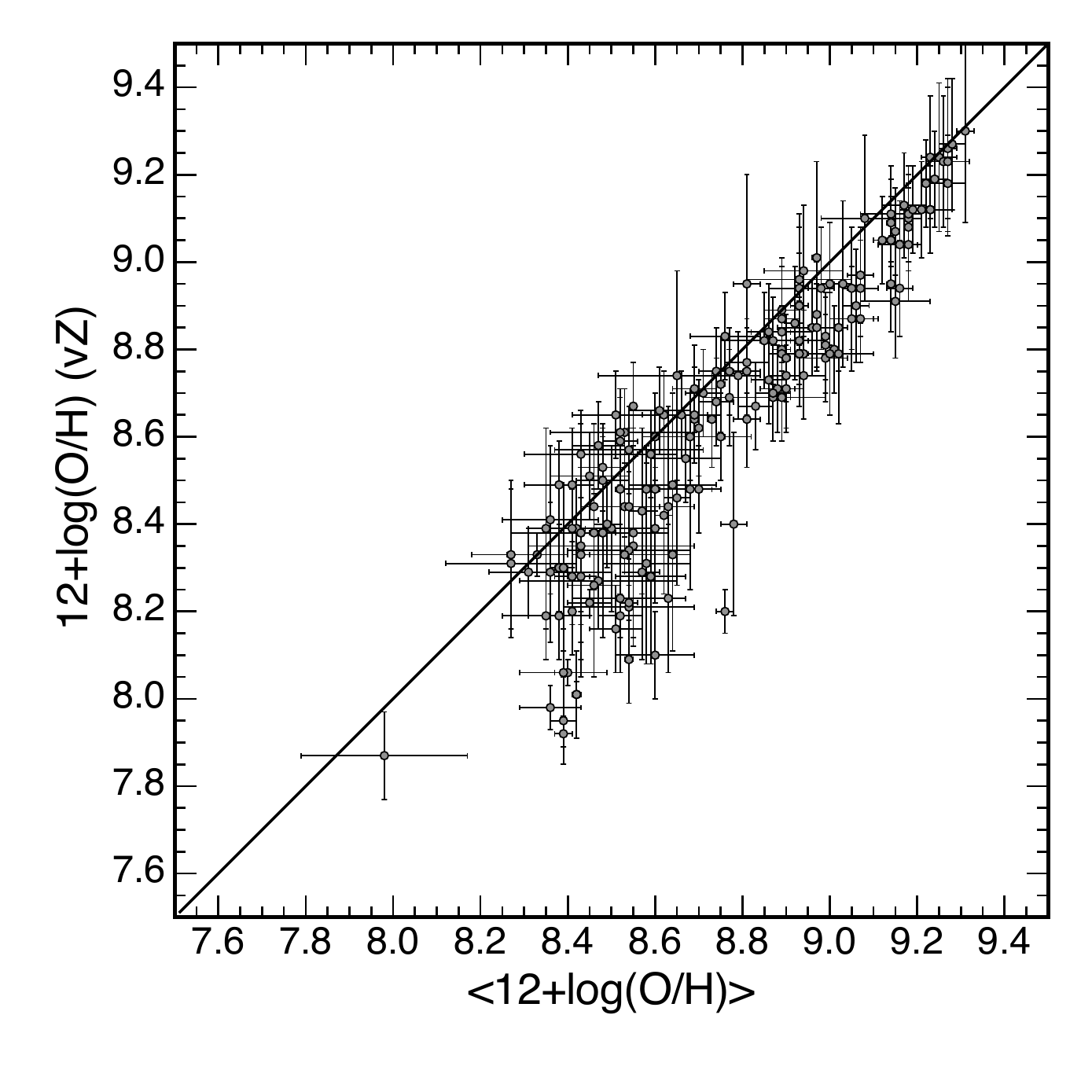}
\caption{ The abundances derived by \citet{vanZee98} for their \HII regions compared with the abundances derived here. Note the close similarity with Figure \ref{fig27}. Here the systematic offset of $\sim 0.2$ dex. can be understood as another manifestation of the systematic offset always found between strong-line and $T_e$-based abundance determinations \citep{LopezSanchez12}. }\label{fig28}
\end{figure}

In their paper, \citet{vanZee98} used a hybrid technique to determine abundances, based upon both strong lines and on a calibration of the excitation with the $T_e$ measured for a small subset of her sample. In essence, therefore, this is essentially a $T_e$-based calibration, similar to those used by Pilyugin and his collaborators \citep{Pilyugin01a,Pilyugin01b,Pilyugin05,Pilyugin11,Pilyugin12}. In Figure \ref{fig28} we show the correlation between the \citet{vanZee98} abundances and those derived in this paper. Note that, although the correlation is very good, there is a small systematic offset between the two in the same sense as as is usually found for strong line methods calibrated using photoionisation models compared with methods based on $T_e$ (see the discussion of this in the Introduction).

\begin{figure}[htpb]
\includegraphics[scale=0.6]{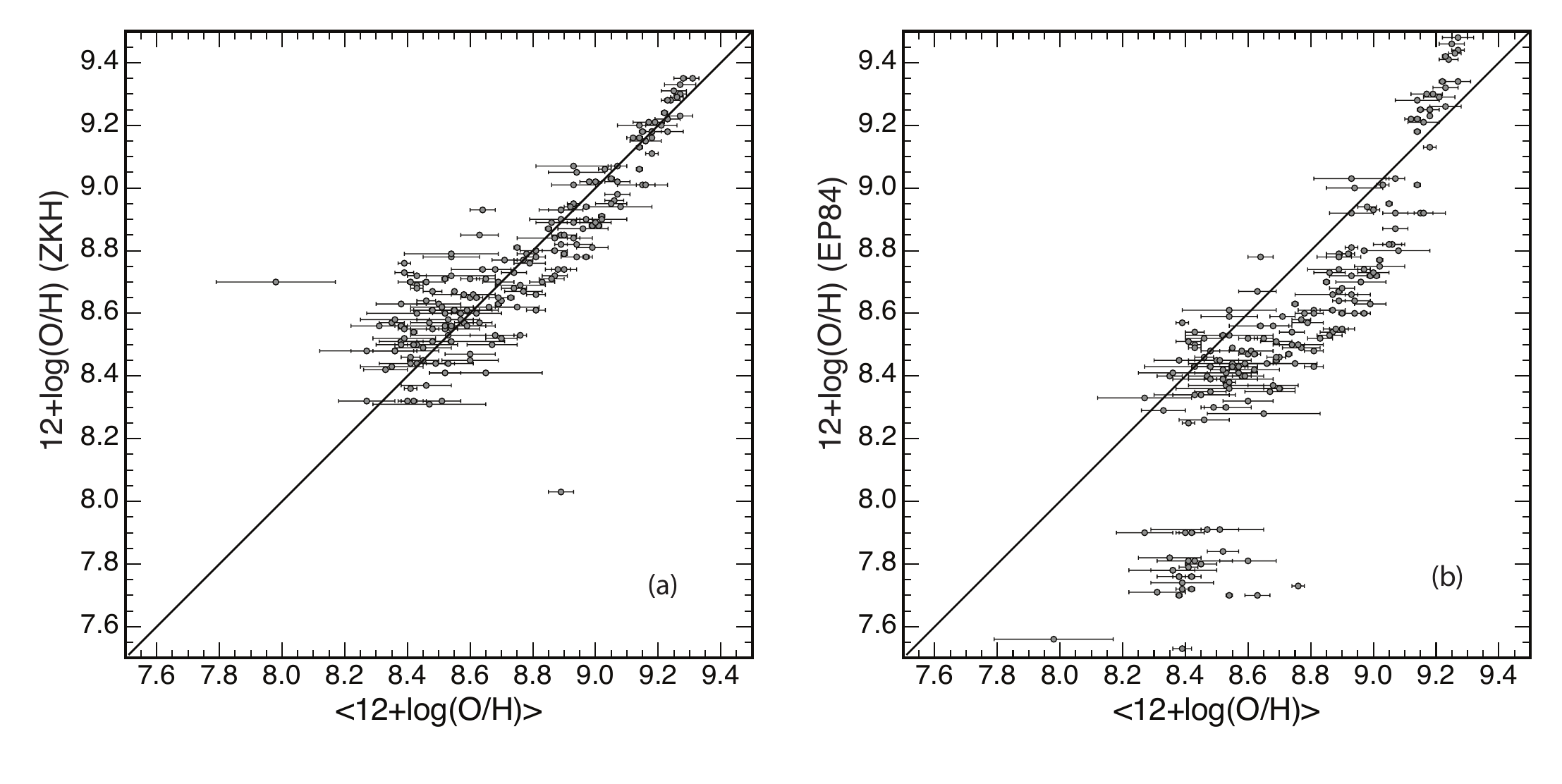}
\caption{ The abundances derived here for the \citet{vanZee98} \HII regions compared with those derived from the methods of \citet{Zaritsky94} and of \citet{EP84}. Both of these are $R_{23}$ techniques, but the \citet{Zaritsky94} method is applicable only to the high abundance branch, which is why the scatter increases below 12+log(O/H)$\sim 8.7$. The \citet{EP84} method is clearly subject to large systematic errors.}\label{fig29}
\end{figure}

\citet{vanZee98} also compared their data with abundances determined by two strong line methods based upon the $R_{23}$ ratio; that of \citet{Zaritsky94} and of \citet{EP84}. The comparison of these two methods with our abundances is shown in Figure \ref{fig29}. Note that the \citet{Zaritsky94} method is applicable only to the high abundance branch, which is why the scatter increases below 12+log(O/H)$\sim 8.7$. Otherwise this method agrees rather closely with our results. As previously found in \citet{Kewley02}, the \citet{EP84} method is subject to large systematic errors.
\FloatBarrier

\subsection{An automated technique to derive abundances}
In this paper we have presented a grid of models covering a wide range of abundance and ionisation parameters typical of \HII regions in galaxies. However, given an observed set of ratios, we need to implement a two dimensional interpolation routine to \emph{read} the diagnostic line ratio grid between the nodes actually computed. For this purpose, we have implemented a dedicated \emph{Python} module to perform this task automatically - the \textsf{pyqz} module. This module relies on the \textsf{griddata} function in the \textsf{scipy.interpolate} module to perform a two dimensional fit to a given diagnostic grid. The  \textsf{griddata} routine allows either a linear or piecewise cubic spline fit to a N-dimensional unstructured dataset. We refer the reader to the \emph{Scipy Reference Guide} for more information on the  \textsf{griddata} function \footnote { The information page for the  \textsf{griddata} function is located at:\newline \url{ http://docs.scipy.org/doc/scipy/reference/generated/scipy.interpolate.griddata.html} .}.

\begin{figure}[htpb]
\includegraphics[scale=0.75]{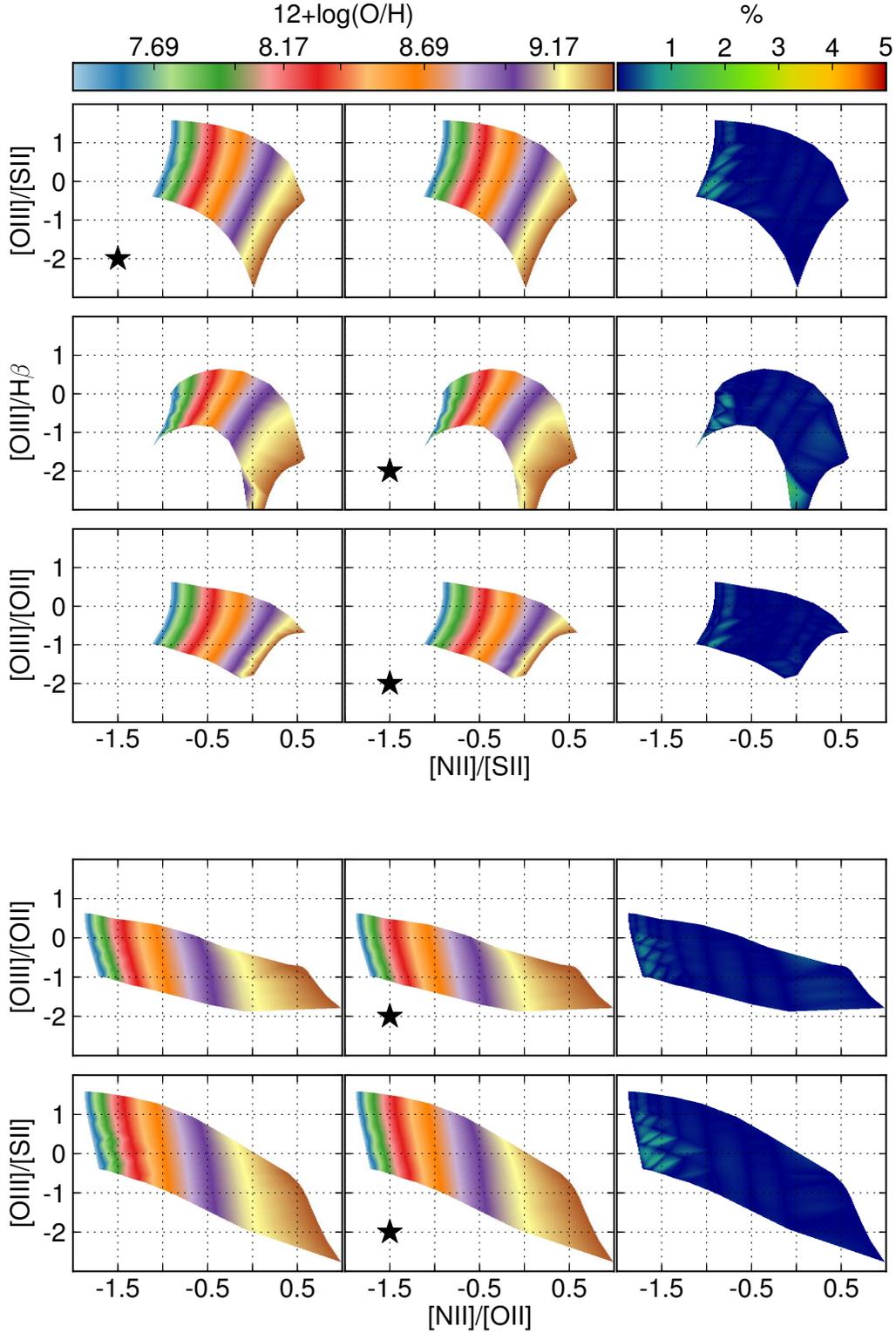}
\caption{Side-by-side comparison between a linear and piecewise cubic interpolation of different diagnostic grids that allow for unambiguous reading of the $\log $O/H value. The third column shows the difference in \% between the two different interpolation methods, and is indicative of how accurately a given diagnostic grid can be read. These grids are for $\kappa=20$.}\label{fig30}
\end{figure}

\begin{figure}[htpb]
\includegraphics[scale=0.75]{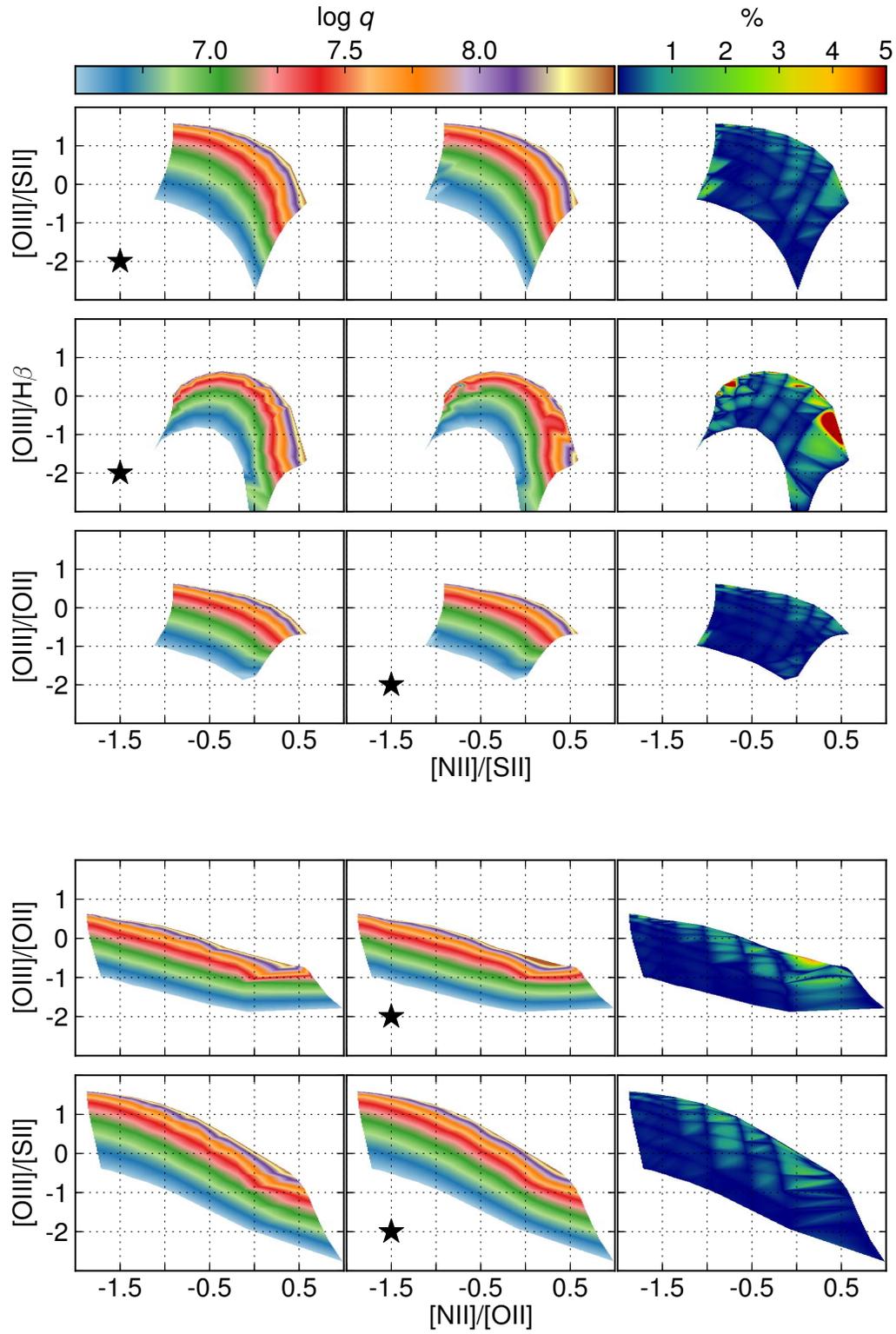}
\caption{As figure \ref{fig30}, but for the ionisation parameter $\log(q)$.}\label{fig31}
\end{figure}

As discussed in Section~\ref{Diagnostics}, several diagnostic grids allow a clear separation of both $\log q$ and 12+$\log$(O/H). In Figure~\ref{fig30} and \ref{fig31}, we use our \textsf{pyqz} module to test how well these different grids can be interpolated to recover the value of $\log q$ or 12+$\log$(O/H), respectively. Each row corresponds to a different diagnostic grid labelled accordingly. In the left and middle column, we show the result of the interpolation performed using the linear or piecewise cubic approach. In the right column, we show the difference, in \%, between the two different interpolation results. The grids in this case are computed for $\kappa=20$.  The error maps in Figure~\ref{fig30} and \ref{fig31} do not represent absolute error on the interpolation results. Nevertheless, they indicate how well a given grid can be read. In most cases, the difference between the two interpolation methods is lower than 5\%. These grids provide consistent results between the two different interpolation methods, with errors below 1\% for most of the interpolation region. 

For both the $\log q$ and 12+$\log$(O/H) grids, we mark with a black star which interpolation method (linear or piecewise cubic) provides the smoothest result, based on a visual comparison of the two different interpolated grid. The absolute grid error associated with the best interpolation method can be expected to be smaller than the error map provided in the right column, which can be used as an upper estimate of the uncertainty associated with reading a given diagnostic grid. We note that the error associated with reading of the grid is much smaller than errors associated with the computation of the grid itself, and observational errors affecting line ratios.

Our \textbf{pyqz} Python module (v0.4) is made freely available for the community to use under the GNU General Public License, and can be downloaded from the Australian National University Data Commons online repository (doi:10.4225/13/516366F6F24ED). This module allows observers to interpolate within any of the line ratio grids listed in Table~\ref{table_7} for $\kappa\in[10,20,50,\infty]$. With any given set of observed line ratios the module returns the corresponding $\log q$ and 12+$\log$(O/H) values for the chosen value of $\kappa$ if the observed ratios lie within a \emph{readable} region of the grid (with no wrapping present). The specific readable regions, for all diagnostic grids and $\kappa$, are listed in Table \ref{table_7} 
\FloatBarrier
\FloatBarrier

\subsection{Implications of $\kappa$ for $T_e$ - based abundance diagnostics}
In a future paper we propose to examine in more detail what are the implications of $\kappa$-distributed electrons on abundances derived by $T_e$ methods. However, here we will give an outline explanation of how  $\kappa$ could help to address the long-standing discrepancy between the abundance scales defined by strong line techniques and that defined by the $T_e$ method.

The $\kappa$-distributed electrons affect the $T_e$ method in three separate ways. First, as pointed out by \citet{Nicholls12,Nicholls13}, the $\kappa$-distribution directly affects the electron temperature measured by the usual temperature-sensitive line ratios such as \O4363 / \OIIIl. This effect is most marked at the high abundance end of the scale (low electron temperature end), as can be seen in Table \ref{table_6}. This effect dies away for abundances below about 0.3 solar (although errors caused by use of the older temperature-averaged collisional strengths persist down to much lower abundances; \citet{Nicholls13}). Since the inferred electron temperature is higher than the electron temperature for Maxwell-Boltzmann distributed electrons, the effect of this is to systematically underestimate the true abundance through the $T_e$ method.

The second factor is that, at the temperatures typical of \HII regions, collisional excitation rates for strong lines such as \OIIIl\ and [\ion{O}{2}]$\lambda \lambda 3727,9$ are reduced in a $\kappa$ distribution, while the strengths of the recombination lines are increased. This weakens these lines relative to the Balmer lines, leading to a further underestimate of the ionic abundances and adding to the systematic offset between strong-line techniques and the $T_e$ method. The weakening of these forbidden lines with respect to the hydrogen recombination lines is more significant at the low end of the abundance scale, as can be clearly seen in Figures \ref{fig9} or \ref{fig19}.

We have estimated the size of these three effects for $\kappa = 20$ using the ratio of the collisional excitation rates implied by the inferred electron temperatures given in Table \ref{table_6} for  $\kappa = 20$ and  $\kappa = \infty$; the Maxwell-Boltzmann case. This correction factor has to be further corrected by multiplying it with  the ratio of the forbidden line considered - in this case,  \OIIIl\ and [\ion{O}{2}] $\lambda \lambda 3727,9$ evaluated at  $\kappa = \infty$ and  $\kappa = 20$. In effect, we are assuming that the derived abundance scales as the chosen line ratio with respect to H$\beta$. The line strengths are drawn from Table \ref{table_4}.

\begin{figure}[htpb]
\includegraphics[scale=0.7]{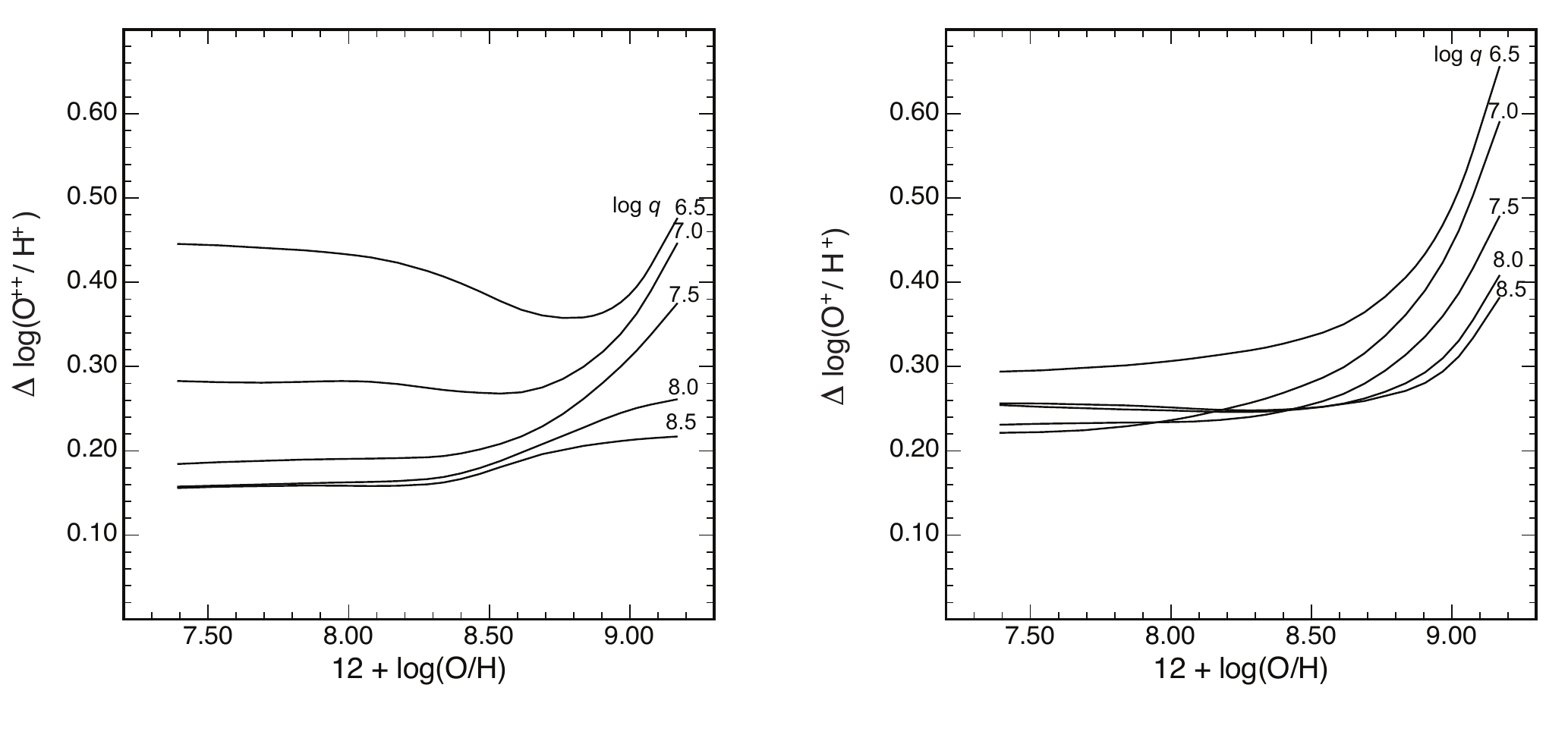}
\caption{The offset in abundance implied between the strong line techniques and the $T_e$ method for O$^{++}$/H$^+$ (left) and O$^{+}$/H$^+$ (right), for an assumed value of $\kappa = 20$. The $x$-axis is the true nebular abundance. The theoretical offset is close to the actual difference observed for \HII regions \citep{LopezSanchez12}, suggesting that $\kappa$-distributed electrons might be capable of resolving this long-standing abundance discrepancy problem.}\label{fig32}
\end{figure}

The result of these computations is shown in Figure \ref{fig32}, which gives the estimated offset between the model-based strong-line method and the $T_e$ method for the ionic ratios O$^{++}$/H$^+$ and O$^{+}$/H$^+$, which are fundamental for deriving O/H in the $T_e$ method. Typical offsets lie between 0.2 and 0.4 dex, which are very similar to the observed offset - see (for example) \citet{LopezSanchez12}, fig 12. We conclude that $\kappa$-distributed electrons may well provide the key to resolving the long-standing abundance discrepancy problem in \HII regions.
\FloatBarrier

\section{Conclusions}
In this paper we have investigated the consequences of the assumption of $\kappa$-distributed electrons rather than Maxwell-Boltzmann distributed electrons on strong line abundance diagnostics. These models also account for the impact of new atomic data on collisional excitation rates and transition probabilities, and the effect of the revised solar abundance scale \citep{Grevesse10}. 

We have treated $\kappa$ as a free variable in the grid of models presented here, so that observers can elect either to use or not to use $\kappa$. However, with a $\kappa \sim 20$, or somewhat larger, the observed offset between the recombination temperature of bright \HII regions and the electron temperatures inferred for both the high- and low-excitation zones can be explained. 

With  $\kappa \sim 20$, the UV lines of high-abundance, low-electron temperature \HII regions are predicted to be very strongly enhanced, whereas the effect of $\kappa$ on the mid- and far-IR lines is weak, ranging from $2-25$\%. Our models clearly have some issues in their predictions of the intensities of some of the mid-IR lines, which is likely to be due to our choice of low density, and zero age \citep{Snijders07}. 

For the strong lines at optical wavelengths, we have developed a new set of diagnostic diagrams which rely on the ratios of two forbidden lines rather than the ratio of a forbidden line to a recombination line of hydrogen, as has mostly been used hitherto. These new diagnostics cleanly separate the two parameters which principally determine the strong line emission spectrum: the chemical abundance set and the ionisation parameter. 

However, the derived abundance scale derived in this paper suffers from the weakness of relying on the ratio of [\ion{N}{2}] to either of the $\alpha$-process ions, [\ion{O}{2}]  or [\ion{S}{2}] . Thus, it is highly sensitive to how well the N/O vs. O/H calibration shown in Figure \ref{fig3} can be made. This relationship clearly has scatter, especially at the low abundance end, and the reasons for this have been discussed by many authors \citep{Matteucci85, Henry00, Contini02, Lopez-Sanchez10} - see the recent summary by \citet{Pilyugin11b}. For a given change in the N/O ratio at fixed O/H, the calibrations involving the [\ion{N}{2}]/[\ion{S}{2}] ratio will be more affected than those which depend upon the [\ion{N}{2}]/I [\ion{O}{2}] ratio, since the total range in the [\ion{N}{2}]/[\ion{S}{2}] ratio is more restricted than that of the [\ion{N}{2}]/[\ion{O}{2}] ratio. In addition, the ratio of the secondary nucleosynthetic production of nitrogen to the primary component is sensitive to the IMF, which may change between galaxies. Nonetheless, the fact that the derived abundance is monotonic with the abundance sensitive ratio used is a notable advantage compared to the use of the ratio of a forbidden line to a recombination line of hydrogen, which must always be a two-valued function of abundance. The latter ratios then have to be calibrated with an assumption of which solution branch applies, and furthermore there is a wide range of abundance over which the ratio of a forbidden line to a recombination line of hydrogen is insensitive to changes in the abundance.

The prime effect of the new atomic data and the self-consistency between the abundance set used in the stellar atmospheres and the abundance set used in the \HII region models is to produce, for the first time, a fully consistent solution for the nebular abundances and the nebular ionisation parameter between some half dozen strong-line diagnostics. This greatly increases confidence in their use, as well as in the N/O vs. O/H calibration used here.

$\kappa \sim 20 $ assists in resolving the long-standing abundance discrepancy between the strong-line and $T_e$ based techniques of deriving the nebular abundance. At the high abundance end,  $\kappa$ increases the  electron temperature measured from the ratio of two forbidden lines, which leads to the $T_e$ method delivering too low abundances. At the low abundance end, the effect of $\kappa$  is to decrease the forbidden lines relative to the recombination lines of hydrogen. This also will lead to the $T_e$ method delivering too low abundances. These effects seem, in principal, to account for all of the abundance discrepancy between the strong-line and $T_e  +$ ICF based techniques. This important point will be examined in greater detail in a future paper.

\bibliographystyle{aa}

\newpage
\begin{deluxetable}{lcc}
\tabletypesize{\small}
\tablecaption{Solar abundance set ($Z_{\odot}$) and the logarithmic depletion factors (D) adopted
for each element.\label{Z_table}}
\tablewidth{300pt}
\centering
\tablehead{
\colhead{Element}
& \colhead{$\log({\rm Z_{\odot}})$}
& \colhead{$\log({\rm D})$}\\
}
\startdata
H & 0.00 & 0.00 \\
He & -1.01 & 0.00 \\
C & -3.57 & -0.30 \\
N & -4.60 & -0.05 \\
O & -3.31 & -0.07 \\
Ne & -4.07 & 0.00 \\
Na & -5.75  & -1.00\\
Mg & -4.40 & -1.08 \\
Al & -5.55 & -1.39\\
Si & -4.49 & -0.81 \\
S & -4.86 & 0.00 \\
Cl & -6.63 & -1.00\\
Ar & -5.60 & 0.00 \\
Ca & -5.66 & -2.52 \\
Fe & -4.50 & -1.31 \\
Ni & -5.78 & -2.00\\
\enddata
\end{deluxetable}

\begin{deluxetable}{lcc}
\tabletypesize{\small}
\tablecaption{The C and N abundances as a function of O/H. (Solar: $12+ \log({\rm O/H})=8.69)$ \label{Z_CN}}
\tablewidth{300pt}
\centering
\tablehead{
 \colhead{$12+ \log({\rm O/H})$}
& \colhead{$\log({\rm N/H})$}
& \colhead{$\log({\rm C/H})$}\\
}
\startdata
7.39 &  -6.61& -5.58 \\
7.50 & -6.47 & -5.44 \\
7.69 & -6.23 & -5.20 \\	
7.99 & -5.79 & -4.76 \\
8.17 & -5.51 & -4.48 \\
8.39 & -5.14 & -4.11 \\
8.69 & -4.60 & -3.57 \\
8.80 & -4.40 & -3.37 \\
8.99 & -4.04 & -3.01 \\
9.17 & -3.67 & -2.64 \\
9.39 & -3.17 & -2.14 \\
\enddata
\end{deluxetable}

\begin{deluxetable}{ccccccccccccccc}
\rotate
\small
 \centering
\tabletypesize{\scriptsize}
\tablecaption{The `blue' line fluxes given by the MAPPINGS 4 models relative to $F_{\rm H\beta} = 1.0$. \label{table_4}}
\tablewidth{600pt}
\tablehead{
\colhead{} &\colhead{[O II]} & \colhead{[N II]} & \colhead{[Ne III]} & \colhead{[S II]} & \colhead{H$\gamma$} & \colhead{[O III]} & \colhead{He I} &  \colhead{H$\beta$} & \colhead{[O III]}& \colhead{[O III]}  & \colhead{HeI} & \colhead{[Ar III]} & \colhead{[N I]} & \colhead{[N II]}\\
\colhead{$\log(q)$} & \colhead{3727} & \colhead{3729} & \colhead{3869} & \colhead{4068,78} & \colhead{4340} & \colhead{4363} & \colhead{4471} & \colhead{4861} & \colhead{4959} & \colhead{5007} & \colhead{5016} &\colhead{5192} & \colhead{5198,00} &\colhead{5755}
}
\startdata
$5.0Z_{\odot}$ ; $\kappa=\infty$ \\
\hline
8.50 & 0.0105 & 0.0148 & 0.0002 & 0.0003 & 0.4447 & 0.0000 & 0.0941 & 1.0000 & 0.0021 & 0.0061 & 0.0529 & 0.0000 & 0.0000 & 0.0001 \\
8.25 & 0.0067 & 0.0095 & 0.0001 & 0.0002 & 0.4458 & 0.0000 & 0.0923 & 1.0000 & 0.0014 & 0.0039 & 0.0498 & 0.0000 & 0.0000 & 0.0000 \\
8.00 & 0.0048 & 0.0068 & 0.0001 & 0.0002 & 0.4470 & 0.0000 & 0.0907 & 1.0000 & 0.0010 & 0.0029 & 0.0472 & 0.0000 & 0.0000 & 0.0000 \\
7.75 & 0.0039 & 0.0055 & 0.0001 & 0.0002 & 0.4481 & 0.0000 & 0.0889 & 1.0000 & 0.0008 & 0.0023 & 0.0452 & 0.0000 & 0.0000 & 0.0000 \\
7.50 & 0.0030 & 0.0043 & 0.0001 & 0.0002 & 0.4488 & 0.0000 & 0.0873 & 1.0000 & 0.0006 & 0.0017 & 0.0438 & 0.0000 & 0.0000 & 0.0000 \\
7.25 & 0.0019 & 0.0027 & 0.0000 & 0.0001 & 0.4486 & 0.0000 & 0.0841 & 1.0000 & 0.0003 & 0.0010 & 0.0423 & 0.0000 & 0.0000 & 0.0000 \\
7.00 & 0.0008 & 0.0012 & 0.0000 & 0.0001 & 0.4472 & 0.0000 & 0.0793 & 1.0000 & 0.0001 & 0.0004 & 0.0407 & 0.0000 & 0.0000 & 0.0000 \\
\enddata
\end{deluxetable}

\begin{deluxetable}{ccccccccccccccc}
\rotate
\small
 \centering
\tabletypesize{\scriptsize}
\tablecaption{The `red' line fluxes given by the MAPPINGS 4 models relative to $F_{\rm H\beta} = 1.0$. \label{table_5}}
\tablewidth{600pt}
\tablehead{
\colhead{} &\colhead{He I} & \colhead{[O I]} & \colhead{[S III]} & \colhead{[N II]} & \colhead{H$\alpha$} & \colhead{[N II]} & \colhead{He I} &  \colhead{[S II]} & \colhead{[S II]}& \colhead{[Ar III]}  & \colhead{[O II]} & \colhead{[Ar III]} & \colhead{[S III]} & \colhead{[S III]}\\
\colhead{$\log(q)$} & \colhead{5875} & \colhead{6300} & \colhead{6312} & \colhead{6548} & \colhead{6563} & \colhead{6584} & \colhead{6678} & \colhead{6717} & \colhead{6731} & \colhead{7136} & \colhead{7318,30} &\colhead{7751} & \colhead{9068} &\colhead{9532}
}
\startdata
$5.0Z_{\odot}$ ; $\kappa=\infty$ \\
\hline
8.50 & 0.3218 & 0.0012 & 0.0001 & 0.0565 & 3.3420 & 0.1661 & 0.0880 & 0.0267 & 0.0188 & 0.0052 & 0.0000 & 0.0013 & 0.0559 & 0.1403 \\
8.25 & 0.3130 & 0.0010 & 0.0001 & 0.0472 & 3.3210 & 0.1387 & 0.0863 & 0.0259 & 0.0182 & 0.0044 & 0.0000 & 0.0011 & 0.0488 & 0.1227 \\
8.00 & 0.3037 & 0.0010 & 0.0000 & 0.0429 & 3.2950 & 0.1263 & 0.0844 & 0.0277 & 0.0194 & 0.0042 & 0.0000 & 0.0010 & 0.0457 & 0.1147 \\
7.75 & 0.2942 & 0.0011 & 0.0000 & 0.0422 & 3.2730 & 0.1243 & 0.0823 & 0.0322 & 0.0226 & 0.0045 & 0.0000 & 0.0011 & 0.0447 & 0.1122 \\
7.50 & 0.2865 & 0.0012 & 0.0000 & 0.0416 & 3.2580 & 0.1223 & 0.0804 & 0.0375 & 0.0263 & 0.0047 & 0.0000 & 0.0011 & 0.0422 & 0.1059 \\
7.25 & 0.2758 & 0.0011 & 0.0000 & 0.0355 & 3.2660 & 0.1044 & 0.0775 & 0.0378 & 0.0266 & 0.0040 & 0.0000 & 0.0010 & 0.0339 & 0.0851 \\
7.00 & 0.2620 & 0.0008 & 0.0000 & 0.0235 & 3.3000 & 0.0692 & 0.0733 & 0.0298 & 0.0210 & 0.0026 & 0.0000 & 0.0006 & 0.0209 & 0.0525 \\
\enddata
\end{deluxetable}

\begin{deluxetable}{ccccccccc}
 \centering
\tabletypesize{\scriptsize}
\tablecaption{The recombination and forbidden line temperatures of the MAPPINGS 4 models. \label{table_6}}
\tablewidth{370pt}
\tablehead{
\colhead{$\log(q)$} &\colhead{$T_{\rm H}$} & \colhead{$T_{\rm He}$} & \colhead{$T_{\rm [O III]}$} & \colhead{$T_{\rm [Ar III]}$} & \colhead{$T_{\rm [S III]}$} & \colhead{$T_{\rm [O II]}$} & \colhead{$T_{\rm [N II]}$} &  \colhead{$T_{\rm [S II]}$} \\
}
\startdata
$5.0Z_{\odot}$ ; $\kappa=\infty$ \\
\hline
8.50 & 2706 & 2506 & -- &  -- &  3814 & 3721 & 3953 & 3599 \\
8.25 & 2729 & 2562 & -- & -- & 3586 & 3508 & 3720 & 3423 \\
8.00 & 2799 & 2682 & -- & -- &  3407 & 3359 & 3538 & 3279 \\
7.75 & 2889 & 2824 & -- & -- & 3266 & 3007 & 3397 & 3168 \\
7.50 & 2950 & 2928 & -- & -- & 3133& 2987 & 3248 & 3052 \\
7.25 & 2918 & 2932 & -- & -- & 2977  & -- & 3051 & 2898 \\
7.00 & 2756 & 2807 & -- & -- & 2804 & -- & 2833 & 2717 \\
\enddata
\end{deluxetable}

\begin{table*}[htb!]
\caption{Line ratios diagnostics used with the \textbf{pyqz} Python module v0.2 and their associated regions of validity in  $\log Z = $12+$\log$(O/H) and $\log q$ space.}\label{table_7}
	\begin{center}
	\footnotesize{
		\begin{tabular}{l@{ vs. }lcccccccc}
			\hline
			\hline
			\multicolumn{2}{c}{Line Ratios}& \multicolumn{2}{c}{$\kappa=10$} & \multicolumn{2}{c}{$\kappa=20$} & \multicolumn{2}{c}{$\kappa=50$} & \multicolumn{2}{c}{$\kappa=\infty$}\\
			\multicolumn{2}{c}{$ $}&$\log Z$&$\log q$&$\log Z$& $\log q$&$\log Z$& $\log q$&$\log Z$& $\log q$\\
			\hline
			$$[NII]/[SII] & [OIII]/[SII]             &7.39--9.39& 6.5--8.5 & 7.39--9.39 & 6.5--8.5 & 7.39--9.39 & 6.5--8.5   & 7.39--9.39 & 6.5--8.5\\
			$$[NII]/[SII] & [OIII]/H$\beta$   &7.39--9.39& 6.5--8.5 & 7.39--9.39 & 6.5--8.5 & 7.39--9.39 & 6.5--8.25 & 7.39--9.39 &  6.5--8.5 \\
			$$[NII]/[SII] & [OIII]/[OII] &7.39--9.39&  6.5--8.5 & 7.39--9.39 & 6.5--8.5 & 7.39--9.39 & 6.5--8.5  & 7.39--8.99 &  6.5--8.5 \\
			$$[NII]/[OII] & [OIII]/[SII] &7.39--9.39 &  6.5--8.5  & 7.39--9.39 &  6.5--8.5  & 7.39--9.39 & 6.5--8.5  & 7.39--9.39 &  6.5--8.5 \\
			$$[NII]/[OII] & [OIII]/H$\beta$ &7.39--9.39 & 6.5--7.5 & 7.39--9.39 & 6.5--7.5 & 7.39--9.39 & 6.5--7.5 & 7.39--9.39 & 6.5--7.5\\
			$$[NII]/[OII] & [OIII]/[OII] &7.39--9.39 & 6.5--8.25 & 7.39--9.39 & 6.5--8.0 & 7.39--9.39 & 6.5--7.75 & 7.39--9.39 & 6.5--7.75\\
			$$[NII]/H$\alpha$ & [OIII]/H$\beta$ &7.39--8.69 &  6.5--8.5  & 7.39--8.39 &  6.5--8.5 & 7.39--8.39 &  6.5--8.25 & 7.39--8.39 &  6.5--8.5 \\
			$$[NII]/H$\alpha$ & [OIII]/[OII] &7.39--8.69 &  6.5--8.5 & 7.39--8.39 &  6.5--8.5  & 7.39--8.17 &  6.5--8.25 & 7.39--7.99 &  6.5--8.5 \\
			\hline
\end{tabular}
}
\end{center}
\end{table*}

\end{document}